\documentclass[aps,prl,twocolumn,superscriptaddress,etalmode=truncate,etaldisplay=10]{revtex4-2}
\usepackage[english]{babel}
\usepackage[utf8]{inputenc}
\usepackage[colorinlistoftodos, color=green!40, prependcaption]{todonotes}
\usepackage{amsthm}
\usepackage{mathtools}
\usepackage{physics}
\usepackage{xcolor}
\usepackage{graphicx}
\usepackage[left=23mm,right=13mm,top=35mm,columnsep=15pt]{geometry} 
\usepackage{adjustbox}
\usepackage{placeins}
\usepackage[T1]{fontenc}
\usepackage{lipsum}
\usepackage{csquotes}
\usepackage{siunitx}
\usepackage{booktabs}

\usepackage{xr}

\usepackage[pdftex, pdftitle={Article}, pdfauthor={Author}]{hyperref} 

\begin{document}

\title{Strong coupling between a dielectric nanocavity and a monolayer transition metal dichalcogenide}

\author{F. Schröder}
    \affiliation{Department of Electrical and Photonics Engineering, Technical University of Denmark, {\O}rsteds Plads 343, 2800 Kgs. Lyngby, Denmark}
    \affiliation{NanoPhoton - Center for Nanophotonics, Technical University of Denmark, {\O}rsteds Plads 345A, 2800 Kgs. Lyngby, Denmark}
\author{P. Wyborski}
    \affiliation{Department of Electrical and Photonics Engineering, Technical University of Denmark, {\O}rsteds Plads 343, 2800 Kgs. Lyngby, Denmark}    
\author{M. Xiong}
    \affiliation{Department of Electrical and Photonics Engineering, Technical University of Denmark, {\O}rsteds Plads 343, 2800 Kgs. Lyngby, Denmark}
    \affiliation{NanoPhoton - Center for Nanophotonics, Technical University of Denmark, {\O}rsteds Plads 345A, 2800 Kgs. Lyngby, Denmark}
\author{G. Kountouris}
    \affiliation{Department of Electrical and Photonics Engineering, Technical University of Denmark, {\O}rsteds Plads 343, 2800 Kgs. Lyngby, Denmark}
    \affiliation{NanoPhoton - Center for Nanophotonics, Technical University of Denmark, {\O}rsteds Plads 345A, 2800 Kgs. Lyngby, Denmark}
\author{B. Munkhbat}
    \affiliation{Department of Electrical and Photonics Engineering, Technical University of Denmark, {\O}rsteds Plads 343, 2800 Kgs. Lyngby, Denmark} 
\author{M. Wubs}
    \affiliation{Department of Electrical and Photonics Engineering, Technical University of Denmark, {\O}rsteds Plads 343, 2800 Kgs. Lyngby, Denmark}
    \affiliation{NanoPhoton - Center for Nanophotonics, Technical University of Denmark, {\O}rsteds Plads 345A, 2800 Kgs. Lyngby, Denmark}
\author{P. T. Kristensen}
    \affiliation{Department of Electrical and Photonics Engineering, Technical University of Denmark, {\O}rsteds Plads 343, 2800 Kgs. Lyngby, Denmark}
    \affiliation{NanoPhoton - Center for Nanophotonics, Technical University of Denmark, {\O}rsteds Plads 345A, 2800 Kgs. Lyngby, Denmark}
\author{J. M{\o}rk}
    \affiliation{Department of Electrical and Photonics Engineering, Technical University of Denmark, {\O}rsteds Plads 343, 2800 Kgs. Lyngby, Denmark}
    \affiliation{NanoPhoton - Center for Nanophotonics, Technical University of Denmark, {\O}rsteds Plads 345A, 2800 Kgs. Lyngby, Denmark}
\author{N. Stenger}
    \email[Correspondence email address: ]{niste@dtu.dk}
    \affiliation{Department of Electrical and Photonics Engineering, Technical University of Denmark, {\O}rsteds Plads 343, 2800 Kgs. Lyngby, Denmark}
    \affiliation{NanoPhoton - Center for Nanophotonics, Technical University of Denmark, {\O}rsteds Plads 345A, 2800 Kgs. Lyngby, Denmark}
    
    
\date{\today} 

\begin{abstract}
We demonstrate strong coupling between light in a dielectric nanocavity with deep sub-wavelength confinement and excitons in a monolayer of molybdenum ditelluride.  
Avoided crossing is demonstrated by both photoluminescence and reflection measurements, from which we extract a light-matter interaction strength of $g_{\mathrm{PL}} =\SI{5.3\pm0.3}{\milli\eV}$ and $g_{\mathrm{R}} =\SI{4.7\pm0.7}{\milli\eV}$, respectively. The associated Rabi splitting is twice as large as the system's losses. These values are in good agreement with values \textcolor{black}{obtained by a novel exciton reaction} coordinate formalism, 
yielding \textcolor{black}{$g_{\mathrm{theory}} = \SI{5.2\pm0.7}{\milli\eV}$}. The strong light-matter interaction, combined with low losses and sub-wavelength confinement of light, \textcolor{black}{demonstrates a new regime of light-matter interactions where strong nonlinearities at the single-photon level are expected}. 
\end{abstract}


\maketitle
Monolayer (ML) transition-metal dichalcogenides (TMDCs), consisting of one molybdenum atom and two chalcogen atoms per unit cell, are direct bandgap semiconductors in the 2H phase~\cite{Splendiani2010, Mak2010}. Their unique excitons, featuring high binding energy~\cite{He2014, Chernikov2014b}, high absorption coefficient~\cite{Li2014a, Munkhbat2022}, and large oscillator strength~\cite{Robert2016a, Robert2016}, make them compelling platforms for exploring strong light-matter interactions~\cite{Schneider2018}. Strong coupling occurs when the energy exchange between light and excitons is fast enough to surpass the system's decay rate, resulting in Rabi oscillations and hybridized polariton states~\cite{Tormo2015, Goncalves2020, Tserkezis2020}. 
\textcolor{black}{The regime of strong light-matter interaction has been demonstrated with excitons in ML TMDCs coupled to optical microcavities, where light is confined on length scales similar to or larger than the wavelength of the light~\cite{Liu2015a,Dufferwiel2015,Sidler2017,Gogna,Lackner2021,Shan2022}. As these structures are based on dielectric materials, cavity-induced losses are small, and cavity linewidths are typically on the order of $\SI{0.1}{\milli\eV}$ to $\SI{10}{\milli\eV}$. Other ways to achieve strong light-matter interactions with TMDCs and periodically-patterned dielectric structures are bound states in the continuum~\cite{Qin2022,Zheng2023,Maggiolini2023} and meta-surfaces~\cite{Danielsen2025a, Sortino}.} Similarly, TMDCs integrated with plasmonic structures, such as surface plasmon-polaritons~\cite{B.Iyer2022, Casses2024} and plasmonic nanocavities~\cite{Wen2017, Stuhrenberg2018, Geisler2019, Munkhbat2020}, can exhibit strong light-matter interactions. In the latter case, light is confined on deep sub-wavelength length scales, but the confinement is accompanied by significant metal-induced losses \textcolor{black}{and plasmonic linewidths typically exceed $\SI{100}{\milli\eV}$}.

Integrating dielectric cavities, such as photonic crystal cavities~\cite{Painter1999, Akahane2003} and nanobeam cavities~\cite{Ota2018} with ML molybdenum ditelluride (MoTe$_2$) is particularly promising due to its excitonic transition energy in the near-infrared~\cite{Ruppert2014}. 
This has led, for example, to weak coupling and enhanced spontaneous emission~\cite{Fang2019}. While recent advances have explored the light-matter interactions of ML TMDCs with dielectric nanobeam cavities~\cite{Rosser2022, Qian2022}, strong coupling in these systems has not yet been achieved. 

Recently, confinement of light on sub-wavelength scales using dielectric materials was realized~\cite{Robinson2005, Hu2016, Choi2017, Hu2018, Babar2023,Albrechtsen2022, Xiong2024}. Inverse design by topology-optimization has been demonstrated as a versatile tool for designing these cavities with extreme dielectric confinement (EDC)~\cite{Jensen2011,Molesky2018,Wang2018b}, and topology-optimized EDC cavities have been experimentally realized in silicon~\cite{Albrechtsen2022} and indium phosphide (InP)~\cite{Xiong2024}. The sub-wavelength confinement, which was previously achievable only with plasmonic structures~\cite{Wang2006,Naik2013,Khurgin2015}, comes without the associated ohmic and absorption losses intrinsic to plasmonic materials and therefore enables intriguing possibilities~\cite{Gurrieri2024,Abutoama2024,Dong2024,Kountouris2024}. In particular, a recent study demonstrated lasing from a quantum well embedded in such an EDC cavity~\cite{Xiong2024a}. 

\begin{figure*}[ht!]
    \centering
    \includegraphics[width=\linewidth]{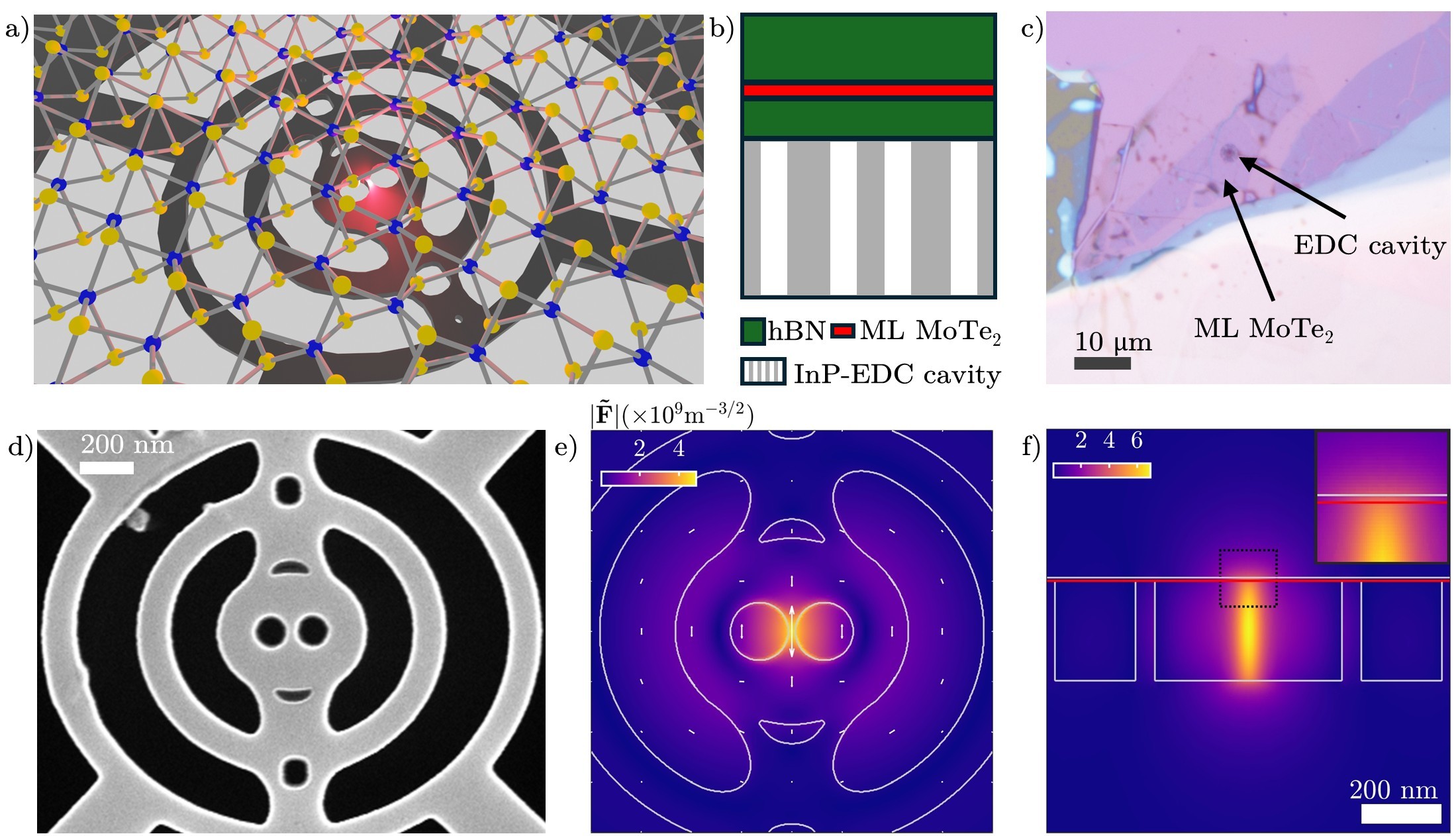}
    \caption{a) Artistic representation of the system. Light is confined in an EDC cavity. The ML MoTe$_2$ lies on top of the EDC cavity ~\cite{Momma2011, Jain2013}. The hBN flakes are not depicted. b) Sketch of the side view of the structure. c) Microscope image of the hBN-encapsulated ML MoTe$_2$ on the EDC cavity. d) SEM image of the passive cavity. Simulated electric-field distribution of the structure in e) the $xy$-plane at $z = z_{\mathrm{2D}}$ and f) the $yz$-plane at $x = 0$. Contour lines show the topography of the cavity and the hBN layer. Arrows in (e) depict the in-plane polarization of the electric-field. The red line in (f) depicts the ML MoTe$_2$.}
    \label{fig:sample}
\end{figure*}

\textcolor{black}{Similarly, the integration of EDC cavities with excitons in monolayer TMDCs unlocks new possibilities for exciton dynamics and device applications. 
Exciton-polaritons confined by a cavity mode to a single, sub-wavelength-scale region have been shown to exhibit strong exciton-exciton interactions~\cite{Besga2015, Munoz-Matutano2019, Denning2022a}. The combination of near-plasmonic sub-wavelength confinement, low-loss dielectric cavities, and the strong oscillator strength of excitons in monolayer TMDCs enables a novel regime for exploring strong light-matter interactions. Studying polaritons in this regime is of significant interest, as a nonlinearity in an interacting system yields photon antibunching~\cite{Bamba2011, Ryou2018}.
In particular, recent theoretical work~\cite{Denning2022} has shown that enhanced quantum nonlinearities at the single-photon level could be achieved, enabling a polariton blockade~\cite{Kyriienko2020, Delteil2019a, Denning2022}.
}
Moreover, the spin-valley selection rules of ML TMDCs support the study of chiral light-matter interactions~\cite{Xiao2012, Mak2012a, Yang2015a, Fong2021}, while the system also serves as a promising platform for exploring condensation effects~\cite{Gurrieri2024}. 
Unlike plasmonic nanostructures, dielectric cavities integrate seamlessly with existing technology~\textcolor{black}{\cite{Cheben2018, Wang2020, Dong2024, Babar2023}}, offering significant scalability and application benefits.

\textcolor{black}{In this work, we experimentally demonstrate strong coupling between an InP-based EDC cavity with sub-wavelength confinement and excitons in ML MoTe$_2$.} The avoided crossing of the system is demonstrated with temperature-resolved photoluminescence (PL) and reflection measurements, in which the excitonic transition and the cavity mode are resonant at $T = \SI{40}{\kelvin}$. The extracted light-matter interaction strength is approximately $\SI{5}{\milli\eV}$. 
\textcolor{black}{The extracted cavity linewidth is $\SI{3.3\pm0.1}{\milli\eV}$, more than an order of magnitude smaller than typical plasmonic linewidths, while a lateral field confinement length of $\sigma\approx \SI{70}{\nm}$ is maintained, more than a factor of two smaller than $\lambda/(2n)$.}
The obtained Rabi-splitting is twice as large as the losses of the system, and the experimentally obtained value is in agreement with numerical simulations based on the reaction-coordinate formalism~\cite{Denning2022a}. \textcolor{black}{This marks an important step towards the observation of exciton-exciton interactions and single-photon nonlinearities~\cite{Denning2022} in integrable photonic devices operating close to the O-band.}


An artistic representation of the ML MoTe$_2$ on an EDC cavity is depicted in Fig.~\ref{fig:sample}(a). 
ML MoTe$_2$ is sandwiched in hexagonal Boron Nitride (hBN) to prevent degradation of the monolayer~\cite{Chen2015} and to reduce the linewidth~\cite{Ajayi2017, Cadiz2017, Han2018, Fang2019}. The resulting hBN/MoTe$_2$/hBN heterostructure is placed on the EDC cavity, cf. Fig.~\ref{fig:sample}(b). The MoTe$_2$ layer is separated by \textcolor{black}{$\SI{1.5\pm1.3}{\nano\metre}$} from the surface of the structured InP, and we refer to the $z$ position of the ML MoTe$_2$ as $z_{\mathrm{2D}}$. Please see Sec.~\ref{sec:SI fabrication} in the Supplementary Information for details on the 2D material fabrication process. A microscope image of the resulting sample is depicted in Fig.~\ref{fig:sample}(c). Due to the large extent of the ML MoTe$_2$ compared to the cavity mode, the excitons are considered delocalized in the plane but strictly confined in the $z$-direction~\cite{Denning2022a}. Nevertheless, the exciton-polaritons are laterally confined due to interaction with the sub-wavelength cavity mode.

\textcolor{black}{The EDC cavity design results from a simplified version of a topology-optimized (TO) structure, where the TO structure is approximated by only ellipses and tangents. This smoothes the topography while maintaining the overall structure of the original TO design, which has been shown to successfully retain the properties of the cavity~\cite{Kountouris2022}.
A scanning electron microscope (SEM) image of a similar cavity on the same chip can be found in Fig.~\ref{fig:sample}(d).
The outer rings of the cavity increase the $Q$ factor by suppressing radiative losses, similar to circular Bragg reflectors, while the holes in the center provide extreme dielectric confinement on sub-wavelength scales.
The confinement is largely determined by the spacing of the void regions in the cavity center.
A void spacing of $\SI{20}{\nano\metre}$ was targeted, safely above what has been determined as a fabrication limitation~\cite{Xiong2024} while yielding a sub-wavelength field confinement (see below).
The previous design targeted a resonance in the Telecom-C band. 
To acquire a design where the cavity resonance energy matches the excitonic transition of the A-exciton in ML MoTe$_2$ at cryogenic temperatures, the design was therefore scaled appropriately. 
The EDC cavity is fabricated in a $\SI{245}{\nano\metre}$ thick InP membrane~\cite{Xiong2024}. This material system is chosen because the excitonic transition in MoTe$_2$ is below the bandgap of InP. Therefore, the system is not limited by absorption losses of the dielectric material. 
}

The dominant response of the optical nanocavity can be accurately described by a single quasinormal mode (QNM)~\cite{Ching1998, Muljarov2010, Kristensen2013, Lalanne2018, Kristensen2020, Both2022} with complex eigenenergy of the form $\tilde{E}_{\mathrm{cav}} = E_{\mathrm{cav}}-i \Gamma_{\mathrm{cav}}/2$, in which $E_{\mathrm{cav}}$ denotes center position and $\Gamma_{\mathrm{cav}}$ is the full width at half maximum (FWHM) of the emission spectrum in the single-mode approximation. From numerical calculations, as detailed in the Supplementary Information, we find that the structure supports a QNM with nominal resonance energy of $E_{\mathrm{cav}} = \SI{1.187}{\eV}$, a nominal quality $Q$ factor of \textcolor{black}{$Q = 571(7)$}, and an effective mode volume of \textcolor{black}{$V_{\mathrm{eff}} = 0.060(2)(\lambda/n)^3$}. We note, however, that dielectric nanocavities often exhibit significantly lower $Q$ factors than predicted by simulations~\cite{Albrechtsen2022, Xiong2024}\textcolor{black}{, and we extract an experimental $Q$ factor of $358(11)$ (see Sec.~\ref{sec:SI fabrication} in the Supplementary Information). This reduction is primarily attributed to fabrication imperfections, such as surface roughness and geometric deviations~\cite{Dimopoulos2022}. Owing to their extreme field confinement, EDC cavities exhibit pronounced sensitivity to such imperfections~\cite{Xiong2024}. Moreover,} the resonance frequency is highly sensitive to small variations of the geometry~\cite{Kountouris2022}. Fig.~\ref{fig:sample}(e) and (f) show the magnitude of the normalized field profile $|\mathbf{\tilde{F}}(\mathbf{r})|$ in the $xy$ plane at $z_{\mathrm{2D}}$ and in the $yz$ plane at $x = 0$, respectively. The arrows in Fig.~\ref{fig:sample}(e) depict $\Re[{F_{x,y}(\mathbf{r})}]$, highlighting the linear polarization of the cavity $\hat{E}_{\mathrm{cav}}$ along the $y$ axis~\cite{Kountouris2022, Schroder2025a}. \textcolor{black}{The lateral confinement length of the field in the MoTe$_2$ plane is estimated as $\sigma \approx \SI{70}{\nano\metre}$ (see Sec.~\ref{sec:SI Simulation} in the Supplementary Information).}
The structure also supports another mode with orthogonal polarization, lower resonance energy, and reduced quality factor—rigorously analyzed in Ref.~\cite{Schroder2025a} and referred to as the low-$Q$ mode.

For excitons in ML TMDCs coupled to dielectric nanocavities, the light-matter interaction strength $g$ can be calculated using the QNMs in combination with material-specific properties of the ML TMDC~\cite{Carlson2021,Denning2022a}. In this study, we follow Ref.~\cite{Denning2022a} and describe the light-matter interaction in a reaction-coordinate formalism, in which $g_{\mathrm{theory}}$ can be expressed as
\begin{equation}
    g_{\mathrm{theory}} = \sqrt{\dfrac{\hbar^2 e_0^2}{\pi \epsilon_0 m_0^2 E_{\mathrm{cav}} a^2_{\mathrm{B}}}\sum_\alpha\int d^2\mathbf{r}\left|\mathbf{\tilde{F}}(\mathbf{r}_{xy}, z_{\mathrm{2D}})\cdot\mathbf{p}^{\alpha}_{\mathrm{cv}}\right|^2 },
    \label{eq:Light matter coupling}
\end{equation}
where $\mathbf{p}^{\alpha}_{\mathrm{cv}}$ denotes the transition dipole moment of the ML TMDC, $\alpha$ denotes the valley index $K, K'$~\cite{Xiao2012}, and $a_{\mathrm{B}}$ denotes the excitonic Bohr radius of the A-exciton. 
\textcolor{black}{The relevant material parameters are extracted from Refs.~\cite{Pettit1965, McCaulley1994, Ruppert2014, Gerber2018, Meckbach2020, Edalati-Boostan2020,Grudinin2023}.}
The integral is performed in the $xy$ plane of the ML MoTe$_2$, cf. Fig.~\ref{fig:sample}(e), as indicated with the red line in Fig.~\ref{fig:sample}(f). From Eq.~\ref{eq:Light matter coupling} it follows that the light-matter interaction strength strongly depends on the polarization of the cavity mode. If we consider a linearly-polarized cavity mode with $\mathbf{\tilde{F}}(\mathbf{r}_{xy}, z_0) = F_0(\mathbf{r}_{xy}, z_0) \hat{e}_{y}$, then $g$ reaches its maximal value of $\mathbf{p}^{\alpha}_{\mathrm{cv}} \parallel \hat{e}_y$ and vanishes for $\mathbf{p}^{\alpha}_{\mathrm{cv}} \perp \hat{e}_y$. Eq.~\ref{eq:Light matter coupling} is solved numerically, yielding \textcolor{black}{$g_{\mathrm{theory}} = \SI{5.2\pm0.7}{\milli\eV}$}, see Sec.~\ref{sec:SI Simulation} in the Supplementary Information for details on the numerical calculations.

\begin{figure*}[ht!]
  \includegraphics[width=\textwidth]{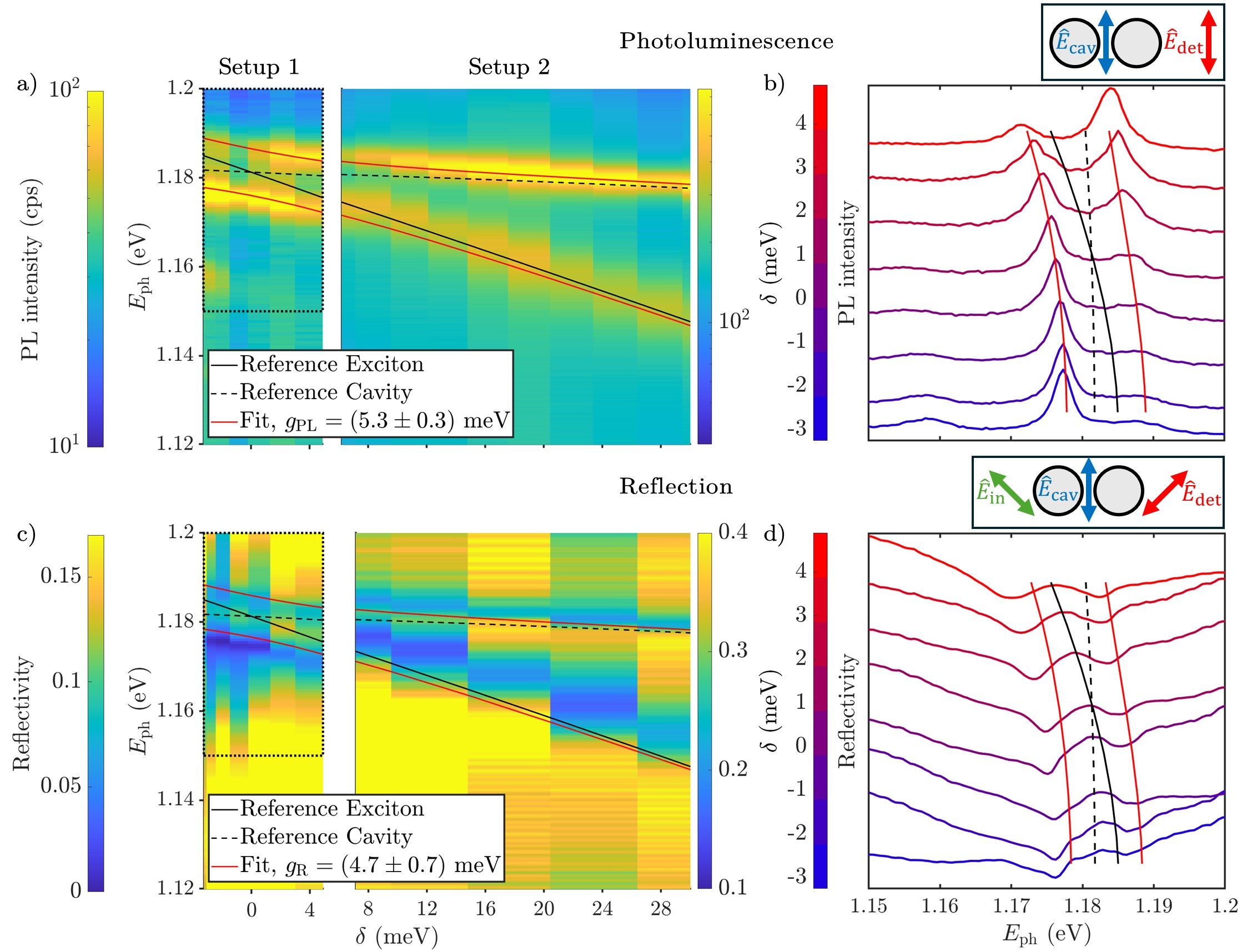}
  \caption{a) and b) PL spectra as a function of detuning, recorded with the detection polarization aligned with the cavity mode. The dotted box in (a) depicts the region of the spectra depicted in (b). The solid and dashed black line depicts the reference for the exciton and for the cavity mode, respectively, see Sec.~\ref{sec:SI reference} in the Supplementary Information. c) and d) Reflectivity spectra recorded in a cross-polarization configuration. Setups 1 and 2 are explained in the main text.}
  \label{fig:PL spectra heatmap peak}
\end{figure*}

We investigate PL spectra as a function of temperature $T$ with the detection polarization $\hat{E}_{\mathrm{det}}$ oriented parallel to $\hat{E}_{\mathrm{cav}}$. 
Temperature variations primarily affect the excitonic emission energy $E_{\mathrm{exc}}$ through the bandgap temperature dependence and modify the exciton linewidth $\Gamma_{\mathrm{exc}}$ due to phonon interactions. \textcolor{black}{At $T = \SI{4}{\kelvin}$, we extract $\Gamma_{\mathrm{exc}} = \SI{5.1\pm0.9}{\milli\eV}$, in good agreement with Ref.~\cite{Kutrowska-Girzycka2022}}. 
In contrast, the cavity resonance energy $E_{\mathrm{cav}}$ is influenced by the temperature dependence of the refractive index, but its variation is much smaller than the changes in $E_{\mathrm{exc}}$. 
\textcolor{black}{The cavity linewidth is determined as $\Gamma_{\mathrm{cav}} = \SI{3.3\pm0.1}{\milli\eV}$.}
We explore these dependencies systematically in order to calculate the detuning $\delta = E_{\mathrm{cav}} - E_{\mathrm{exc}}$, see Sec.~\ref{sec:SI reference} in the Supplementary Information for details. Fig.~\ref{fig:PL spectra heatmap peak}(a) shows the PL as a function of detuning in a range from $T=\SI{4}{\kelvin}$ to $T=\SI{163}{\kelvin}$. In addition, Fig.~\ref{fig:PL spectra heatmap peak}(b) shows the associated line plots 
in a region of interest close to zero detuning, as indicated in panel (a). In both panels (a) and (b), fits from reference measurements of the uncoupled cavity and of the uncoupled excitons are shown as black dashed and solid lines, respectively. We go from negative detuning to positive values in the measured temperature range, with $T = \SI{40}{\kelvin}$ corresponding to $\delta = \SI{-0.2 \pm 0.7}{\milli\eV}$. Clearly, avoided crossing is observed, which is a hallmark of strong light-matter coupling.

We use two different setups for different temperature regimes as indicated in Fig.~\ref{fig:PL spectra heatmap peak}(a), see Appendix for details. Setup 1 contains a liquid-helium cryostat for $T \leq \SI{70}{\kelvin}$, corresponding to $\delta \leq \SI{4.9\pm0.7}{\milli\eV} $. Setup 2 contains a liquid-nitrogen cryostat for $T \geq \SI{79}{\kelvin}$, corresponding to $\delta \geq \SI{6.2\pm1.0}{\milli\eV} $. \textcolor{black}{ 
A small spectral offset is often observed when comparing data acquired with different setups due to imperfect alignment and calibration of the spectrometers. We carefully compare the reference measurements acquired with both setups (cf. Sec. \ref{sec:SI reference} in the Supplementary Information) and deduce an offset of approximately $\SI{3}{\milli\eV}$ between the data from Setups 1 and 2; see Fig.~\ref{fig:off cav labels} in the Appendix.} The data from Setup 2 in Fig.~\ref{fig:PL spectra heatmap peak}(a) and (c) have been corrected accordingly. \textcolor{black}{We note that Setup 1 provides a range from $\delta = \SI{-3.3\pm0.7}{\milli\eV}$ to $\delta = \SI{4.9\pm0.7}{\milli\eV}$, and therefore the data from Setup 1 alone already provide evidence of avoided crossing.}
To assess the light-matter interaction strength, we fit the extracted peak positions of the lower ($E_-$) and of the upper ($E_+$) polariton with the real part of a coupled-oscillator model, see Sec.~\ref{sec:SI Fits 40K} and~\ref{sec:SI COM fit} in the Supplementary Information for details. The fit result, indicated as the red lines in Fig.~\ref{fig:PL spectra heatmap peak}(a) and (b), yields $g_{\mathrm{PL}} = \SI{5.3\pm 0.3}{\milli\eV}$. \textcolor{black}{This value is on the same order as obtained for a ML TMDC on a nanobeam cavity~\cite{Qian2022, Rosser2022}, despite the much tighter light confinement in this work. This is in line with the prediction that the light-matter interaction strength is largely determined by the out-of-plane confinement~\cite{Denning2022a} and not by the mode volume, as for example for quantum dots in microcavities~\cite{Andreani1999}. 
However, as shown in Ref.~\cite{Denning2022}, exciton-exciton interactions are largely determined by the lateral confinement, highlighting the potential of the reported results for future studies on nonlinearities.
}

The PL measurements are supplemented with reflection measurements in a cross-polarization configuration, see Appendix 
for details~\cite{Schroder2025a}. Fig.~\ref{fig:PL spectra heatmap peak}(b) and (c) depict reflection spectra as a function of $\delta$. As observed from PL measurements, the avoided crossing is clearly visible as two dips in the reflection spectra, confirming hybridization of excitons and photons. A fit of the reflection measurements with the coupled-oscillator model yields $g_{\mathrm{R}} = \SI{4.7\pm 0.7}{\milli\eV}$, which agrees well with the value obtained from PL measurements. Notably, the experimentally obtained values $g_{\mathrm{R}}$ and $g_{\mathrm{PL}}$ are in excellent agreement with the value obtained with the exciton-reaction coordinate formalism \textcolor{black}{$g_{\mathrm{theory}} = \SI{5.2\pm0.7}{\milli\eV}$}, as discussed above.

Strong coupling is usually identified when the number of oscillations in the polariton states $N_{\mathrm{Rabi}}$ exceeds one~\cite{Todisco2020}. This is achieved when~\cite{Tormo2015, Geisler2019, Goncalves2020}
\begin{eqnarray}
N_{\mathrm{Rabi}} = \dfrac{2E_{\mathrm{Rabi}}}{\Gamma_{\mathrm{exc}}+\Gamma_{\mathrm{cav}}} \geq 1 \Leftrightarrow  E_{\mathrm{Rabi}} \geq \dfrac{\Gamma_{\mathrm{exc}} +\Gamma_{\mathrm{cav}}}{2},
\end{eqnarray}
with $E_{\mathrm{Rabi}} = \sqrt{4g^2-(\Gamma_{\mathrm{cav}}-\Gamma_{\mathrm{exc}})^2/4}$ denoting the Rabi splitting. To assess the Rabi splitting, the cavity and exciton linewidth are evaluated at $T = \SI{40}{\kelvin}$/ $\delta = \SI{-0.2 \pm 0.7}{\milli\eV}$, yielding $\Gamma_{\mathrm{cav}} = \SI{3.3\pm0.1}{\milli\eV}$ and $\Gamma_{\mathrm{exc}} = \SI{6.0\pm0.7}{\milli\eV}$ see Sec.~\ref{sec:SI Fits 40K} in the Supplementary information for details. Together with the aforementioned values for $g$, we find $E_{\mathrm{Rabi}} = \SI{10.6\pm0.7}{\milli\eV}$ and $\SI{9.4\pm1.5}{\milli\eV}$ as deduced from PL and reflection measurements, respectively. This value agrees well with the energy splitting of the polaritonic emission determined from the peak positions at $\delta = \SI{-0.2\pm0.7}{\milli\eV}$, $E_+ - E_- = \SI{10.1 \pm 0.3}{\milli\eV}$ and $\SI{9.4 \pm 0.7}{\milli\eV}$ for PL and reflection measurements, respectively. As $ E_{\mathrm{Rabi}} > \left(\Gamma_{\mathrm{exc}} +\Gamma_{\mathrm{cav}} \right)/2 = \SI{4.7 \pm 0.4}{\milli\eV}$, $E_{\mathrm{Rabi}}$ overcomes the averaged losses in the system by more than a factor of two. We deduce $N_{\mathrm{Rabi}} = \SI{2.3\pm0.1}{}$ and $\SI{2.0\pm0.2}{}$ for PL and reflection measurements, respectively, and we conclude that the system is clearly in the strong-coupling regime. 

\begin{figure}[t!]
  \includegraphics[width=\linewidth]{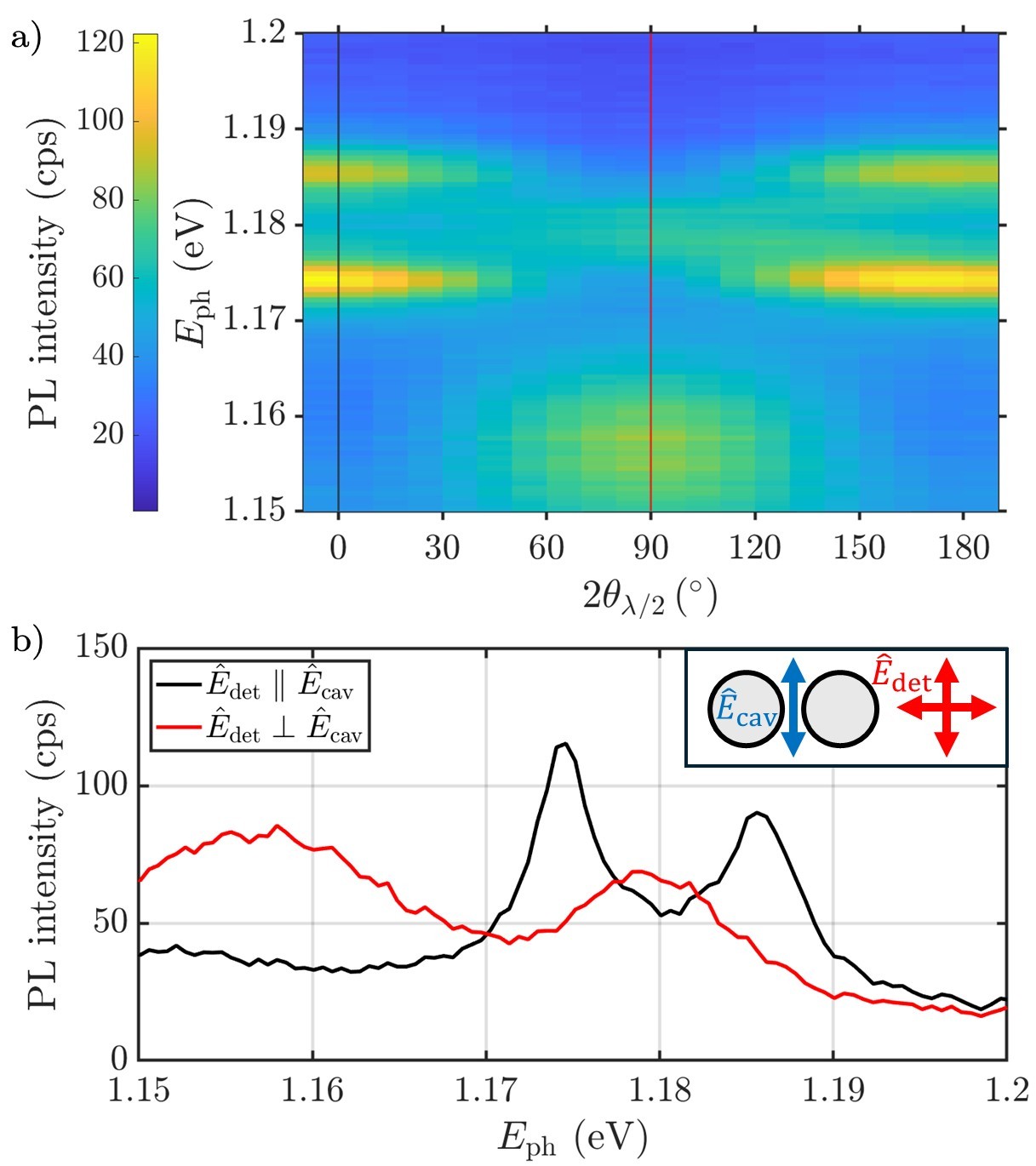}
  \caption{a) PL spectra at $T = \SI{50}{\kelvin}$ as a function of $\theta_{\lambda/2}$ in front of the analyzer. b) PL spectra from (a) with the analyzer oriented parallel and perpendicular to the cavity mode for black and red lines, respectively.}
  \label{fig:Pol spectra series}
\end{figure}

We note that in a coupled-oscillator model, the light-matter interaction strength is in general complex-valued~\cite{Carlson2021, Abutoama2024, Bleu2024}. To evaluate the magnitude of the complex part of the light-matter interaction strength, we fit the extracted peak positions and the linewidths, corresponding to the real and the imaginary parts of the eigenvalues, respectively, see Sec.~\ref{sec:SI COM fit} in the Supplementary Information for details. We find that the coupled-oscillator model also fits the linewidths accurately, which is additional confirmation of the reported strong light-matter interaction. Moreover, we find that the imaginary part of the light-matter interaction strength is less than $\SI{6}{\percent}$ of the real part and is almost zero within the error bars.

Finally, we turn our attention to the polarization properties of the photoluminescence signal. 
As the cavity mode is linearly polarized, orienting the detection polarization parallel to the cavity mode polarization enables the study of the strongly coupled system. 
On the other hand, by orienting the detection polarization perpendicular to the cavity mode, primarily residual excitons, that do not couple to the cavity mode are detected.
Those are excited in the ML MoTe$_2$ in the vicinity of the cavity mode.
The emission is studied at $T = \SI{50}{\kelvin}$, corresponding to $\delta = \SI{1.4 \pm 0.7}{\milli\eV}$.
Fig.~\ref{fig:Pol spectra series} show the PL spectra as a function of the angle of the halfwave ($\lambda/2$) plate $\theta_{\lambda/2}$ in front of the analyzer.
Together with the analyzer, rotating the $\lambda/2$ plate by $\theta_{\lambda/2}$ effectively rotates the detection polarization state by twice that value. 
Two peaks, separated by $\SI{11}{\milli\eV}$ and associated with the polaritonic emission, are identified for $\hat{E}_{\mathrm{cav}} \parallel \hat{E}_{\mathrm{det}}$, corresponding to $2\theta_{\lambda/2} = \SI{0}{\degree}$ and $\SI{180}{\degree}$, see Fig.~\ref{fig:Pol spectra series}.
Choosing $2\theta_{\lambda/2} = \SI{90}{\degree}$ yields a detection polarization perpendicular to the orientation of the cavity mode.
One peak centered between the two polaritonic peaks is detected, corresponding to emission from residual excitons that do not couple to the cavity mode.
Another peak is observed at $E_{\mathrm{ph}} = \SI{1.157}{\eV}$ for $\hat{E}_{\mathrm{det}} \perp \hat{E}_{\mathrm{cav}}$. 
This is well explained by another mode in the cavity. 
In Ref.~\cite{Schroder2025a}, we demonstrate that in addition to the mode of interest in this work, the EDC cavity supports a mode with slightly lower eigenenergy, lower $Q$ factor, and orthogonal polarization. The excitation of this low-$Q$ mode could be due to trions, which are known to exist at a lower transition energy than excitons in the ML MoTe$_2$~\cite{Helmrich2018}.

In summary, we have demonstrated strong coupling between excitons in a ML MoTe$_2$ and 
light confined by an EDC cavity on sub-wavelength length scales. The light-matter interaction strength is deduced both from photoluminescence and reflection measurements, yielding $g_{\mathrm{PL}} = \SI{5.3\pm0.3}{\milli\eV}$ and $g_{\mathrm{R}} = \SI{4.7\pm0.7}{\milli\eV}$. 
\textcolor{black}{
These values are consistent with our numerical calculation exploiting the exciton-reaction coordinate-formalism~\cite{Denning2022a}, for which we find the value $g_{\mathrm{theory}} = \SI{5.2\pm0.7}{\milli\eV}$.
Polarization-projection of the fluorescence signal confirms the polaritonic nature.
This reported system serves as an ideal testbed for studying nanoscale confined light-matter interactions.
It combines the small losses of dielectric nanocavities ($\Gamma_{\mathrm{cav}} = \SI{3.3\pm0.1}{\milli\eV}$) and sub-wavelength lateral field confinement ($\sigma \approx \SI{70}{\nm} < \lambda/(2n)$).
Hence, the polaritons are confined on sub-wavelength scales by both the out-of-plane confinement of the ML and the in-plane confinement of the dielectric nanocavity.
The tight confinement is expected to facilitate strong nonlinearities~\cite{Besga2015, Munoz-Matutano2019, Denning2022}, and combined with small losses, is expected to give rise to the polariton-blockade effect~\cite{Bamba2011, Ryou2018, Kyriienko2020, Denning2022}.
Moreover, it supports research on condensation effects~\cite{Gurrieri2024} and chiral interactions~\cite{Xiao2012, Yang2015a}.}

\section*{Acknowledgements} \label{sec:acknowledgements}
    We thank Dorte R. Danielsen, Duc Hieu Nguyen, Philip Holm, and Manh-Ha Dohan for the fruitful discussion about 2D material fabrication, as well as Peter Bøggild and Timothy J. Booth for their support.
    This work was supported by the \href{http://dx.doi.org/10.13039/501100001732}{\underline{Danish National Research Foundation}} through NanoPhoton - Center for Nanophotonics, grant number DNRF147. N. S. acknowledges funding by the Novo Nordisk Foundation NERD Programme (project QuDec NNF23OC0082957). P.W. and B.M. acknowledge support from the European Research Council (ERC-StG \textit{TuneTMD}, grant no. 101076437), and the Villum Foundation (grant no. VIL53033). P.W. and B.M. also acknowledge the European Research Council (ERC-CoG \textit{Unity}, grant no. 865230) and Carlsberg Foundation (grant no. CF21-0496).

\appendix

\section*{Appendix: Experimental setups} \label{sec:appendix Setups}
\renewcommand{\thefigure}{A\arabic{figure}}
\setcounter{figure}{0}  

\begin{figure}[htb!]
    \centering
    \includegraphics[width=\linewidth]{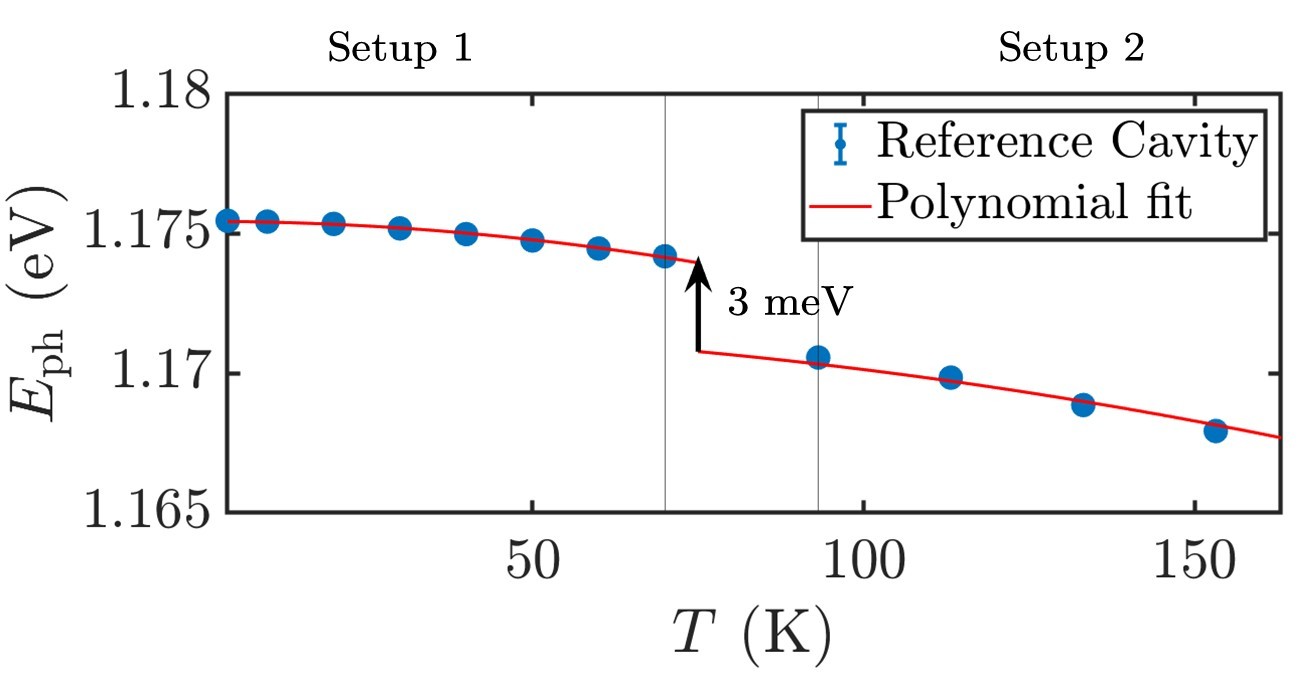}
    \caption{\textcolor{black}{Extracted peak energy from the reference cavity as a function of temperature. The red lines denote a fit with a parabola. The temperature regions of Setup 1 and 2 are indicated. The deduced offset is marked by the black arrow.}}
    \label{fig:off cav labels}
\end{figure}

Two setups are used to access the temperature regime from room temperature to $\SI{4}{\kelvin}$. Setup 1 contains a closed-cycle liquid-helium cryostat (AttoDRY800XS, \textit{Attocube}), giving us access to $T \leq \SI{70}{\kelvin}$. Setup 2 contains a liquid-nitrogen cryostat (HFS600, \textit{Linkam Scientific}), allowing us to measure $T \geq \SI{79}{\kelvin}$.   
In both setups, the system is excited with a red laser ($\lambda_{\mathrm{laser}} = 650$ and $\SI{637.5}{\nano\metre}$ for Setups 1 and 2, respectively). The excitation light beam is directed via a dichroic mirror (Setup 1)/ a $50/50$ beamsplitter (Setup 2) to a microscope objective, focusing the light on the sample. 
The emitted light is, after polarization projection with a $\lambda/2$ plate and a linear polarizer (= analyzer), detected by a spectrometer. A longpass filter ensures the filtering of the excitation light beam. We note that small deviations of the recorded wavelengths are likely to occur when comparing spectra recorded in both setups due to imperfect alignment of the spectrometers. \textcolor{black}{Such an offset is evident in all measurements, see Sec.~\ref{sec:SI reference} and Sec.~\ref{sec:SI COM fit}. We extract the offset from the reference measurements (cf. Sec.~\ref{sec:SI reference}). Fig.~\ref{fig:off cav labels} shows the peak energy of the reference cavity as a function of temperature. The fit of the peak position with a parabola is indicated as the red lines. A shift in photon energy is observed between $T \leq \SI{70}{\kelvin}$ and $T \geq \SI{79}{\kelvin}$, arising from the imperfect alignment of the respective spectrometers used in Setups 1 and 2. The $E_{\mathrm{cav,ref}}(T)$ from Setup 1 and 2 is extrapolated to $\SI{75}{\kelvin}$, and the offset is calculated as $E^{\mathrm{Setup 1}}_{\mathrm{cav,ref}}( \SI{75}{\kelvin})-E^{\mathrm{Setup 2}}_{\mathrm{cav,ref}}( \SI{75}{\kelvin})$. This yields an offset of $\SI{3}{\milli\eV}$ (cf. Fig.~\ref{fig:off cav labels}), and the data from Setup 2 in Fig.~\ref{fig:PL spectra heatmap peak} have been shifted accordingly.}

Adaptions to our PL setups allow us to carry out reflection measurements. For that, we change the light source to a superluminescent diode (SLD1050P, \textit{Thorlabs}). In addition to the analyzer and the $\lambda/2$ plate in the detection path, we insert a linear polarizer and a $\lambda/2$ plate in the excitation path, as well as a quarter-wave ($\lambda/4$) plate in the detection path. In Setup 1, the dichroic mirror is exchanged with a $50/50$ beamsplitter. The reflection measurements are carried out in a cross-polarization configuration, where the polarization of the incoming light $\hat{E}_{\mathrm{in}}$ is perpendicular to the detection polarization $\hat{E}_{\mathrm{det}}$, for example $\hat{E}_{\mathrm{in}} = \hat{x}$, $\hat{E}_{\mathrm{det}} = \hat{y}$. The sample is oriented so that the linear polarization of the cavity mode is oriented $\SI{45}{\degree}$ with respect to both of them, e.g. $\hat{E}_{\mathrm{cav}} = \frac{1}{\sqrt{2}}\left(\hat{x}+\hat{y}\right)$~\cite{Dimopoulos2022, Schroder2025a}. The quarter-wave plate is carefully adjusted to suppress elliptically polarized elements induced by scattering at the glass window of the cryostat. All spectra have been normalized with a reference spectrum recorded on a gold surface. 



\pagebreak
\widetext
\begin{center}
\textbf{\large Supplementary Information: Strong coupling between a dielectric nanocavity and a monolayer transition metal dichalcogenide}
\end{center}
\setcounter{equation}{0}
\setcounter{section}{0}
\setcounter{figure}{0}
\setcounter{table}{0}
\setcounter{enumiv}{0}
\makeatletter
\renewcommand{\theequation}{S\arabic{equation}}
\renewcommand{\thesection}{S\arabic{section}}
\renewcommand{\thefigure}{S\arabic{figure}}

\section{2D material fabrication}\label{sec:SI fabrication}
\begin{figure}[htb!]
         \centering
         \includegraphics[width = 0.9\linewidth]{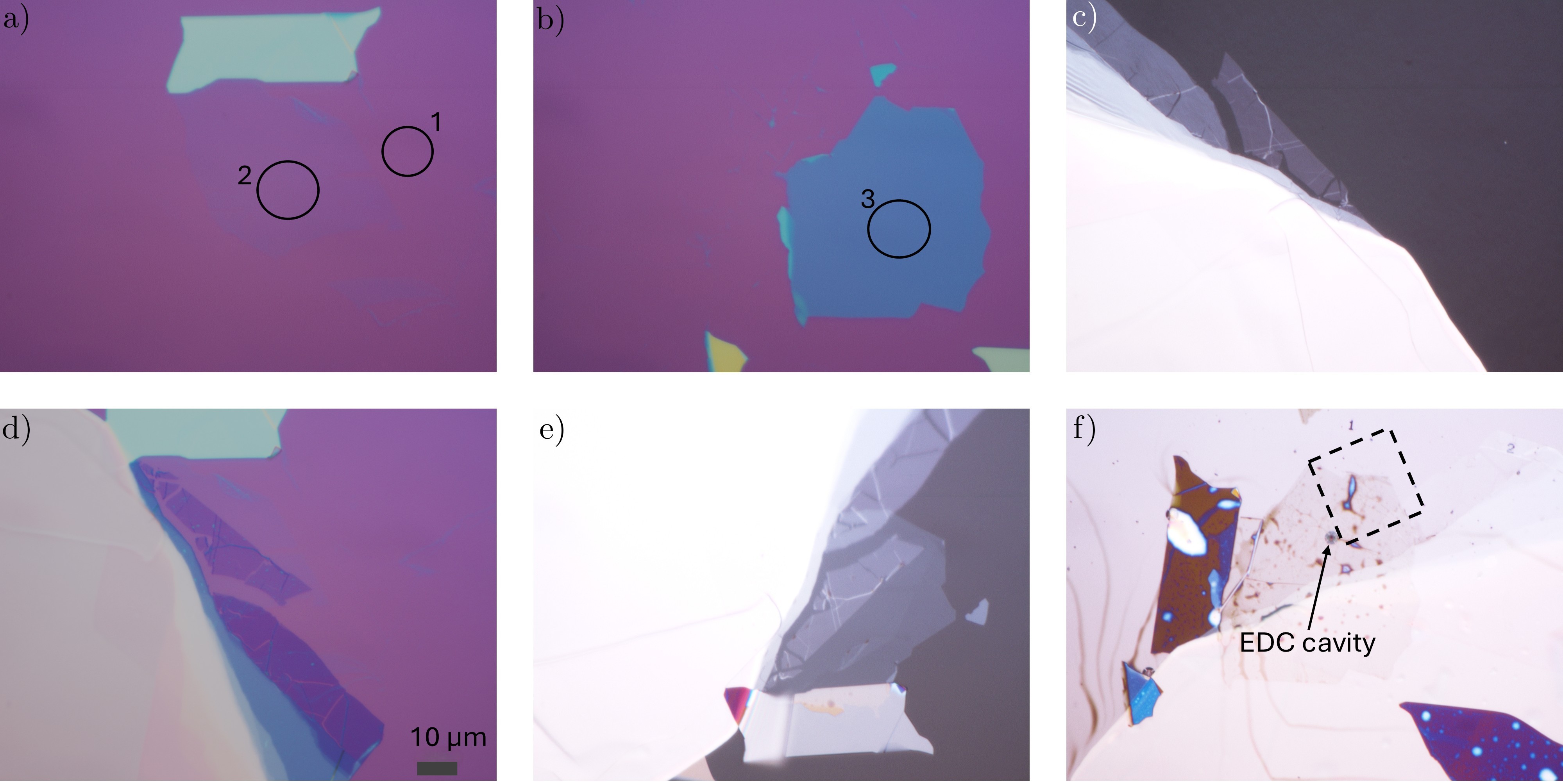}
         \caption{\textcolor{black}{Microscope image with 100x magnification after each fabrication step. a) Lower hBN flake on a $\SI{300}{\nano\metre}$ SiO$_2$/Si substrate. The flake of interest is the faint purple region in the center of the image. b) Upper hBN flake on a $\SI{300}{\nano\metre}$ SiO$_2$/Si substrate. The circles in~(a) and~(b) indicate the regions where the RGB color values are extracted. c) MoTe$_2$ on a PDMS substrate. d) MoTe$_2$ on lower hBN on a $\SI{300}{\nano\metre}$ SiO$_2$/Si substrate. e) hBN/MoTe$_2$/hBN heterostructure on the PC/PDMS stamp. f) Heterostructure on the EDC cavity. The dashed rectangle in~(f) indicates the region where the AFM map was taken, see Fig. \ref{fig:AFM_thickness}.}}
         \label{fig:SI sample fabrication}
\end{figure}
In this Supplementary Information, the 2D material fabrication process is described.  
Firstly, MoTe$_2$ and hBN (\textit{HQ Graphene}) are mechanically exfoliated on polydimethylsiloxane (PDMS) on a glass slide and on a $\SI{300}{\nano\metre}$ silicon dioxide (SiO$_2$) on a silicon (Si) wafer, respectively (cf. Fig.~\ref{fig:SI sample fabrication}(a)-(c)). \textcolor{black}{
The flake depicted in Fig.~\ref{fig:SI sample fabrication}(a) is labeled as the lower hBN flake, while the flake depicted in~(b) is labeled as the upper hBN flake.
The thicknesses of the hBN flakes are determined with the optical contrast method~\cite{Gorbachev2011a}.}
\textcolor{black}{
\begin{table}[htb!]
    \centering
    \begin{tabular}{lrrrrrrr}
    \toprule
                  & Marker &  R &   G &  B   &  $t_{\mathrm{R}}$ (nm) &  $t_{\mathrm{G}}$ (nm) &  $t_{\mathrm{B}}$ (nm) \\
    \midrule
      \textbf{Substrate} & 1 & 136 &  82 & 140  & 0   & 0   & 0     \\
      \textbf{lower hBN flake} & 2 & 135 &  88 & 146  & 0   & 2.0 & 2.4   \\
      \textbf{upper hBN} & 3 &  79 & 110 & 166  & 8.5 & 9.3 & 9.3   \\
    \bottomrule
    \end{tabular}
    \caption{\textcolor{black}{Extracted color values for red (R), green (G), and black (B) at markers 1-3 corresponding to the substrate, to the lower, and to the upper hBN flake. The deduced thicknesses from the RGB color values are denoted as $t_{\mathrm{R}}$, $t_{\mathrm{G}}$, and $t_{\mathrm{B}}$ for red, green, and black, respectively.}}
    \label{tab:RGB_values}
\end{table}
}
\begin{figure}[htb!]
    \centering
    \includegraphics[width=0.6\linewidth]{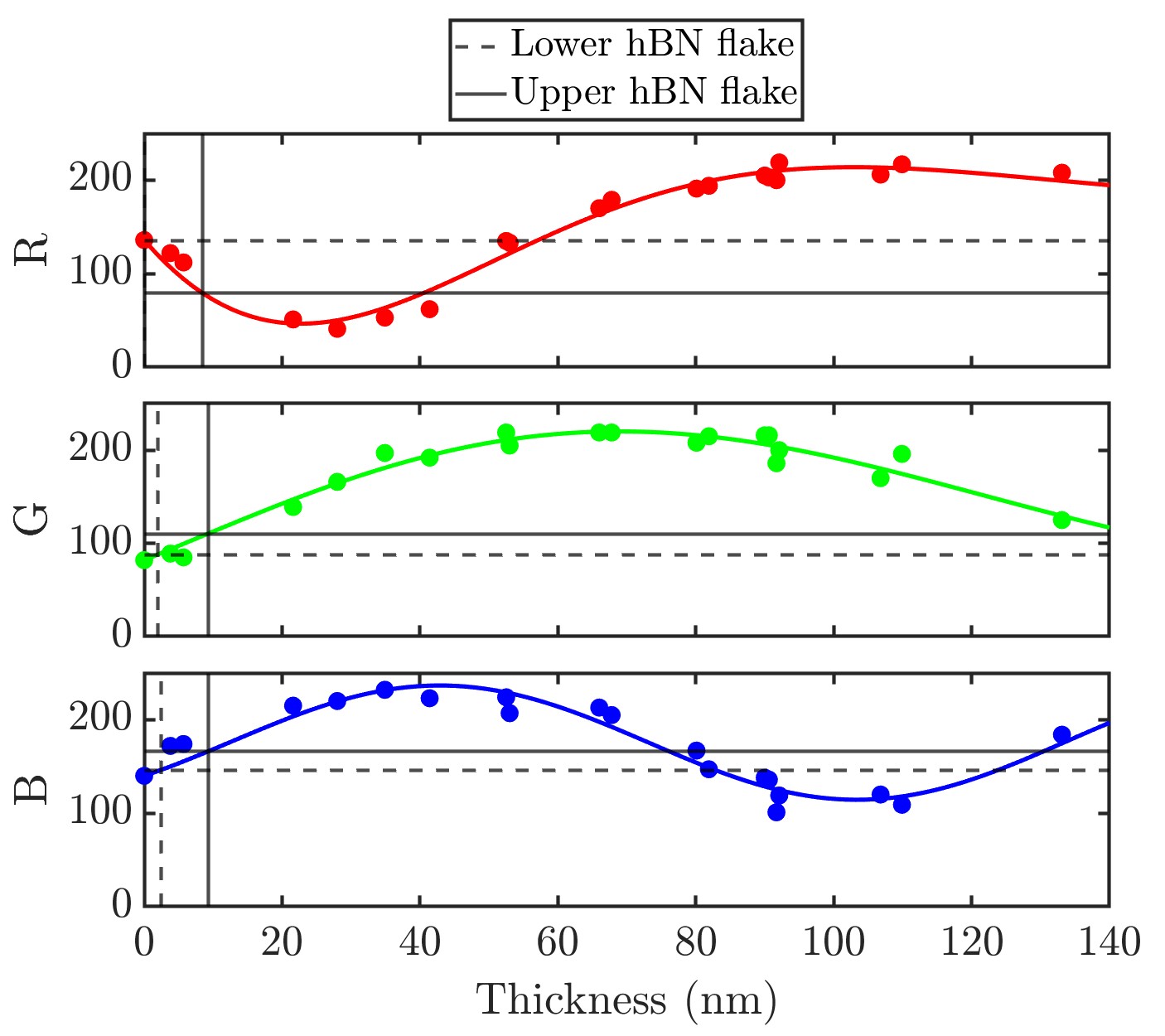}
    \caption{\textcolor{black}{Calibration of the microscope images. RGB color values of hBN flakes as a function of thickness. Data points are from Ref.~\cite{Holm2024} except for the value of the substrate, cf. Tab.~\ref{tab:RGB_values}. The grey dashed and solid lines mark the extracted values for the lower and the upper hBN flakes, respectively.}}
    \label{fig:RGB_Thickness}
\end{figure}

\textcolor{black}{
A calibration of the microscope was pursued in Ref.~\cite{Holm2024}, and the color values for red (R), green (G), and black (B) from the optical images correspond to a certain thickness. 
The RGB color values for the substrate and for the flakes are determined; see markers 1-3 in Fig.~\ref{fig:SI sample fabrication}(a)-(b). These are summarized in Tab.~\ref{tab:RGB_values}.
Fig.~\ref{fig:RGB_Thickness} depicts the calibration function for hBN flakes. 
The data points in Fig.~\ref{fig:RGB_Thickness} are from Ref.~\cite{Holm2024}. In addition, the value for the substrate (cf. Tab. \ref{tab:RGB_values}) is included. The colored solid lines depict a fit of the data points with the general fit function $V(t) = A_1 e^{-at} \sin(k t + \varphi) + A_2$, where $V$ denotes either the R, G or B value, and $t$ denotes thickness. The fit is constrained to match the substrate value, ensuring positive thicknesses. The RGB values read from Tab.~\ref{tab:RGB_values} for the lower and upper hBN flakes are indicated as grey dashed and solid lines, respectively. 
With those values, the intersect with the fit curve yields the thicknesses $t_{\mathrm{R}}$,  $t_{\mathrm{G}}$ and $t_{\mathrm{B}}$ for R, G, and B, respectively (also indicated with grey dashed and solid lines). 
The determined thicknesses are summarized in Tab.~\ref{tab:RGB_values}. 
The average of the three thicknesses is formed to determine the best estimate and the standard deviation for the thickness, yielding $t_{\mathrm{hBN,lower}} = \SI{1.5\pm1.3}{\nano\metre}$ and $t_{\mathrm{hBN,upper}} = \SI{9.0\pm0.5}{\nano\metre}$. 
}

The 2D materials are stacked using a commercially available transfer system (\textit{HQ Graphene}). 
The monolayer MoTe$_2$ is placed on the thin hBN flake with a dry-transfer method~\cite{Castellanos-Gomez2014}. 
The lower hBN flake on the substrate is placed on a hot plate on a three-axis micrometer stage.
Another three-axis micrometer stage holds the glass slide with the MoTe$_2$ flake.
The lower hBN flake and the ML MoTe$_2$ are aligned with an optical microscope and pressed together.
Then, the hotplate is heated up to $\SI{60}{\celsius}$ to ensure that the ML sticks to the hBN flake, see Fig.~\ref{fig:SI sample fabrication}(d).
After that, the upper hBN flake is picked up with a polycarbonate (PC)/polydimethylsiloxane (PDMS) film~\cite{Zomer2014, Pizzocchero2016}.
The stamp is pressed on the upper hBN flake at $\SI{60}{\celsius}$. 
The hotplate is set to $\SI{110}{\celsius}$ when contact is established. 
After 5 minutes, the substrate is moved down, and the hBN flake sticks to the PC film.
The hBN/MoTe$_2$ heterostructure (HS) is picked up by the upper hBN flake on a PC/PDMS film, pressing them together at $\SI{110}{\celsius}$. 
After waiting for 5 minutes and moving the substrate down again, the hBN-encapsulated ML MoTe$_2$ is formed on the PC film, see Fig.~\ref{fig:SI sample fabrication}(e).
The hBN/MoTe$_2$/hBN HS is then pressed on the EDC cavity at $\SI{110}{\celsius}$. 
The temperature is then set to $\SI{190}{\celsius}$ so that the PC film melts. 
After 20-30 seconds, the sample is moved away, and the HS remains on the cavity. 
After that, the sample is cleaned in chloroform for more than 20 minutes and washed with isopropanol (IPA), see Fig.~\ref{fig:SI sample fabrication}(e). 
The image in Fig.~\ref{fig:sample}(c) in the main text depicts the final result, where Fig.~\ref{fig:SI sample fabrication}(d) and (f) are overlayed with each other for better visibility of the ML MoTe$_2$. \textcolor{black}{It is evident from Fig.~\ref{fig:SI sample fabrication}(f) that the sample contains residual polymers from the PC film. As these are likely separated from the MoTe$_2$ and the EDC cavity by the upper hBN flake, they are not expected to influence the optical properties of the system.}

\begin{figure}[htb!]
    \centering
    \includegraphics[width=0.8\linewidth]{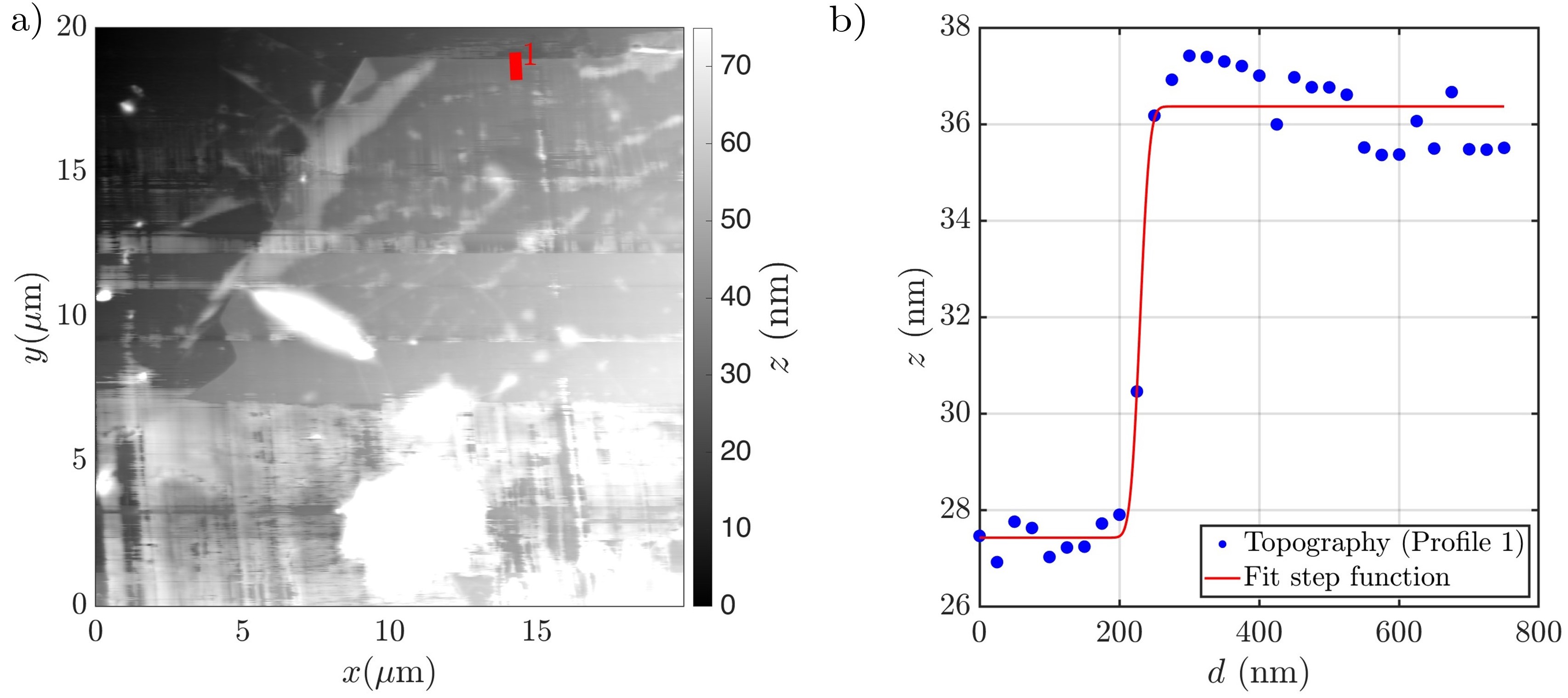}
    \caption{\textcolor{black}{a) AFM map of the final sample, in the region indicated in Fig.~\ref{fig:SI sample fabrication}(f). b) Height profile extracted along line 1 in~(a). The red line depicts the fit with an error function.}}
    \label{fig:AFM_thickness}
\end{figure}

\textcolor{black}{
To confirm the deduced thickness from the optical contrast, an AFM measurement of the final structure is carried out. The AFM map, taken in the region indicated as the dashed rectangle in Fig.~\ref{fig:SI sample fabrication}(f), can be found in Fig.~\ref{fig:AFM_thickness}(a). Polymer residues are also visible in the AFM measurements. An edge of the upper hBN flake is identified at the top of Fig.~\ref{fig:AFM_thickness}(a). The extracted height profile along the red line in~(a) is depicted in Panel~(b). The height profile is fitted with an error function (a convolution of a Gaussian and a step function~\cite{Andrews1997}. 
This yields $t_{\mathrm{hBN, upper}} = \SI{8.9\pm0.6}{\nano\metre}$, in good agreement with the value obtained with the optical contrast method.}

\begin{figure}[ht!]
     \centering
     \includegraphics[width = 0.55\linewidth]{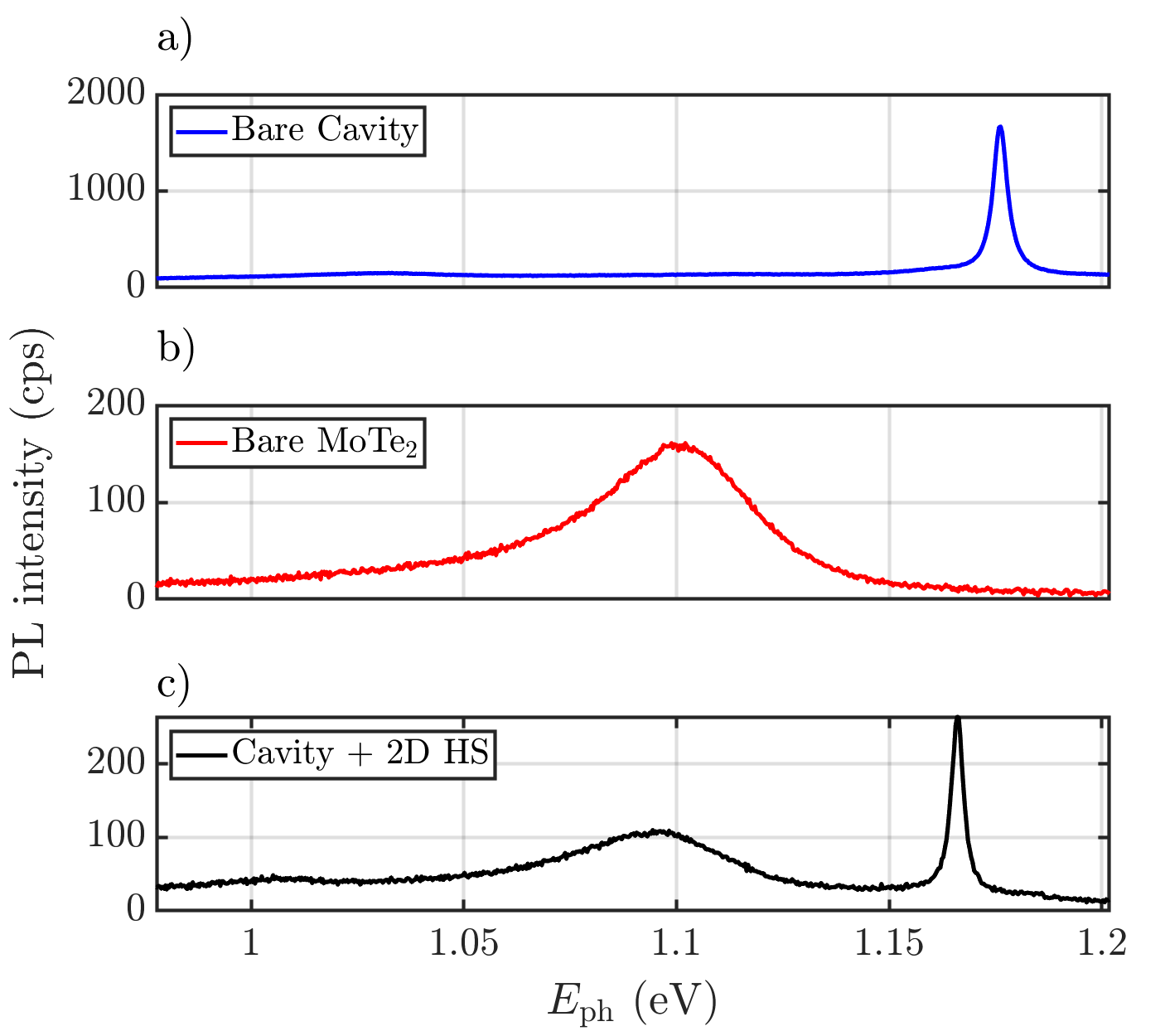}
     \caption{PL spectra at room temperature, of a) the EDC cavity before stacking the hBN/MoTe$_2$/hBN heterostructure on top, of b) a MoTe$_2$ layer on PDMS and of c) the combined system.}
     \label{fig:SI RT spectra MoTe2 EDC SC}
\end{figure}

PL spectra at room temperature of the bare EDC cavity before stacking the HS on top, and of the excitonic emission from a monolayer MoTe$_2$ are compared with the combined system as fabricated, cf. Fig.~\ref{fig:SI RT spectra MoTe2 EDC SC}. Clearly, the excitonic emission and the cavity mode can be observed together, demonstrating the successful deposition of the HS on the EDC cavity. It is evident that even without the excitonic emission from MoTe$_2$, PL signal from the EDC cavity can be observed. We attribute this to spontaneous recombination in the dielectric material due to the above-bandgap excitation of InP. 
A Lorentzian fit of the cavity mode emission of the fabricated sample (cf. Fig.~\ref{fig:SI RT spectra MoTe2 EDC SC}(c)) reveals a resonance energy of $E_{\mathrm{cav}} = \SI{1.166}{\eV}$ and a full-width half-maximum $\Gamma_{\mathrm{cav}} = \SI{3.4}{\milli\eV}$, yielding a quality factor of $Q = \SI{348 \pm 7}{}$ \textcolor{black}{before depositing the hBN/MoTe$_2$/hBN heterostructure on the cavity}. It is evident from Fig.~\ref{fig:SI RT spectra MoTe2 EDC SC} that depositing the 2D HS on the EDC cavity blackshifts the resonance energy by $\approx \SI{10}{\milli\eV}$, resulting from the increase of the effective refractive index of the cavity mode~\cite{Kountouris2022}. Moreover, encapsulating the MoTe$_2$ flake with hBN and depositing the HS on the EDC cavity blackshifts the excitonic transition by $\approx \SI{6}{\milli\eV}$, see Fig.~\ref{fig:SI RT spectra MoTe2 EDC SC}, which can be explained by screening of the Coloumb interaction of the electron-hole pairs in the ML TMDC~\cite{Li2014a, Chernikov2014b, Kutrowska-Girzycka2022}. \textcolor{black}{The cavity linewidth remains unaffected, yielding $\Gamma_{\mathrm{cav}} = \SI{3.3\pm0.1}{\milli\eV}$ after depositing the HS on the cavity, corresponding to $Q = 358\pm11$.}
\newpage
\
\newpage
\section{Details on simulations}\label{sec:SI Simulation}
Subject to suitable radiation conditions~\cite{Kristensen2020}, the electric-field QNMs $\mathbf{f}_j(\mathbf{r})$ are solutions to the source-less wave equation
\begin{equation}
    \nabla \times \nabla \times \tilde{\mathbf{f}}_j(\mathbf{r}) - \tilde{k}^2_{j}\epsilon_{\mathrm{R}}(\mathbf{r})\tilde{\mathbf{f}}_j(\mathbf{r}) = 0
    \label{eq:Helmholtz}
\end{equation}
in which $\epsilon_{\mathrm{R}}(\mathbf{r})$ is the spatially-dependent relative permittivity and the index $j$ lables the different QNMs with corresponding wavenumber $\tilde{k}^2_{j} = \tilde\omega_j/c$, where $\tilde{\omega}_j$ is the complex eigenfrequency. Even for purely real permittivities, the complex frequencies account for a finite dissipation due to radiation, which means that the field leaks away from the cavity and eventually diverges at large distances. Nevertheless, the QNMs can be normalized by one of several complementary formulas~\cite{Kristensen2015}. This also has implications for the expression for the coupling strength in Eq.~(\ref{eq:Light matter coupling}), for which the divergent integral was regularized by use of a quasistatic approximation to the Green tensor in Ref.~\cite{Denning2022a}. For the present calculation, we used a finite area of 2D material an the field was sufficiently localized, that we could ignore this issue and calculate the coupling strength simply by an integral across the extend of the cavity using the normalized field profile 
$\mathbf{\tilde{F}}_j(\mathbf{r}) =\mathbf{\tilde{f}}_j(\mathbf{r})/\sqrt{\langle\langle \mathbf{\tilde{f}}_{j}|\mathbf{\tilde{f}}_{j} \rangle\rangle}$, in which $\sqrt{\langle\langle \mathbf{\tilde{f}}_{j}|\mathbf{\tilde{f}}_{j} \rangle\rangle}$ denotes the normalization.

For the eigenmode analysis, we used a finite-element method calculation with the commercial software COMSOL Multiphysics, solving Eq.~(\ref{eq:Helmholtz}) numerically following the approach in Ref.~\cite{Kountouris2022}. The thicknesses of the hBN layers were determined as described in Sec.~\ref{sec:SI fabrication}. For simplification, the thin ML MoTe$_2$ (thickness of $\SI{0.65}{\nano\metre}$~\cite{Ruppert2014}) was not inlcuded  in the calculations. The simulation domain included the structured $\SI{245}{\nano\metre}$ thick InP membrane and an hBN layer with a thickness of $\SI{10.5}{\nano\metre}$. 
The integral in Eq.~(\ref{eq:Light matter coupling}) of the main text was evaluated at $z_{2D} = \SI{1.5}{\nano\metre}$ above the cavity surface, corresponding to the thickness of the lower hBN flake. The constants $a_{\mathrm{B}}$ and $\mathbf{p}^{\alpha}_{\mathrm{cv}}$ was calculated with values from~\cite{Meckbach2020, Edalati-Boostan2020, Haastrup2018, Ruppert2014}. The refractive index of hBN and InP was taken from Refs.~\cite{Grudinin2023} and~\cite{Pettit1965, McCaulley1994}, respectively.

\begin{figure} [htb!]
    \centering
    \includegraphics[width=\linewidth]{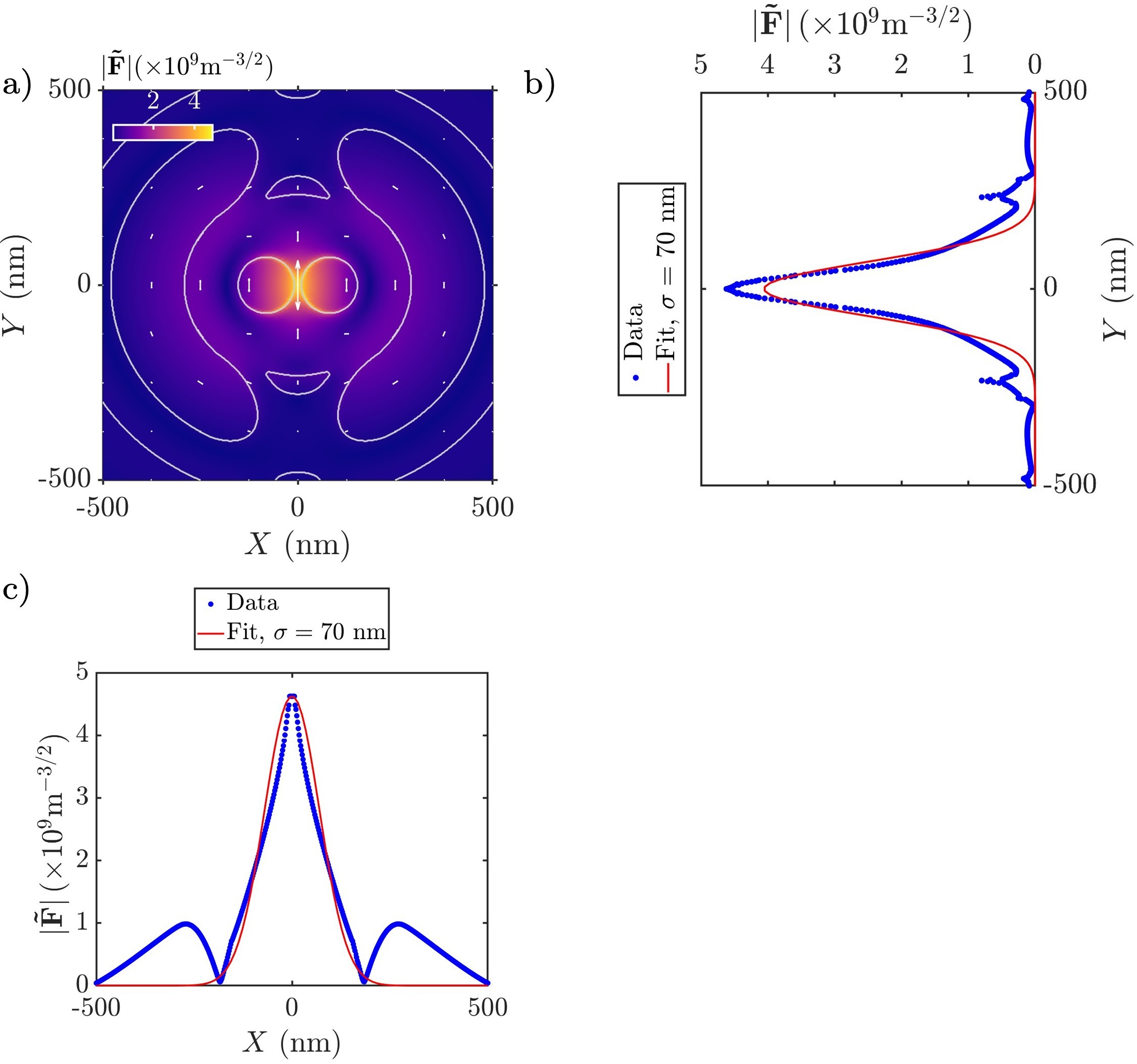}
    \caption{\textcolor{black}{Simulated electric field profile.~a) $|\mathbf{\tilde{F}}|$ in the plane of the MoTe$_2$ layer. The white arrows denote the in-plane polarization. The contour lines show the topography of the cavity at $Z = 0$. Extracted field along the~b) $Y$ axis and the~c) $X$ axis. A fit with a Gaussian function is indicated as by red lines.}}
    \label{fig:ModePlots_Confinement}
\end{figure}

\textcolor{black}{Fig.~\ref{fig:ModePlots_Confinement} depicts the simulated electric field profile. Panel~(a) shows $| \mathbf{\tilde{F}}|$ in the plane of the MoTe$_2$ layer (cf. Fig~\ref{fig:sample}(e) in the main text). Moreover, the field profiles along the $Y$ and $X$ axes are shown in~(b) and~(c), respectively. A fit with a Gaussian function of the form 
\begin{equation}
|\mathbf{F}(x)| = \frac{A}{\sigma \sqrt{2\pi}} e^{-\frac{x^2}{2\sigma^2}}
\end{equation}
is indicated by the red lines. Here, $x$ denotes the spatial coordinate ($X,Y$), and $A$ and $\sigma$ denote fit parameters. This yields a lateral confinement of $\sigma \approx \SI{70}{\nano\metre}$, much smaller than $\lambda_{\mathrm{cav}}/(2n_{\mathrm{InP}}) = \SI{1049}{\nano\metre}/6.4 = \SI{164}{\nano\metre}$.
}

A convergence study was carried out as described in Ref.~\cite{Kountouris2022}, yielding \textcolor{black}{$E^{\mathrm{sim}}_{\mathrm{cav}} = 1.18696 \pm 1\times10^{-5}$ eV, $\Gamma^{\mathrm{sim}}_{\mathrm{cav}} = \SI{2.07 \pm 0.03}{\milli\eV}$ and $V_{\mathrm{eff}} = (0.060\pm0.002)(\lambda/n)^3$.} 

\begin{figure}[htb!]
    \centering
    \includegraphics[width=0.6\linewidth]{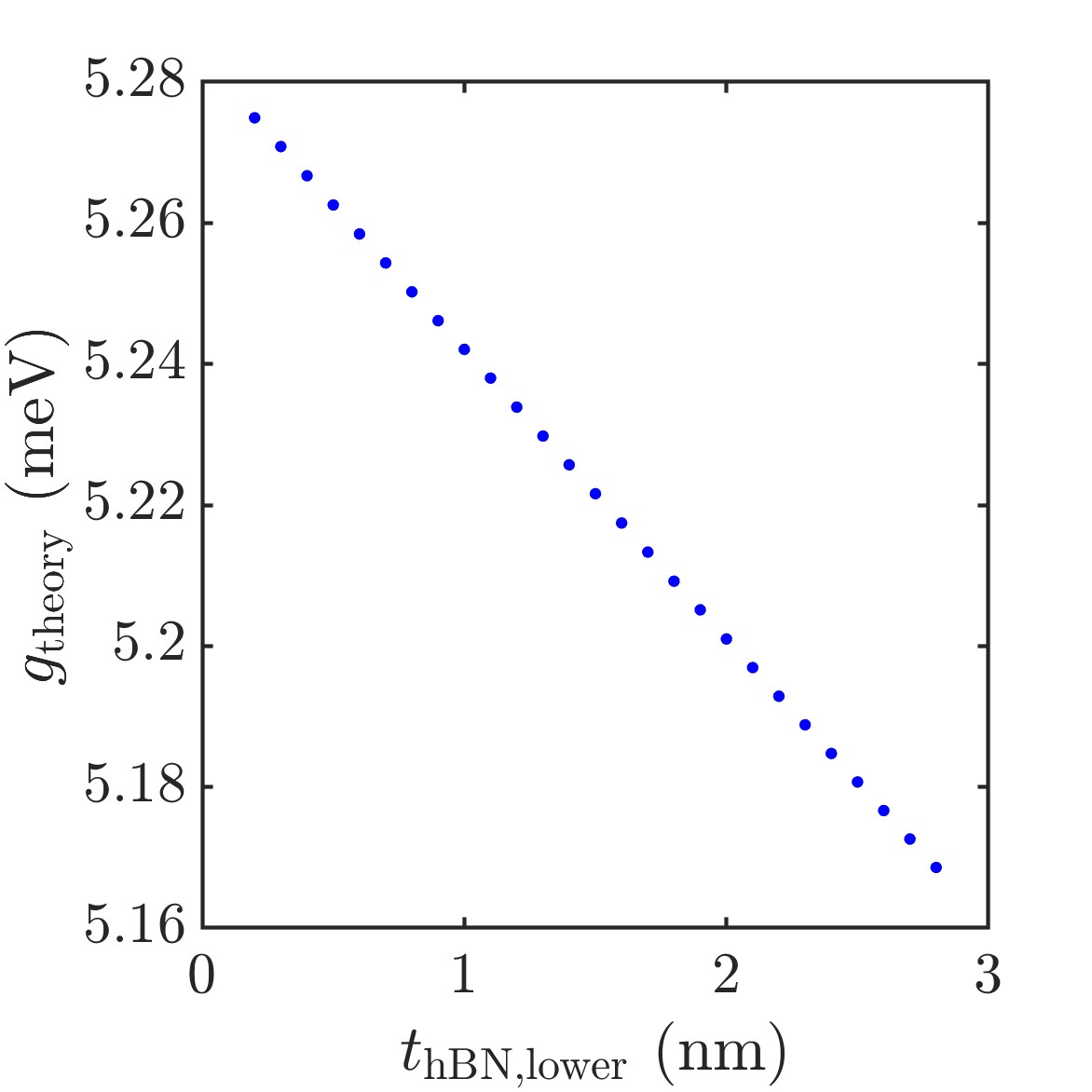}
    \caption{\textcolor{black}{Calculated light-matter interaction strength as a function of the thickness of the lower hBN flake.}}
    \label{fig:g_lowerHBN}
\end{figure}

\textcolor{black}{
The thickness of the lower hBN flake has been determined as $\SI{1.5\pm1.3}{\nm}$ (see Sec.~\ref{sec:SI fabrication}). The lower hBN flake separates the ML MoTe$_2$ from the surface of the InP layer via $z_{\mathrm{2D}}$ in Eq.~\ref{eq:Light matter coupling} in the main text.
To assess the uncertainty on the light-matter interaction strength induced by the uncertainty of the thickness of the lower hBN flake (see Sec.~\ref{sec:SI fabrication}), Eq.~\ref{eq:Light matter coupling} 
is evaluated as a function of $t_{\mathrm{hBN,lower}}$. Fig.~\ref{fig:g_lowerHBN} depicts $g_{\mathrm{theory}}$ as a function of the thickness of the lower hBN flake within the uncertainty range of $t_{\mathrm{hBN,lower}}$. At the limiting points of $t_{\mathrm{hBN,lower}} = \SI{0.2}{\nm}$ and $t_{\mathrm{hBN,lower}} = \SI{2.8}{\nm}$, $g_{\mathrm{theory}}$ does not differ by more than $\SI{0.06}{\milli\eV}$ from the value evaluated at $t_{\mathrm{hBN,lower}} = \SI{1.5}{\nm}$. The total uncertainty for $g$ is determined from the numerical uncertainty, the uncertainty of the material parameters~\cite{Ruppert2014, Gerber2018, Meckbach2020, Edalati-Boostan2020} and the uncertainty induced by $t_{\mathrm{hBN,lower}}$, yielding $g_{\mathrm{theory}} = \SI{5.2\pm0.7}{\milli\eV}$.}

\newpage
\
\newpage
\section{Reference measurements}
\label{sec:SI reference}
In this Supplementary Information, the reference measurements for deducing $E_{\mathrm{exc}}(T)$, $E_{\mathrm{cav}}(T)$, $\Gamma_{\mathrm{exc}}(T)$ and $\Gamma_{\mathrm{cav}}(T)$ are described.

\begin{figure}[htb!]
         \centering
         \includegraphics[width = 0.7\linewidth]{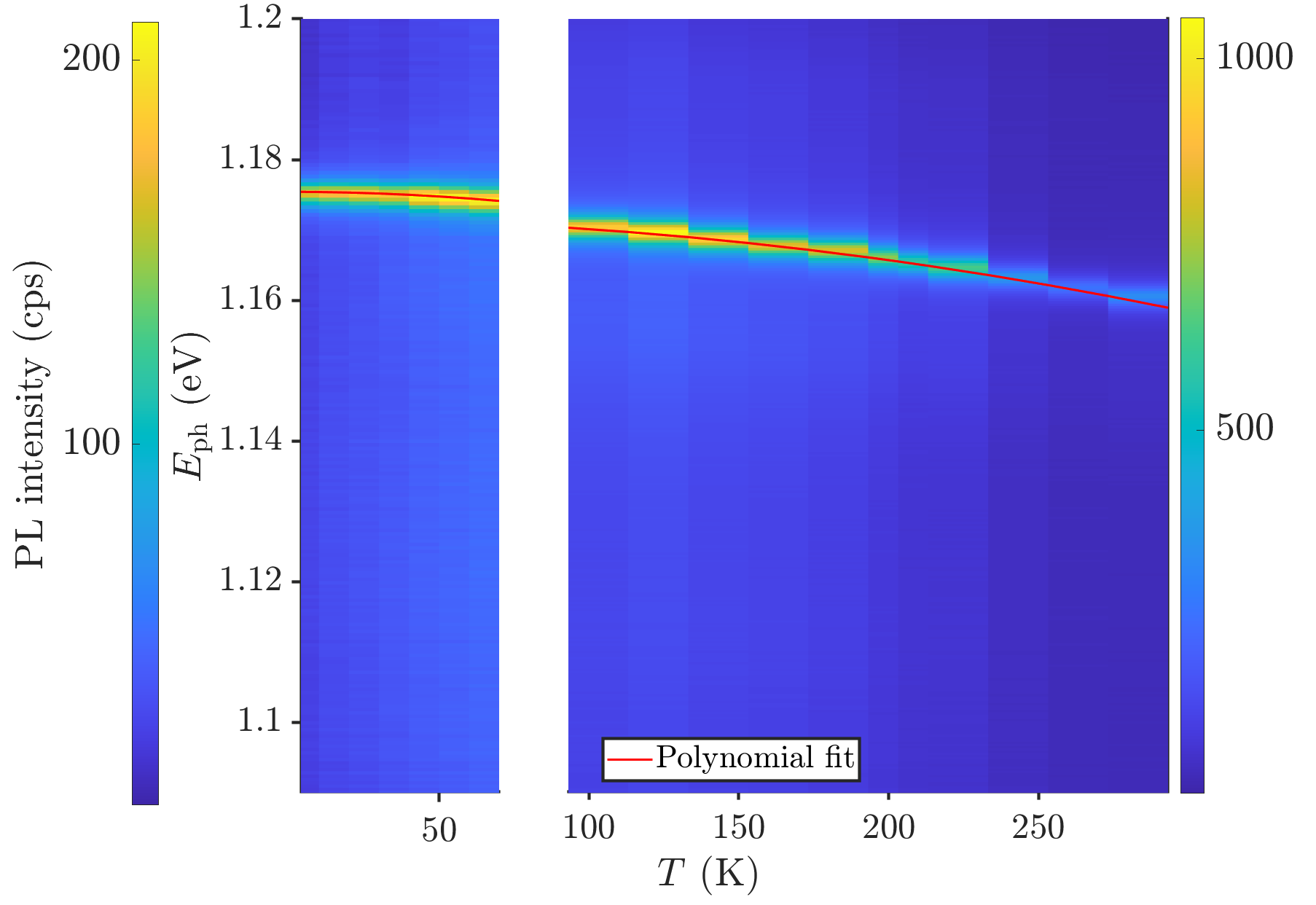}
         \caption{PL spectra as a function of temperature from the reference cavity. The red line indicates a fit of the peak position with a parabola.}
         \label{fig:Cavity reference}
\end{figure}

\begin{figure}[htb!]
         \centering
         \includegraphics[width = \linewidth]{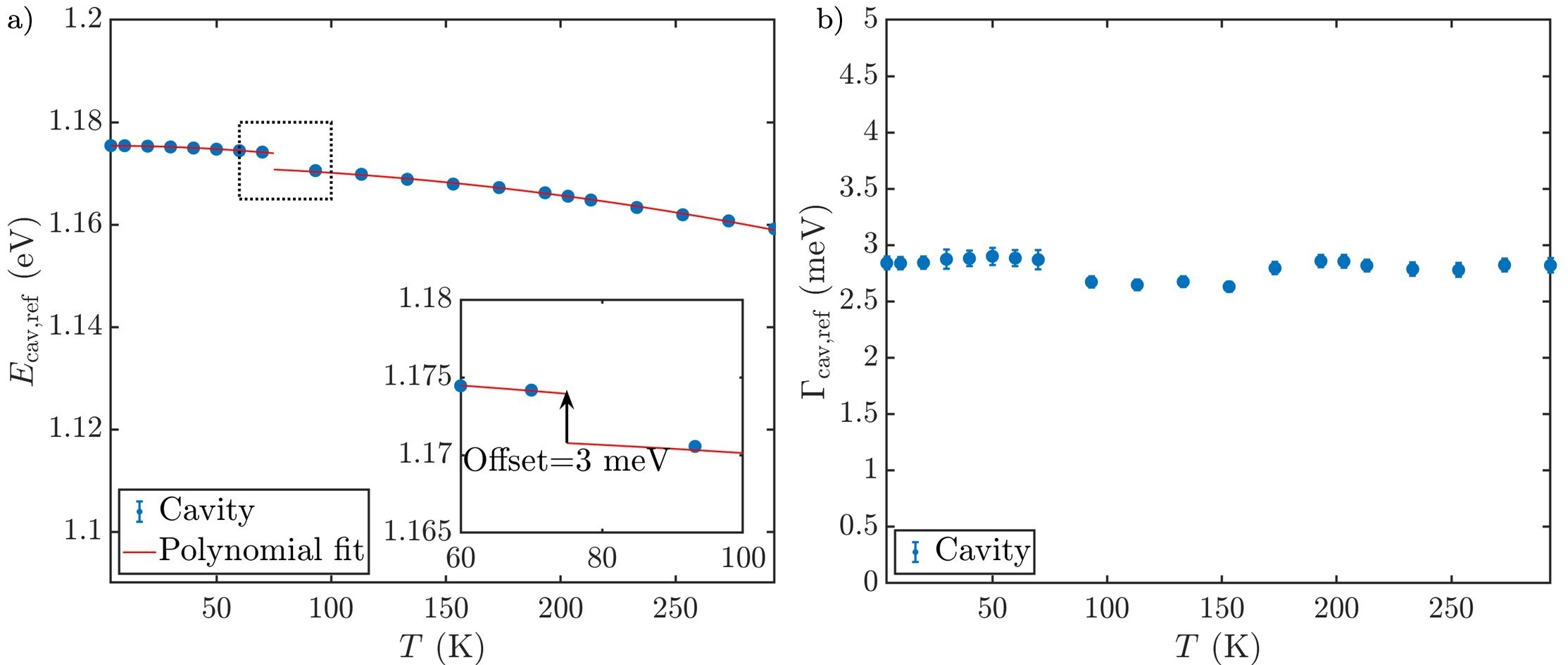}
         \caption{a) Extracted peak positions and b) linewidths of the reference cavity as a function of $T$. The red line in (a) depicts the fits with a parabola.}
         \label{fig:SI Cavity reference fit}
\end{figure}

Fig.~\ref{fig:Cavity reference} depicts spectra as a function of temperatures of a reference cavity. For those measurements, the detection polarization is aligned with the polarization of the cavity mode. Each spectrum is fitted with a single Lorentzian (cf. Fig.~\ref{fig:SI Cav fit}), yielding $E_{\mathrm{cav,ref}}(T)$ and $\Gamma_{\mathrm{cav,ref}}(T)$ (see Fig.~\ref{fig:SI Cavity reference fit}), where the subscript "ref" indicates that these values are obtained from a reference cavity. $E_{\mathrm{cav,ref}}(T)$ is fitted with a parabola, $E_{\mathrm{cav,ref}}(T) = y_0 + b  T^2$. The obtained temperature dependence of the resonance energy is indicated by the red line in Fig.~\ref{fig:Cavity reference} and Fig.~\ref{fig:SI Cavity reference fit}(a). Moreover, as evident from Fig.~\ref{fig:SI Cavity reference fit}(b), the linewidth of the cavity does not depend on temperature as it is not limited by phonon interactions. \textcolor{black}{The deduced offset between $T \leq \SI{70}{\kelvin}$ and $T \geq \SI{79}{\kelvin}$ due to imperfect alignment of the respective spectrometers used in Setups 1 and 2 is indicated in the inset of Fig.~\ref{fig:SI Cavity reference fit} (see Appendix of the main text).}

\begin{figure}[htb!]
         \centering
         \includegraphics[width = 0.5\linewidth]{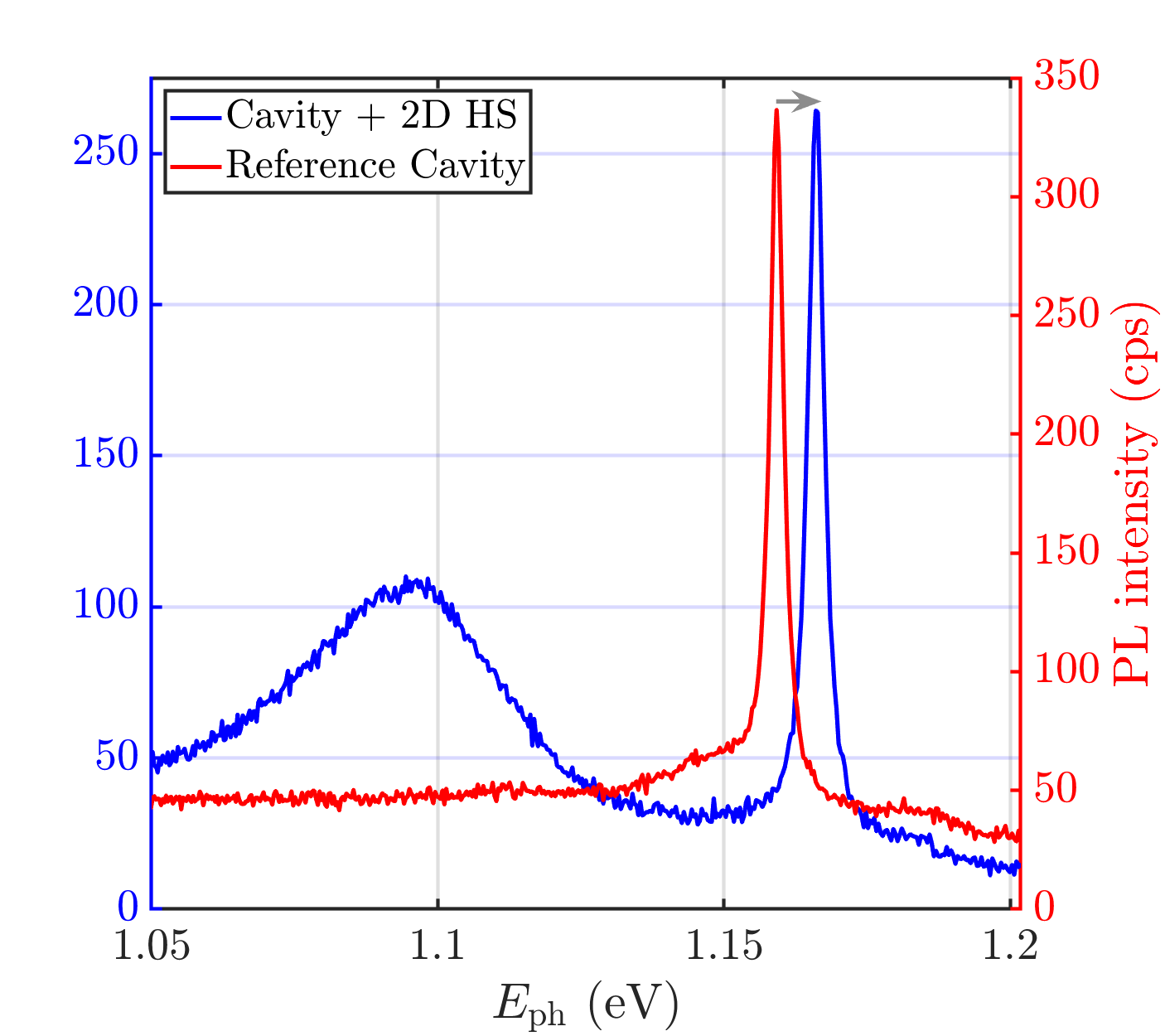}
         \caption{PL spectra at $T = \SI{293}{\kelvin}$ for the reference cavity and the sample consisting of EDC cavity and the HS for red and black lines, respectively. The offset $\Delta$ is indicated as the grey arrow.}
         \label{fig:SI Coupled ReferenceCavity}
\end{figure}

The resonance energy of the reference cavity $E_{\mathrm{cav,ref}}(T)$ deviates from the resonance energy of the cavity with the hBN/MoTe$_2$/hBN heterostructure $E_{\mathrm{cav}}(T)$ (cf. Fig.~\ref{fig:SI Coupled ReferenceCavity}). This is partially due to the presence of the heterostructure, which changes the effective refractive index of the cavity mode \textcolor{black}{(see Sec.~\ref{sec:SI fabrication})}, and partially due to fabricational differences of the cavities, see main text. This overall shift $\Delta = E_{\mathrm{cav,ref}} - E_{\mathrm{cav}}$ is quantified at $T = \SI{293}{\kelvin}$ and $T = \SI{315}{\kelvin}$ for Setups 1 and Setup 2, respectively, corresponding to a detuning $\delta > \SI{70}{\milli\eV}$. From the coupled-oscillator model it follows that for large detuning, the polaritonic eigenenergies equal the eigenenergies of the respective oscillators. In other words, at this temperature, the influence of the excitonic transition on the cavity mode is negligible. This yields $\Delta = \SI{6.89\pm0.03}{\milli\eV}$ and $\Delta = \SI{6.3\pm0.1}{\milli\eV}$ for Setup 1 and Setup 2, respectively, see the gray arrow in Fig.~\ref{fig:SI Coupled ReferenceCavity}. The temperature dependence of the cavity can be deduced from the temperature dependence of the reference cavity $E_{\mathrm{cav}}(T) =E_{\mathrm{cav,ref}}(T) + \Delta $. Moreover, since the linewidth of the reference cavity does not depend on the temperature, we use the value for $\Gamma_{\mathrm{cav}} = \SI{3.3\pm0.1}{\milli\eV}$ deduced from the cavity with the hBN/MoTe$_2$/hBN heterostructure at $T = \SI{293}{\kelvin}$ (cf. Fig.~\ref{fig:SI Coupled ReferenceCavity}) as the cavity's reference linewidth for all temperatures.

\begin{figure}[htb!]
         \centering
         \includegraphics[width = 0.7\linewidth]{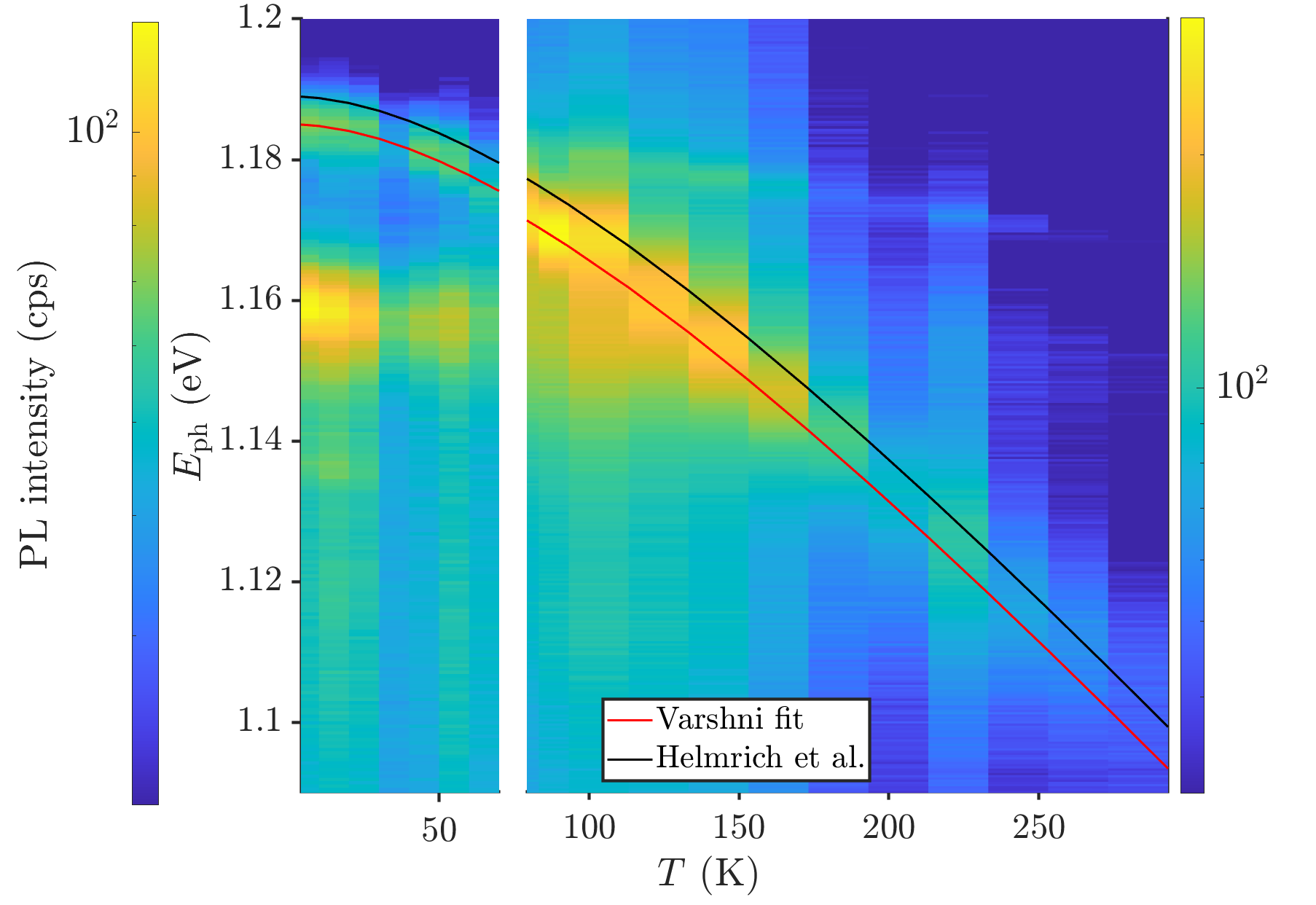}
         \caption{PL spectra as a function of temperature from the hBN/MoTe$_2$/hBN heterostructure on the EDC cavity with the detection polarization perpendicular to the cavity mode. The red line indicates a Varshni fit of the peak position of the excitonic transition. The black line depicts the temperature dependence of the exciton in MoTe$_2$ on silicon dioxide found in Ref.~\cite{Helmrich2018}.}
         \label{fig:MoTe2 reference 1}
\end{figure}

\begin{figure}[htb!]
         \centering
         \includegraphics[width = \linewidth]{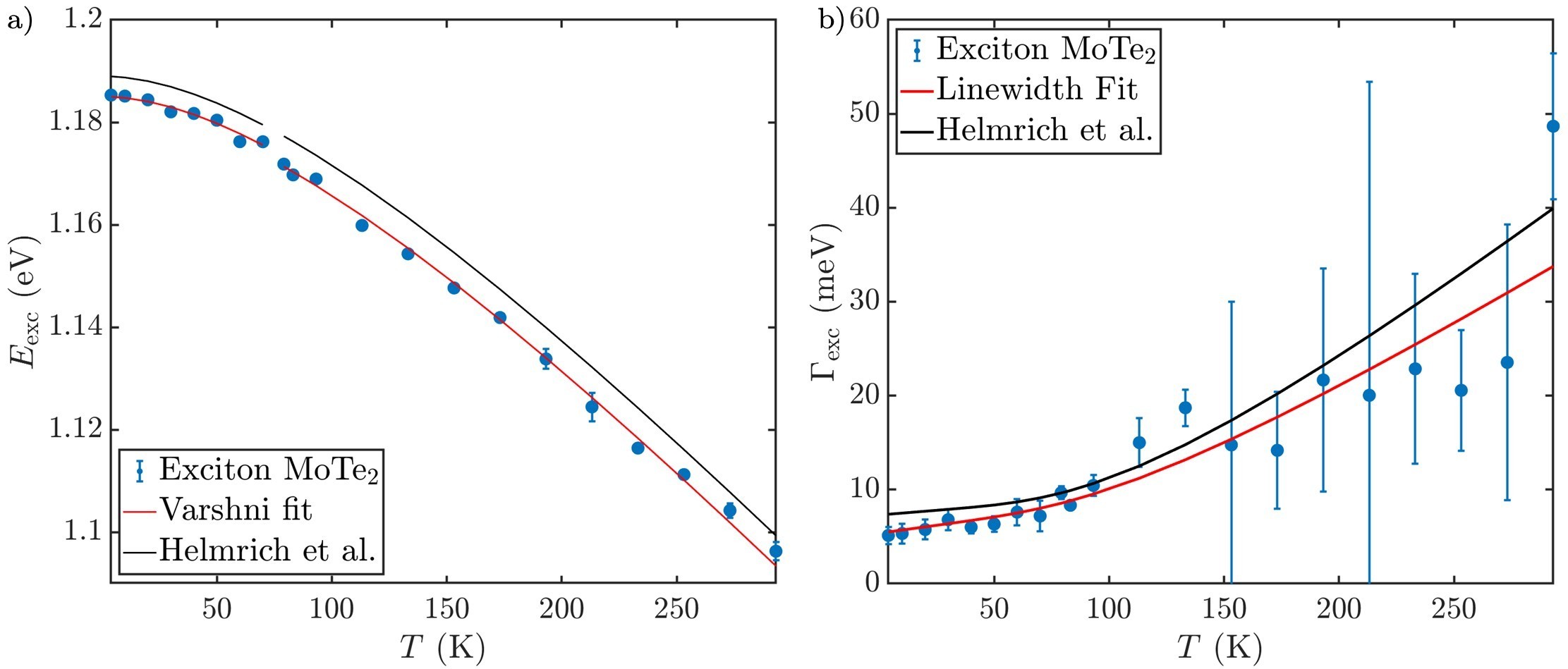}
         \caption{a) Extracted peak positions and b) linewidths of the uncoupled excitons as a function of $T$. The red lines depict fit results as described in the main text. The black line depicts the temperature dependence of the exciton in MoTe$_2$ on silicon dioxide found in Ref.~\cite{Helmrich2018}.}
         \label{fig:SI MoTe2 reference fit}
\end{figure}

Fig.~\ref{fig:MoTe2 reference 1} depicts PL emission spectra of the hBN/MoTe$_2$/hBN heterostructure on the EDC cavity with the detection polarization perpendicular to the cavity mode. Again, a shift in photon energy is observed between $T \leq \SI{70}{\kelvin}$ and $T \geq \SI{79}{\kelvin}$ as described above. As described in the main text, this way we obtain emission from uncoupled excitons in the vicinity of the cavity mode. Each spectrum is fitted with two Lorentzian peaks (see also Fig.~\ref{fig:SI Exc fit} for a fit at $T = \SI{40}{\kelvin}$). One peak fits the excitonic transition, and the other fits the trion transition stimulating the low-$Q$ mode. This way, the excitonic transition energy $E_{\mathrm{exc}}(T)$ and the linewidth $\Gamma_{\mathrm{exc}}(T)$ are obtained for each temperature, cf. Fig.~\ref{fig:SI MoTe2 reference fit}. We observe that signatures of the cavity mode remain faintly visible for higher temperatures, possibly due to non-perfect polarization projection due to scattering from objects in the optical beam path, mainly from the glass window of the cryostat. The low-$Q$ mode~\cite{Schroder2025a} is visible throughout the whole measurement series and gets dominant at lower temperatures probably because of excitation by trions~\cite{Helmrich2018}.

As the temperature dependence of the excitonic transition energy is given by the temperature dependence of the band gap, the extracted $E_{\mathrm{exc}}(T)$ are fitted with the Varshni equation~\cite{Helmrich2018}
\begin{equation}
E_{\mathrm{exc}} (T) = E_{\mathrm{exc}}(0) - \dfrac{\alpha T^2}{T+\beta}
    \label{eq:SI Varshni}
\end{equation}
where $E_{\mathrm{exc}}(0)$, $\alpha$ and $\beta$ are material parameters. The temperature dependence of the exciton is expected to follow previously reported values, so $\alpha$ and $\beta$ are fixed~\cite{Helmrich2018}. This leaves $E_{\mathrm{exc}}(0)$ as the only free fitting parameter. The fit result is indicated as the red line in Fig.~\ref{fig:MoTe2 reference 1} and Fig.~\ref{fig:SI MoTe2 reference fit}(a), showing good agreement with the obtained PL spectra and deviates by the values found in Ref.~\cite{Helmrich2018} by a constant offset. A small but constant offset is likely to occur for different samples and can be explained by a different dielectric environment, yielding differences due to screening of the Coulomb interaction~\cite{Chernikov2014b}. The data presented in Ref.~\cite{Helmrich2018} are measured from a monolayer MoTe$_2$ on a SiO$_2$ substrate, whereas our flake is encapsulated in hBN and on a structured InP region (the EDC cavity). This shows that for a quantitative assessment of the reference excitonic energy and, in turn, for the extraction of the light-matter interaction strength, it is highly beneficial to extract the reference from the same sample. 

The temperature dependence of the exciton linewidth $\Gamma_{\mathrm{exc}}$ is given as~\cite{Helmrich2018}
\begin{equation}
    \Gamma_{\mathrm{exc}}(T) = \Gamma_0 + \gamma_{\mathrm{LA}} \, T + \dfrac{\Gamma_{\mathrm{LO}}}{\exp({E_{\mathrm{LO}}}/k_B T) -1},
    \label{eq:SI linewidth formular}
\end{equation}
where $\Gamma_0$ denotes the intrinsic linewidth, $\gamma_{\mathrm{LA}}$ the exciton-acoustic phonon strength, and the last term accounts for interaction with longitudinal optical phonons. The extracted linewidth from the spectra in Fig.~\ref{fig:MoTe2 reference 1} is depicted in Fig.~\ref{fig:SI MoTe2 reference fit}(b). A fit with Eq.~\ref{eq:SI linewidth formular} is indicated as the red line, and compared to the values obtained in Ref.~\cite{Helmrich2018}, indicated as the black line. The fit results agree qualitatively, and small deviations are likely to occur due to slightly altered exciton-phonon interaction in different samples. The fit result for $\Gamma_{\mathrm{exc}}$ has a large uncertainty at $T > \SI{150}{\kelvin}$ due to the spectral overlap of exciton and trion emission, but gives accurate result close to the region of interest $4 - \SI{140}{\kelvin}$ were we observe the avoided crossing (see main text). \textcolor{black}{At $T = \SI{4}{\kelvin}$, we extract an exciton linewidth of $\Gamma_{\mathrm{exc}} = \SI{5.1\pm0.9}{\milli\eV}$, in good agreement with the value observed in Ref.~\cite{Kutrowska-Girzycka2022}. This value is smaller than for ML MoTe$_2$ on a SiO$_2$ substrate (around $\SI{7}{\milli\eV}$~\cite{Robert2016, Helmrich2018}), confirming the benefit of hBN encapsulation. While the linewidth is larger than the value reported in~\cite{Han2018} ($\approx \SI{3}{\milli\eV}$ for an hBN encapsulated monolayer), a variation on the order of a few meV is often observed for different samples, and explained by local inhomogeneities~\cite{Kutrowska-Girzycka2022}.  }

The obtained temperature-dependencies for $E_{\mathrm{exc}}(T)$, $E_{\mathrm{cav}}(T)$, $\Gamma_{\mathrm{exc}}(T)$ and $\Gamma_{\mathrm{cav}}(T)$ allow fitting with a coupled-oscillator model with the light-matter interaction strength as the only free fitting parameter. The calculated detuning $\delta(T) = E_{\mathrm{cav}}(T)-E_{\mathrm{exc}}(T)$ is summarized in Tab.~\ref{tab:SI T det values}. The uncertainty for $T$ is assumed to be $\sigma_T = \pm 0.5$ K. The uncertainty in the detuning $\sigma_\delta$ is calculated from the uncertainty of $E_{\mathrm{exc}}$, $E_{\mathrm{cav}}$ and $T$ with Gaussian error propagation.

\begin{table}[h!]
\centering
\begin{tabular}{|c|c|c|}
\toprule
 $T \pm 0.5 $ (K) &  $\delta$ (meV) &  $\sigma_{\delta}$ (meV) \\
\midrule
4   & -3.3  & 0.7  \\
10  & -3.1  & 0.7  \\
20  & -2.4  & 0.7  \\
30  & -1.5  & 0.7  \\
40  & -0.2  & 0.7  \\
50  & 1.3   & 0.7  \\
60  & 3.0   & 0.7  \\
70  & 4.9   & 0.7  \\
79  & 6.2   & 1.0  \\
83  & 7.1   & 1.0  \\
93  & 9.6   & 1.0  \\
103 & 12.1  & 1.1  \\
113 & 14.8  & 1.1  \\
123 & 17.6  & 1.1  \\
133 & 20.4  & 1.1  \\
143 & 23.4  & 1.1  \\
153 & 26.4  & 1.1  \\
163 & 29.5  & 1.1  \\
173 & 32.6  & 1.1  \\
193 & 39.0  & 1.1  \\
213 & 45.5  & 1.1  \\
233 & 52.1  & 1.1  \\
253 & 58.8  & 1.1  \\
273 & 65.6  & 1.1  \\
293 & 72.4  & 1.1  \\
\bottomrule
\end{tabular}
\caption{Measured values for $T$ and corresponding detuning $\delta$ and uncertainty $\sigma_{\delta}$.}
\label{tab:SI T det values}
\end{table}

\newpage

\section{Fits at $T = \SI{40}{\kelvin}$}
\label{sec:SI Fits 40K}
In this section, we present fits of the spectra of the coupled system and of the references, exemplary at $T = \SI{40}{\kelvin}$, corresponding to $\delta = \SI{-0.2 \pm 0.8}{\milli\eV}$.

\begin{figure}[htb!]
         \centering
         \includegraphics[width = 0.7\linewidth]{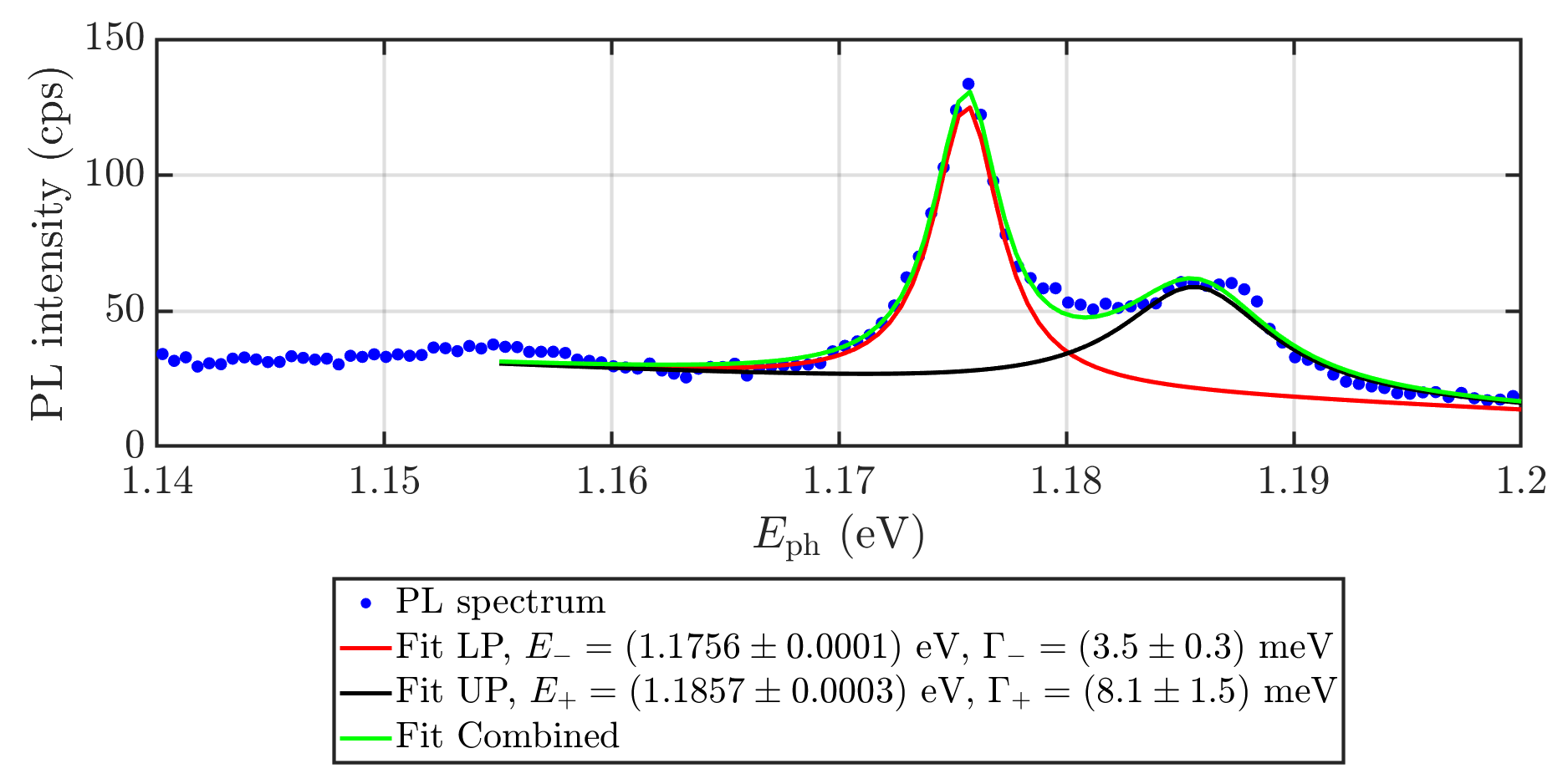}
         \caption{PL spectrum of the hBN/MoTe$_2$/hBN heterostructure on the EDC cavity at $T = \SI{40}{\kelvin}$ with the detection polarization oriented parallel to the cavity mode. Fitting the polaritonic emission is carried out as described in the main text.}
         \label{fig:SI Coupled fit}
\end{figure}

Fig.~\ref{fig:SI Coupled fit} depicts the PL spectrum from the sample consisting of the EDC cavity and the HS with the detection polarization aligned with the polarization of the cavity mode. The lower and upper polariton peaks on a linearly-sloped background are fitted with two Lorentzian lineshapes simultaneously, indicated by the red and black lines in Fig.~\ref{fig:SI Coupled fit}, respectively. The overall fit is indicated as the green line. From the fit, the resonance energies $E_- = \SI{1.176}{\eV}$ and $E_+ = \SI{1.186}{\eV}$, as well as the linewidths $\Gamma_- = \SI{3.5\pm0.3}{\milli\eV}$ and $\Gamma_+ = \SI{8.1\pm1.5}{\milli\eV}$ are obtained.

\begin{figure}[htb!]
         \centering
         \includegraphics[width = 0.7\linewidth]{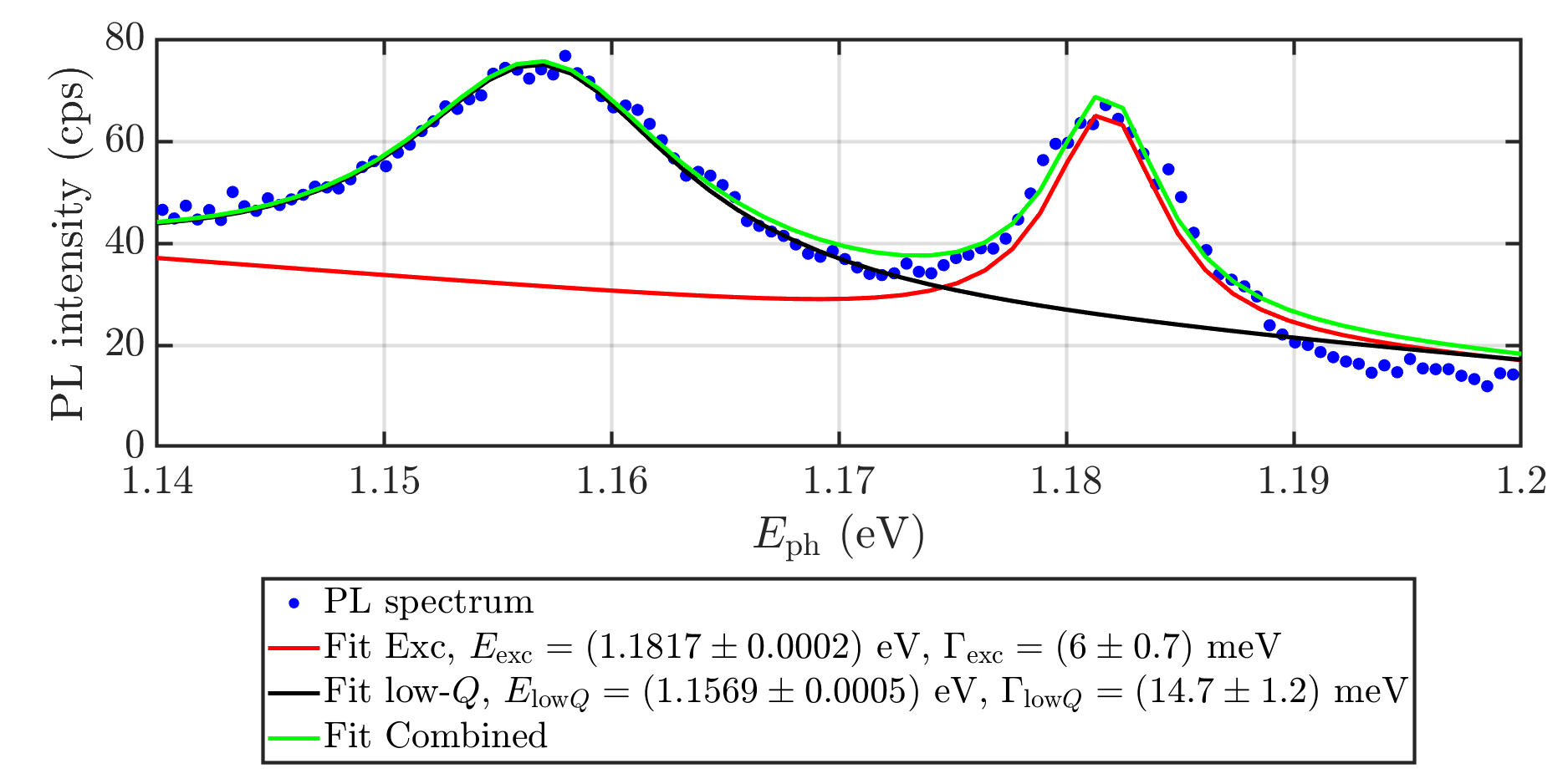}
         \caption{PL spectrum of the hBN/MoTe$_2$/hBN heterostructure on the EDC cavity at $T = \SI{40}{\kelvin}$ with the detection polarization oriented perpendicularly to the cavity mode. Fitting the excitonic emission and the low-$Q$ mode is carried out as described in the main text.}
         \label{fig:SI Exc fit}
\end{figure}

Fig.~\ref{fig:SI Exc fit} depicts the PL spectrum from the sample consisting of the EDC cavity and the HS with the detection polarization perpendicular to the polarization of the cavity mode. The excitonic emission and the low-$Q$ mode are fitted with two Lorentzian peaks on a linearly-sloped background simultaneously, indicated by the red and black lines in Fig.~\ref{fig:SI Exc fit}, respectively. The fit yields $E_{\mathrm{exc}} = \SI{1.182}{eV}$ and $\Gamma_{\mathrm{exc}} = \SI{6.0\pm0.7}{\milli\eV}$.

\begin{figure}[htb!]
         \centering
         \includegraphics[width = 0.7\linewidth]{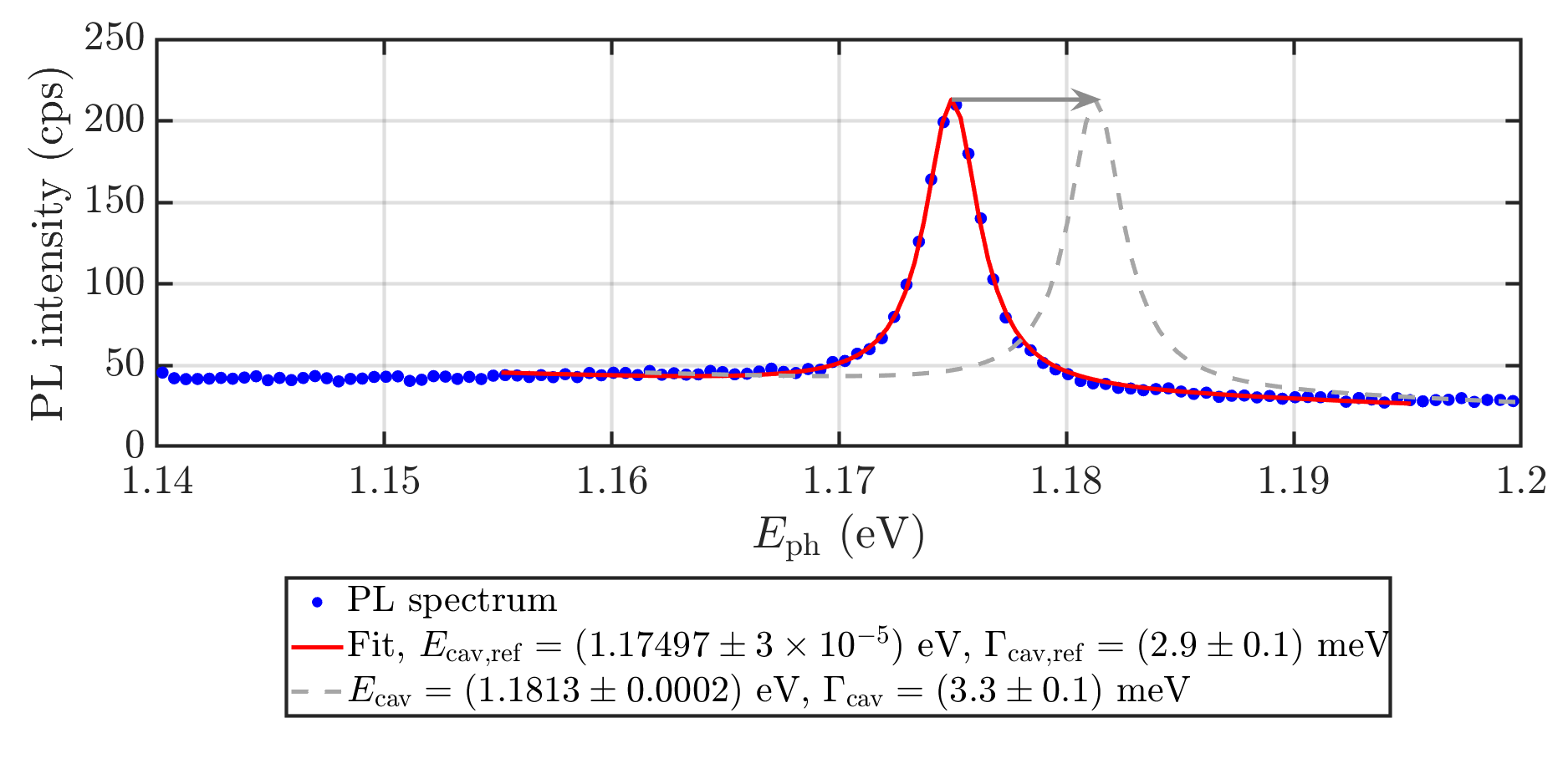}
         \caption{PL spectrum of the reference EDC cavity with the detection polarization parallel to the polarization of the cavity mode. Fitting the cavity mode is carried out as described in the main text. Moreover, the deduced lineshape for $E_{\mathrm{cav}}$ and $\Gamma_{\mathrm{cav}}$ (see Sec.~\ref{sec:SI reference}), is indicated as the dashed grey line. The length of the grey arrow equals the shift in photon energy $\Delta$, cf. Fig.~\ref{fig:SI Coupled ReferenceCavity}.}
         \label{fig:SI Cav fit}
\end{figure}

Fig.~\ref{fig:SI Cav fit} depicts the PL spectrum from the reference EDC cavity with the detection polarization parallel to the polarization of the cavity mode. The emission from the cavity is fitted with a Lorentzian peak on a linearly sloped background, indicated by the red line in Fig.~\ref{fig:SI Cav fit}. The fit yields $E_{\mathrm{cav,ref}} = \SI{1.175}{\eV}$, from which $E_{\mathrm{cav}}$ is deduced as $\SI{1.181}{\eV}$ is deduced, see Sec. \ref{sec:SI reference}.

\begin{figure}[htb!]
         \centering
         \includegraphics[width = 0.7\linewidth]{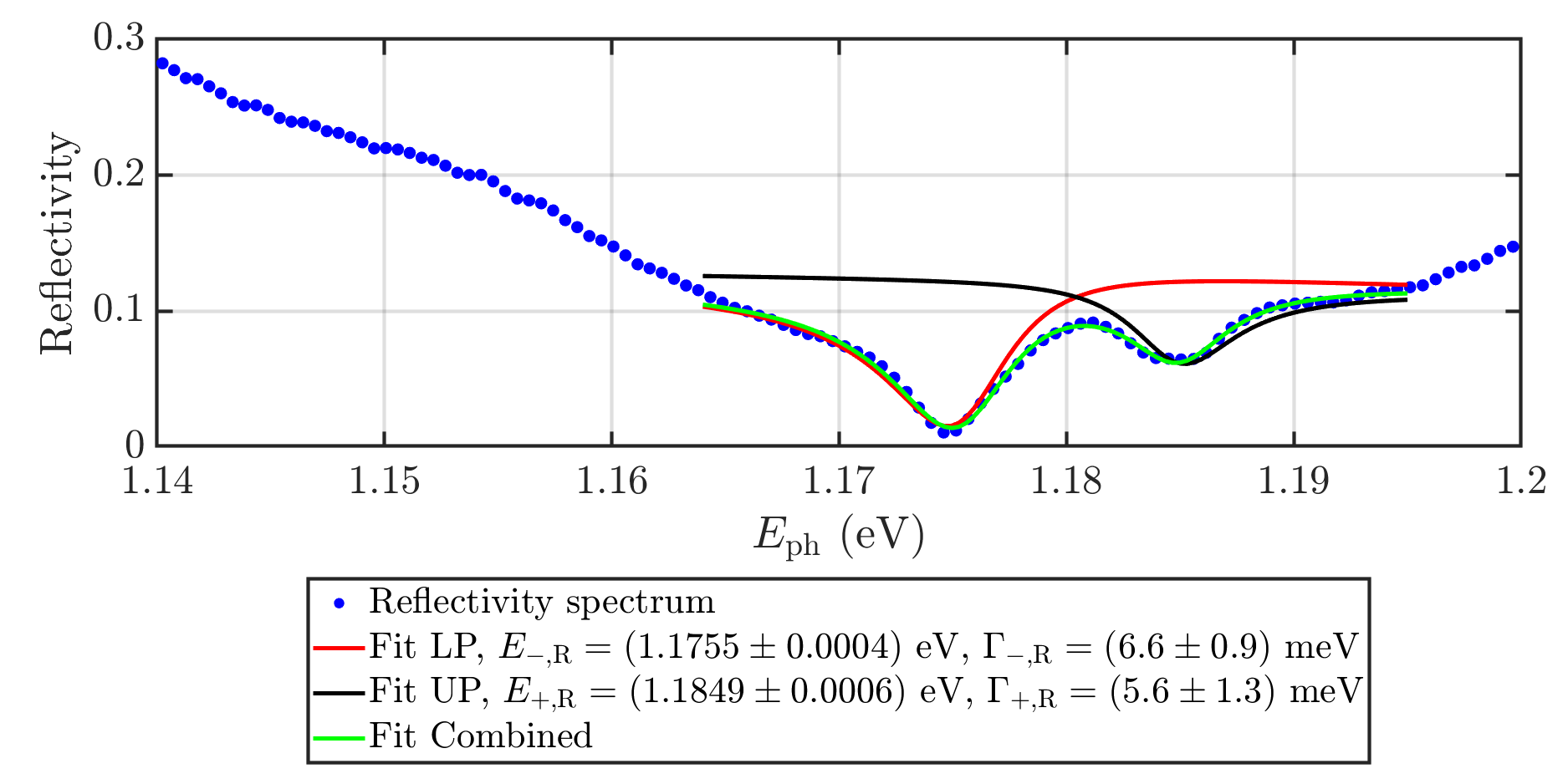}
         \caption{Reflectivity spectrum of the hBN/MoTe2/hBN heterostructure on the EDC cavity at $T = \SI{40}{K}$ in a cross-polarization configuration. Fitting the polaritonic features is carried out as described in the main text.}
         \label{fig:SI Coupled fit Fano}
\end{figure}

Fig.~\ref{fig:SI Coupled fit Fano} depicts the reflectivity spectrum from the sample consisting of the EDC cavity and the HS in a cross-polarization configuration. In reflection measurements, resonances often exhibit Fano lineshapes, resulting from a superposition of a Lorentzian resonance with a slowly-varying background~\cite{Schroder2025a}. Therefore, we fit the reflectivity $R$ in a spectral range close to $E_-$ and $E_+$ with the equation \cite{Danielsen2025a}
\begin{equation}
    R(E_{\mathrm{ph}}) = \left|\dfrac{a_- \, \Gamma_-/2\, \mathrm{e}^{i \varphi_1}}{i(E_{\mathrm{ph}} - E_-) + \Gamma_-/2} + \dfrac{a_+ \, \Gamma_+/2 \,\mathrm{e}^{i \varphi_2}}{i(E_{\mathrm{ph}} - E_+) + \Gamma_+/2} + \dfrac{a_{\mathrm{BG}} \, \Gamma_{\mathrm{BG}}/2}{i(E_{\mathrm{ph}} - E_{\mathrm{BG}}) + \Gamma_{\mathrm{BG}}/2} \right|^2 .
    \label{eq:SI Double Fano}
\end{equation}

Here, $a_-$, $a_+$ and $a_{\mathrm{BG}}$ denote amplitudes for the lower and the upper polaritons and the broad background, respectively, for the chosen polarization settings, and $\varphi_1$ and $\varphi_2$ take into account potential phase differences between the resonances. From the fit, the resonance energies $E_{-, \mathrm{R}} = \SI{1.176}{\eV}$ and $E_{+, \mathrm{R}} = \SI{1.185}{\eV}$, as well as the linewidths $\Gamma_{-, \mathrm{R}} = \SI{6.6\pm0.9}{\milli\eV}$ and $\Gamma_{-, \mathrm{R}} = \SI{5.6\pm1.3}{\milli\eV}$ are found. The subscript "R" indicates that the values are deduced from reflection measurements. These values are in good agreement with the values determined from the PL measurements.

\newpage

\section{Fit with coupled-oscillator model}\label{sec:SI COM fit}
The interaction of light and matter can be described by a coupled-oscillator model, where the electromagnetic field in the cavity and the excitonic transition are modeled by two individual harmonic oscillators. The eigenenergies of the coupled states are determined by the eigenvalues of the matrix~\cite{Schneider2018, Carlson2021}
\begin{equation}
E_{\pm} = \mathrm{EV}\left(\begin{bmatrix}
E_{\mathrm{cav}} - \dfrac{i \Gamma_{\mathrm{cav}}}{2} & g  \\
g & E_{\mathrm{exc}} - \dfrac{i \Gamma_{\mathrm{exc}}}{2}
\end{bmatrix}\right)
\label{eq:Oscillator Matrix Eph}
\end{equation}
Diagonalizing Eq.~\ref{eq:Oscillator Matrix Eph} yields the dispersion relation for the upper ($+$) and lower ($-$) polariton
\begin{equation}
\begin{split}
    E_{\pm} = \dfrac{1}{2} (E_{\mathrm{cav}}+E_{\mathrm{exc}}) - \dfrac{i}{4}(\Gamma_{\mathrm{cav}} - \Gamma_{\mathrm{exc}}) \\ 
    \pm \dfrac{1}{2} \sqrt{4g^2+ \left(\delta-\dfrac{i}{2}(\Gamma_{\mathrm{cav}}-\Gamma_{\mathrm{exc}})\right)^2}.
    \end{split}
    \label{eq:Dispersion Relation}
\end{equation}

From the dispersion relation in Eq.~\ref{eq:Dispersion Relation}, the Rabi splitting is given by the energy difference at zero detuning
\begin{equation}
    E_{\mathrm{Rabi}} = \left. E_+ - E_- \right|_{\delta = 0} = \sqrt{4g^2-\dfrac{(\Gamma_{\mathrm{cav}}-\Gamma_{\mathrm{exc}})^2}{4}}.
    \label{eq:Rabi frequency}
\end{equation}
The system is referred to as strongly coupled if $E_{\mathrm{Rabi}} > (\Gamma_{\mathrm{cav}}+\Gamma_{\mathrm{exc}})/2$, see for example Ref.~\cite{Goncalves2020, Hu2020} and references therein. 

\begin{figure}[htb!]
         \centering
         \includegraphics[width = \linewidth]{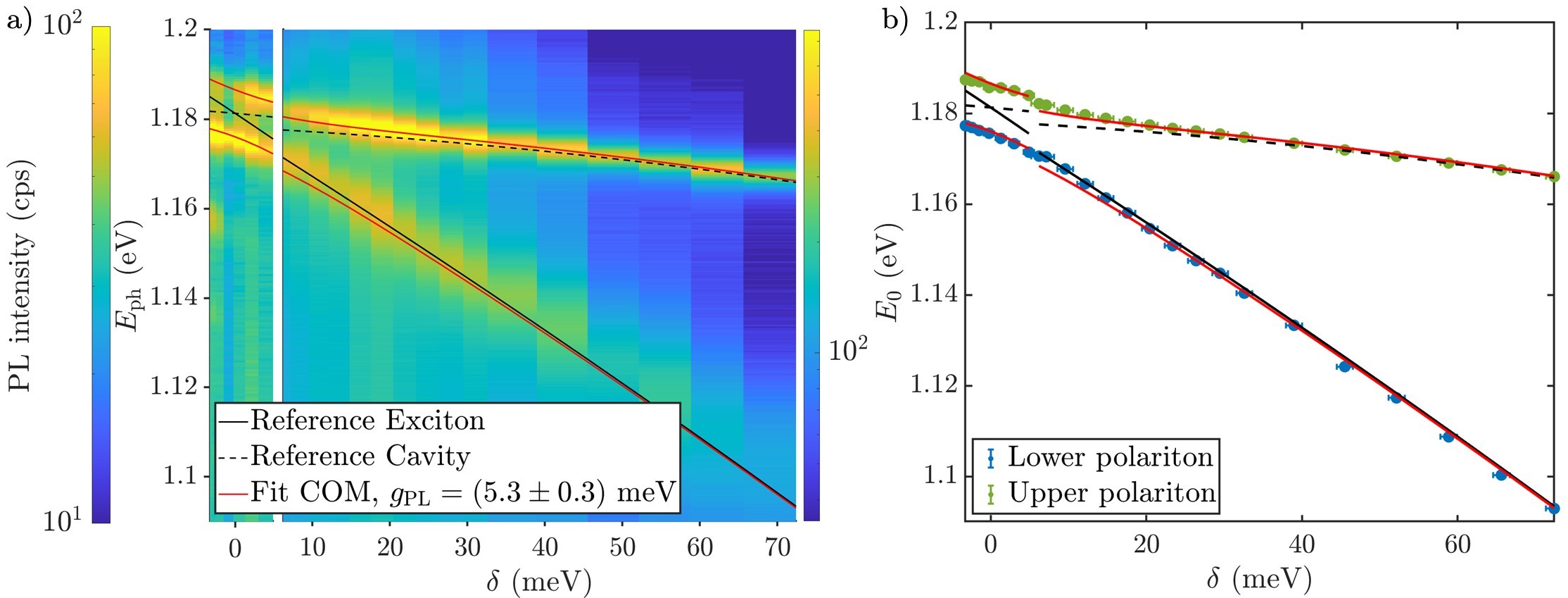}
         \caption{a) PL spectra of the coupled system as a function of detuning over the full temperature range, recorded with the detection polarization aligned with the cavity mode. b) Extracted peak positions from the PL measurements. The solid and dashed black lines depict the reference for the exciton and for the cavity mode, respectively, see Sec.~\ref{sec:SI reference}. A fit with the real part of the coupled-oscillator model is indicated by the red lines.}
         \label{fig:SI Fit COM PL}
\end{figure}

Fig.~\ref{fig:SI Fit COM PL}(a) depicts PL spectra shown in Fig.~\ref{fig:PL spectra heatmap peak} in the main text over the full temperature range, together with fits from reference measurements of the uncoupled cavity and of the uncoupled excitons. As described in Sec.~\ref{sec:SI reference}, a shift in photon energy of $\approx \SI{3}{\milli\eV}$ is observed between $T \leq \SI{70}{\kelvin}$ and $T \geq \SI{79}{\kelvin}$, due to imperfect alignment of the respective spectrometers used in Setups 1 and 2. The extracted peak positions from a Lorentzian fit (cf. Fig.~\ref{fig:SI Coupled fit}) are depicted in Fig.~\ref{fig:SI Fit COM PL}(b). Together with the obtained values for $E_{\mathrm{cav}}$, $E_{\mathrm{exc}}$, $\Gamma_{\mathrm{cav}}$ and $\Gamma_{\mathrm{exc}}$ (see Sec.~\ref{sec:SI reference}), these are fitted with the real part of the dispersion relation (Eq.~\ref{eq:Dispersion Relation}), leaving $g$ as the only free fitting parameter. The fit of the PL measurements, indicated by the red lines, yields $g_{\mathrm{PL}} = \SI{5.3\pm0.3}{\milli\eV}$.

\begin{figure}[htb!]
         \centering
         \includegraphics[width = 0.9\linewidth]{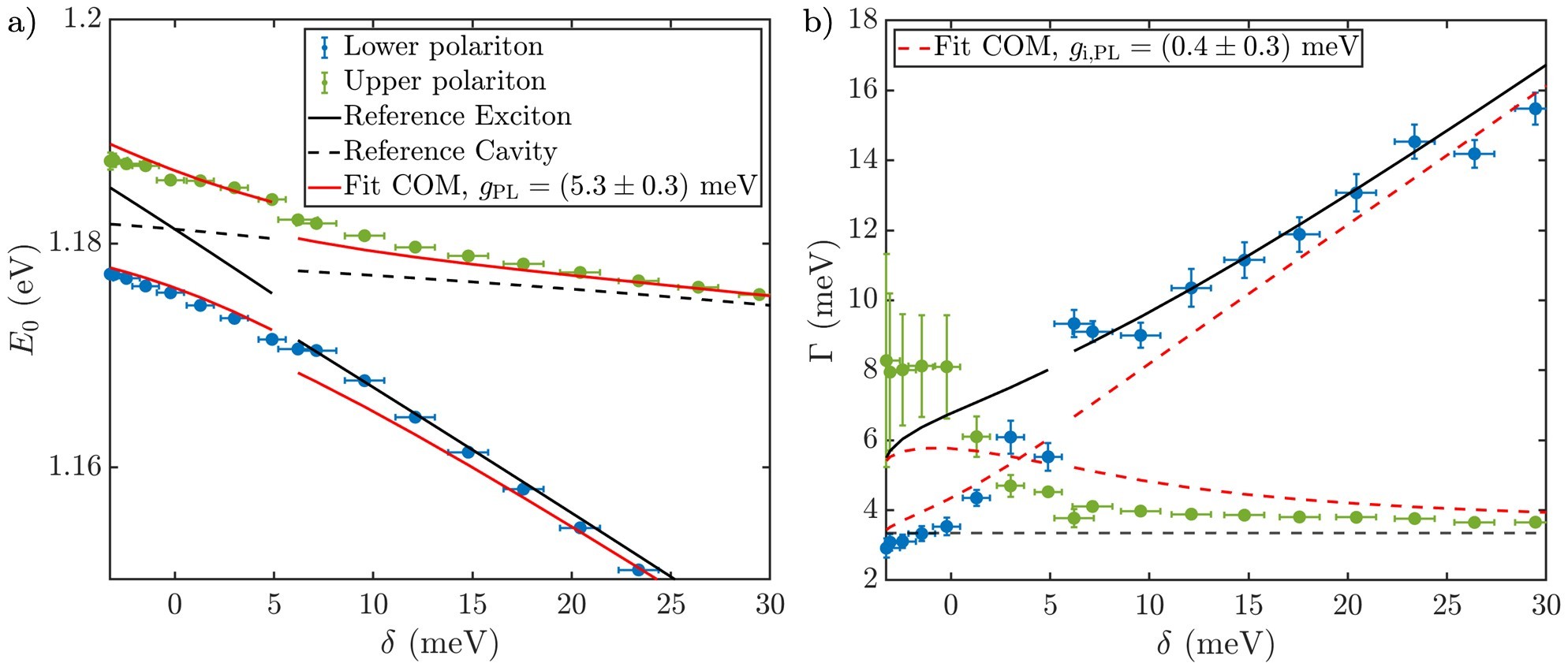}
         \caption{a) Extracted peak positions and b) FWHM of the PL measurements of the coupled system in a region of interest around zero detuning. A fit with the full complex coupled-oscillator model is indicated as solid and dashed red lines for the real and the imaginary parts, respectively.}
         \label{fig:SI Fit COM PL complex}
\end{figure}

In general, the light-matter interaction strength is complex $\tilde{g} = g - ig_{i}$~\cite{Abutoama2024, Bleu2024}. To evaluate the magnitude $g_{i}$, the complex eigenvalues $\tilde{E}_-= E_- - i \Gamma_-/2$ and $\tilde{E}_+=E_+ - i \Gamma_+/2$ are fitted with the real and imginary parts of the dispersion relation (Eq.~\ref{eq:Dispersion Relation}) simultaneously with a complex fit parameter $\tilde{g}$, exemplary for the PL measurements. Fig.~\ref{fig:SI Fit COM PL complex}(a) depicts the extracted peak positions corresponding to $\Re({\tilde{E_\pm}})$ and (b) shows the extracted linewidth, corresponding to $-2\, \Im({\tilde{E_\pm}})$. The fit, indicated as the red solid and dashed lines in (a) and (b), respectively, yields $\tilde{g} = ( (5.3 -0.4i) \pm (0.3 + 0.3i) ) \si{\meV} $. The coupled-oscillator model describes the imaginary part of the spectrum very well, which is additional confirmation of the reported strong light-matter interaction. As the imaginary part of $\tilde{g}$ is less than $\SI{6}{\percent}$ of the real part and is almost zero within the error bars, it is justified to focus on the real part of the coupled-oscillator model in this case.

\begin{figure}[htb!]
         \centering
         \includegraphics[width = \linewidth]{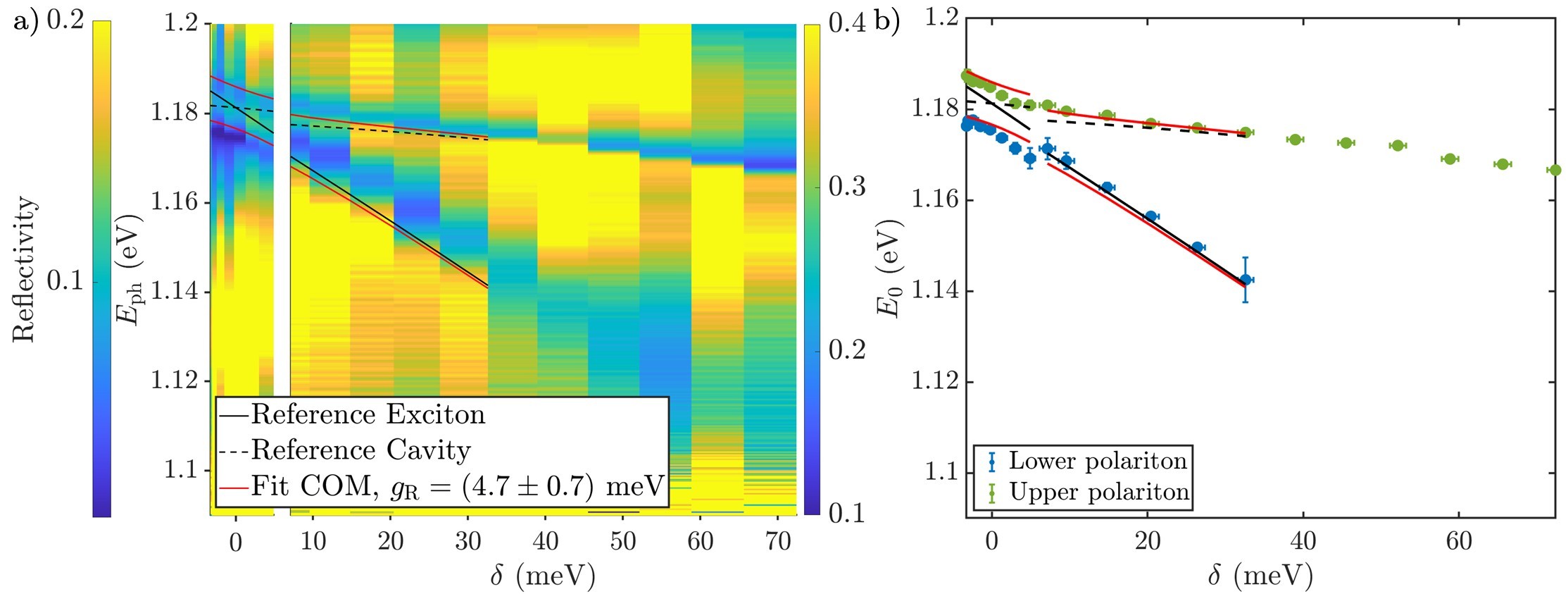}
         \caption{a) Reflectivity spectra of the coupled system as a function of detuning over the full temperature range, recorded with the detection polarization aligned with the cavity mode. b) Extracted peak positions from the reflectivity measurements. The solid and dashed black lines depict the reference for the exciton and for the cavity mode, respectively, see Sec.~\ref{sec:SI reference}. A fit with the real part of the coupled-oscillator model is indicated by the red lines.}
         \label{fig:SI Fit COM Ref}
\end{figure}

Fig.~\ref{fig:SI Fit COM Ref}(a) depicts reflectivity spectra shown in Fig.~\ref{fig:PL spectra heatmap peak} in the main text over the full temperature range. The extracted peak positions from a Fano fit (cf. Fig.~\ref{fig:SI Coupled fit Fano}) are depicted in Fig.~\ref{fig:SI Fit COM Ref}(b). A fit with the real part of the coupled-oscillator model is carried out similarly as for the PL measurements, yielding $g_{\mathrm{R}} = \SI{4.7\pm 0.7}{\milli\eV}$. For the reflection measurements, the fit is restricted to $T < \SI{180}{\kelvin}$ since the signal-to-noise ratio of the lower polariton becomes small, which is associated with large uncertainties in the fit parameters.

\pagebreak

\bibliography{Lit_StrongCoupling}

\begin{thebibliography}{104}%
\makeatletter
\providecommand \@ifxundefined [1]{%
 \@ifx{#1\undefined}
}%
\providecommand \@ifnum [1]{%
 \ifnum #1\expandafter \@firstoftwo
 \else \expandafter \@secondoftwo
 \fi
}%
\providecommand \@ifx [1]{%
 \ifx #1\expandafter \@firstoftwo
 \else \expandafter \@secondoftwo
 \fi
}%
\providecommand \natexlab [1]{#1}%
\providecommand \enquote  [1]{``#1''}%
\providecommand \bibnamefont  [1]{#1}%
\providecommand \bibfnamefont [1]{#1}%
\providecommand \citenamefont [1]{#1}%
\providecommand \href@noop [0]{\@secondoftwo}%
\providecommand \href [0]{\begingroup \@sanitize@url \@href}%
\providecommand \@href[1]{\@@startlink{#1}\@@href}%
\providecommand \@@href[1]{\endgroup#1\@@endlink}%
\providecommand \@sanitize@url [0]{\catcode `\\12\catcode `\$12\catcode `\&12\catcode `\#12\catcode `\^12\catcode `\_12\catcode `\%12\relax}%
\providecommand \@@startlink[1]{}%
\providecommand \@@endlink[0]{}%
\providecommand \url  [0]{\begingroup\@sanitize@url \@url }%
\providecommand \@url [1]{\endgroup\@href {#1}{\urlprefix }}%
\providecommand \urlprefix  [0]{URL }%
\providecommand \Eprint [0]{\href }%
\providecommand \doibase [0]{https://doi.org/}%
\providecommand \selectlanguage [0]{\@gobble}%
\providecommand \bibinfo  [0]{\@secondoftwo}%
\providecommand \bibfield  [0]{\@secondoftwo}%
\providecommand \translation [1]{[#1]}%
\providecommand \BibitemOpen [0]{}%
\providecommand \bibitemStop [0]{}%
\providecommand \bibitemNoStop [0]{.\EOS\space}%
\providecommand \EOS [0]{\spacefactor3000\relax}%
\providecommand \BibitemShut  [1]{\csname bibitem#1\endcsname}%
\let\auto@bib@innerbib\@empty
\bibitem [{\citenamefont {Splendiani}\ \emph {et~al.}(2010)\citenamefont {Splendiani}, \citenamefont {Sun}, \citenamefont {Zhang}, \citenamefont {Li}, \citenamefont {Kim}, \citenamefont {Chim}, \citenamefont {Galli},\ and\ \citenamefont {Wang}}]{Splendiani2010}%
  \BibitemOpen
  \bibfield  {author} {\bibinfo {author} {\bibfnamefont {A.}~\bibnamefont {Splendiani}}, \bibinfo {author} {\bibfnamefont {L.}~\bibnamefont {Sun}}, \bibinfo {author} {\bibfnamefont {Y.}~\bibnamefont {Zhang}}, \bibinfo {author} {\bibfnamefont {T.}~\bibnamefont {Li}}, \bibinfo {author} {\bibfnamefont {J.}~\bibnamefont {Kim}}, \bibinfo {author} {\bibfnamefont {C.~Y.}\ \bibnamefont {Chim}}, \bibinfo {author} {\bibfnamefont {G.}~\bibnamefont {Galli}},\ and\ \bibinfo {author} {\bibfnamefont {F.}~\bibnamefont {Wang}},\ }\bibfield  {title} {\bibinfo {title} {{Emerging photoluminescence in monolayer MoS2}},\ }\href {https://doi.org/10.1021/nl903868w} {\bibfield  {journal} {\bibinfo  {journal} {Nano Lett.}\ }\textbf {\bibinfo {volume} {10}},\ \bibinfo {pages} {1271} (\bibinfo {year} {2010})}\BibitemShut {NoStop}%
\bibitem [{\citenamefont {Mak}\ \emph {et~al.}(2010)\citenamefont {Mak}, \citenamefont {Lee}, \citenamefont {Hone}, \citenamefont {Shan},\ and\ \citenamefont {Heinz}}]{Mak2010}%
  \BibitemOpen
  \bibfield  {author} {\bibinfo {author} {\bibfnamefont {K.~F.}\ \bibnamefont {Mak}}, \bibinfo {author} {\bibfnamefont {C.}~\bibnamefont {Lee}}, \bibinfo {author} {\bibfnamefont {J.}~\bibnamefont {Hone}}, \bibinfo {author} {\bibfnamefont {J.}~\bibnamefont {Shan}},\ and\ \bibinfo {author} {\bibfnamefont {T.~F.}\ \bibnamefont {Heinz}},\ }\bibfield  {title} {\bibinfo {title} {{Atomically Thin MoS2: A New Direct-Gap Semiconductor}},\ }\href {https://doi.org/10.1103/PhysRevLett.105.136805} {\bibfield  {journal} {\bibinfo  {journal} {Phys. Rev. Lett.}\ }\textbf {\bibinfo {volume} {105}},\ \bibinfo {pages} {136805} (\bibinfo {year} {2010})}\BibitemShut {NoStop}%
\bibitem [{\citenamefont {He}\ \emph {et~al.}(2014)\citenamefont {He}, \citenamefont {Kumar}, \citenamefont {Zhao}, \citenamefont {Wang}, \citenamefont {Mak}, \citenamefont {Zhao},\ and\ \citenamefont {Shan}}]{He2014}%
  \BibitemOpen
  \bibfield  {author} {\bibinfo {author} {\bibfnamefont {K.}~\bibnamefont {He}}, \bibinfo {author} {\bibfnamefont {N.}~\bibnamefont {Kumar}}, \bibinfo {author} {\bibfnamefont {L.}~\bibnamefont {Zhao}}, \bibinfo {author} {\bibfnamefont {Z.}~\bibnamefont {Wang}}, \bibinfo {author} {\bibfnamefont {K.~F.}\ \bibnamefont {Mak}}, \bibinfo {author} {\bibfnamefont {H.}~\bibnamefont {Zhao}},\ and\ \bibinfo {author} {\bibfnamefont {J.}~\bibnamefont {Shan}},\ }\bibfield  {title} {\bibinfo {title} {{Tightly Bound Excitons in Monolayer WSe2}},\ }\href {https://doi.org/10.1103/PhysRevLett.113.026803} {\bibfield  {journal} {\bibinfo  {journal} {Phys. Rev. Lett.}\ }\textbf {\bibinfo {volume} {113}},\ \bibinfo {pages} {026803} (\bibinfo {year} {2014})}\BibitemShut {NoStop}%
\bibitem [{\citenamefont {Chernikov}\ \emph {et~al.}(2014)\citenamefont {Chernikov}, \citenamefont {Berkelbach}, \citenamefont {Hill}, \citenamefont {Rigosi}, \citenamefont {Li}, \citenamefont {Aslan}, \citenamefont {Reichman}, \citenamefont {Hybertsen},\ and\ \citenamefont {Heinz}}]{Chernikov2014b}%
  \BibitemOpen
  \bibfield  {author} {\bibinfo {author} {\bibfnamefont {A.}~\bibnamefont {Chernikov}}, \bibinfo {author} {\bibfnamefont {T.~C.}\ \bibnamefont {Berkelbach}}, \bibinfo {author} {\bibfnamefont {H.~M.}\ \bibnamefont {Hill}}, \bibinfo {author} {\bibfnamefont {A.}~\bibnamefont {Rigosi}}, \bibinfo {author} {\bibfnamefont {Y.}~\bibnamefont {Li}}, \bibinfo {author} {\bibfnamefont {B.}~\bibnamefont {Aslan}}, \bibinfo {author} {\bibfnamefont {D.~R.}\ \bibnamefont {Reichman}}, \bibinfo {author} {\bibfnamefont {M.~S.}\ \bibnamefont {Hybertsen}},\ and\ \bibinfo {author} {\bibfnamefont {T.~F.}\ \bibnamefont {Heinz}},\ }\bibfield  {title} {\bibinfo {title} {{Exciton Binding Energy and Nonhydrogenic Rydberg Series in Monolayer WSe2}},\ }\href {https://doi.org/10.1103/PhysRevLett.113.076802} {\bibfield  {journal} {\bibinfo  {journal} {Phys. Rev. Lett.}\ }\textbf {\bibinfo {volume} {113}},\ \bibinfo {pages} {076802} (\bibinfo {year} {2014})}\BibitemShut {NoStop}%
\bibitem [{\citenamefont {Li}\ \emph {et~al.}(2014)\citenamefont {Li}, \citenamefont {Chernikov}, \citenamefont {Zhang}, \citenamefont {Rigosi}, \citenamefont {Hill}, \citenamefont {van~der Zande}, \citenamefont {Chenet}, \citenamefont {Shih}, \citenamefont {Hone},\ and\ \citenamefont {Heinz}}]{Li2014a}%
  \BibitemOpen
  \bibfield  {author} {\bibinfo {author} {\bibfnamefont {Y.}~\bibnamefont {Li}}, \bibinfo {author} {\bibfnamefont {A.}~\bibnamefont {Chernikov}}, \bibinfo {author} {\bibfnamefont {X.}~\bibnamefont {Zhang}}, \bibinfo {author} {\bibfnamefont {A.}~\bibnamefont {Rigosi}}, \bibinfo {author} {\bibfnamefont {H.~M.}\ \bibnamefont {Hill}}, \bibinfo {author} {\bibfnamefont {A.~M.}\ \bibnamefont {van~der Zande}}, \bibinfo {author} {\bibfnamefont {D.~A.}\ \bibnamefont {Chenet}}, \bibinfo {author} {\bibfnamefont {E.-M.}\ \bibnamefont {Shih}}, \bibinfo {author} {\bibfnamefont {J.}~\bibnamefont {Hone}},\ and\ \bibinfo {author} {\bibfnamefont {T.~F.}\ \bibnamefont {Heinz}},\ }\bibfield  {title} {\bibinfo {title} {{Measurement of the optical dielectric function of monolayer transition-metal dichalcogenides: MoS2, MoSe2, WS2 and WSe2}},\ }\href {https://doi.org/10.1103/PhysRevB.90.205422} {\bibfield  {journal} {\bibinfo  {journal} {Phys. Rev. B}\ }\textbf {\bibinfo {volume} {90}},\ \bibinfo {pages} {205422} (\bibinfo {year}
  {2014})}\BibitemShut {NoStop}%
\bibitem [{\citenamefont {Munkhbat}\ \emph {et~al.}(2022)\citenamefont {Munkhbat}, \citenamefont {Wr{\'{o}}bel}, \citenamefont {Antosiewicz},\ and\ \citenamefont {Shegai}}]{Munkhbat2022}%
  \BibitemOpen
  \bibfield  {author} {\bibinfo {author} {\bibfnamefont {B.}~\bibnamefont {Munkhbat}}, \bibinfo {author} {\bibfnamefont {P.}~\bibnamefont {Wr{\'{o}}bel}}, \bibinfo {author} {\bibfnamefont {T.~J.}\ \bibnamefont {Antosiewicz}},\ and\ \bibinfo {author} {\bibfnamefont {T.~O.}\ \bibnamefont {Shegai}},\ }\bibfield  {title} {\bibinfo {title} {{Optical Constants of Several Multilayer Transition Metal Dichalcogenides Measured by Spectroscopic Ellipsometry in the 300–1700 nm Range: High Index, Anisotropy, and Hyperbolicity}},\ }\href {https://doi.org/10.1021/acsphotonics.2c00433} {\bibfield  {journal} {\bibinfo  {journal} {ACS Photonics}\ }\textbf {\bibinfo {volume} {9}},\ \bibinfo {pages} {2398} (\bibinfo {year} {2022})}\BibitemShut {NoStop}%
\bibitem [{\citenamefont {Robert}\ \emph {et~al.}(2016{\natexlab{a}})\citenamefont {Robert}, \citenamefont {Lagarde}, \citenamefont {Cadiz}, \citenamefont {Wang}, \citenamefont {Lassagne}, \citenamefont {Amand}, \citenamefont {Balocchi}, \citenamefont {Renucci}, \citenamefont {Tongay}, \citenamefont {Urbaszek},\ and\ \citenamefont {Marie}}]{Robert2016a}%
  \BibitemOpen
  \bibfield  {author} {\bibinfo {author} {\bibfnamefont {C.}~\bibnamefont {Robert}}, \bibinfo {author} {\bibfnamefont {D.}~\bibnamefont {Lagarde}}, \bibinfo {author} {\bibfnamefont {F.}~\bibnamefont {Cadiz}}, \bibinfo {author} {\bibfnamefont {G.}~\bibnamefont {Wang}}, \bibinfo {author} {\bibfnamefont {B.}~\bibnamefont {Lassagne}}, \bibinfo {author} {\bibfnamefont {T.}~\bibnamefont {Amand}}, \bibinfo {author} {\bibfnamefont {A.}~\bibnamefont {Balocchi}}, \bibinfo {author} {\bibfnamefont {P.}~\bibnamefont {Renucci}}, \bibinfo {author} {\bibfnamefont {S.}~\bibnamefont {Tongay}}, \bibinfo {author} {\bibfnamefont {B.}~\bibnamefont {Urbaszek}},\ and\ \bibinfo {author} {\bibfnamefont {X.}~\bibnamefont {Marie}},\ }\bibfield  {title} {\bibinfo {title} {{Exciton radiative lifetime in transition metal dichalcogenide monolayers}},\ }\href {https://doi.org/10.1103/PhysRevB.93.205423} {\bibfield  {journal} {\bibinfo  {journal} {Phys. Rev. B}\ }\textbf {\bibinfo {volume} {93}},\ \bibinfo {pages} {1} (\bibinfo {year}
  {2016}{\natexlab{a}})}\BibitemShut {NoStop}%
\bibitem [{\citenamefont {Robert}\ \emph {et~al.}(2016{\natexlab{b}})\citenamefont {Robert}, \citenamefont {Picard}, \citenamefont {Lagarde}, \citenamefont {Wang}, \citenamefont {Echeverry}, \citenamefont {Cadiz}, \citenamefont {Renucci}, \citenamefont {H{\"{o}}gele}, \citenamefont {Amand}, \citenamefont {Marie}, \citenamefont {Gerber},\ and\ \citenamefont {Urbaszek}}]{Robert2016}%
  \BibitemOpen
  \bibfield  {author} {\bibinfo {author} {\bibfnamefont {C.}~\bibnamefont {Robert}}, \bibinfo {author} {\bibfnamefont {R.}~\bibnamefont {Picard}}, \bibinfo {author} {\bibfnamefont {D.}~\bibnamefont {Lagarde}}, \bibinfo {author} {\bibfnamefont {G.}~\bibnamefont {Wang}}, \bibinfo {author} {\bibfnamefont {J.~P.}\ \bibnamefont {Echeverry}}, \bibinfo {author} {\bibfnamefont {F.}~\bibnamefont {Cadiz}}, \bibinfo {author} {\bibfnamefont {P.}~\bibnamefont {Renucci}}, \bibinfo {author} {\bibfnamefont {A.}~\bibnamefont {H{\"{o}}gele}}, \bibinfo {author} {\bibfnamefont {T.}~\bibnamefont {Amand}}, \bibinfo {author} {\bibfnamefont {X.}~\bibnamefont {Marie}}, \bibinfo {author} {\bibfnamefont {I.~C.}\ \bibnamefont {Gerber}},\ and\ \bibinfo {author} {\bibfnamefont {B.}~\bibnamefont {Urbaszek}},\ }\bibfield  {title} {\bibinfo {title} {{Excitonic properties of semiconducting monolayer and bilayer MoTe2}},\ }\href {https://doi.org/10.1103/PhysRevB.94.155425} {\bibfield  {journal} {\bibinfo  {journal} {Phys. Rev. B}\ }\textbf
  {\bibinfo {volume} {94}},\ \bibinfo {pages} {155425} (\bibinfo {year} {2016}{\natexlab{b}})}\BibitemShut {NoStop}%
\bibitem [{\citenamefont {Schneider}\ \emph {et~al.}(2018)\citenamefont {Schneider}, \citenamefont {Glazov}, \citenamefont {Korn}, \citenamefont {H{\"{o}}fling},\ and\ \citenamefont {Urbaszek}}]{Schneider2018}%
  \BibitemOpen
  \bibfield  {author} {\bibinfo {author} {\bibfnamefont {C.}~\bibnamefont {Schneider}}, \bibinfo {author} {\bibfnamefont {M.~M.}\ \bibnamefont {Glazov}}, \bibinfo {author} {\bibfnamefont {T.}~\bibnamefont {Korn}}, \bibinfo {author} {\bibfnamefont {S.}~\bibnamefont {H{\"{o}}fling}},\ and\ \bibinfo {author} {\bibfnamefont {B.}~\bibnamefont {Urbaszek}},\ }\bibfield  {title} {\bibinfo {title} {{Two-dimensional semiconductors in the regime of strong light-matter coupling}},\ }\href {https://doi.org/10.1038/s41467-018-04866-6} {\bibfield  {journal} {\bibinfo  {journal} {Nat. Commun.}\ }\textbf {\bibinfo {volume} {9}},\ \bibinfo {pages} {2695} (\bibinfo {year} {2018})}\BibitemShut {NoStop}%
\bibitem [{\citenamefont {T{\"{o}}rm{\"{a}}}\ and\ \citenamefont {Barnes}(2015)}]{Tormo2015}%
  \BibitemOpen
  \bibfield  {author} {\bibinfo {author} {\bibfnamefont {P.}~\bibnamefont {T{\"{o}}rm{\"{a}}}}\ and\ \bibinfo {author} {\bibfnamefont {W.~L.}\ \bibnamefont {Barnes}},\ }\bibfield  {title} {\bibinfo {title} {{Strong coupling between surface plasmon polaritons and emitters: a review}},\ }\href {https://doi.org/10.1088/0034-4885/78/1/013901} {\bibfield  {journal} {\bibinfo  {journal} {Reports Prog. Phys.}\ }\textbf {\bibinfo {volume} {78}},\ \bibinfo {pages} {013901} (\bibinfo {year} {2015})}\BibitemShut {NoStop}%
\bibitem [{\citenamefont {Gon{\c{c}}alves}\ \emph {et~al.}(2020)\citenamefont {Gon{\c{c}}alves}, \citenamefont {Stenger}, \citenamefont {Cox}, \citenamefont {Mortensen},\ and\ \citenamefont {Xiao}}]{Goncalves2020}%
  \BibitemOpen
  \bibfield  {author} {\bibinfo {author} {\bibfnamefont {P.~A.}\ \bibnamefont {Gon{\c{c}}alves}}, \bibinfo {author} {\bibfnamefont {N.}~\bibnamefont {Stenger}}, \bibinfo {author} {\bibfnamefont {J.~D.}\ \bibnamefont {Cox}}, \bibinfo {author} {\bibfnamefont {N.~A.}\ \bibnamefont {Mortensen}},\ and\ \bibinfo {author} {\bibfnamefont {S.}~\bibnamefont {Xiao}},\ }\bibfield  {title} {\bibinfo {title} {{Strong Light–Matter Interactions Enabled by Polaritons in Atomically Thin Materials}},\ }\href {https://doi.org/10.1002/adom.201901473} {\bibfield  {journal} {\bibinfo  {journal} {Adv. Opt. Mater.}\ }\textbf {\bibinfo {volume} {8}},\ \bibinfo {pages} {1} (\bibinfo {year} {2020})}\BibitemShut {NoStop}%
\bibitem [{\citenamefont {Tserkezis}\ \emph {et~al.}(2020)\citenamefont {Tserkezis}, \citenamefont {Fern{\'{a}}ndez-Dom{\'{i}}nguez}, \citenamefont {Gon{\c{c}}alves}, \citenamefont {Todisco}, \citenamefont {Cox}, \citenamefont {Busch}, \citenamefont {Stenger}, \citenamefont {Bozhevolnyi}, \citenamefont {Mortensen},\ and\ \citenamefont {Wolff}}]{Tserkezis2020}%
  \BibitemOpen
  \bibfield  {author} {\bibinfo {author} {\bibfnamefont {C.}~\bibnamefont {Tserkezis}}, \bibinfo {author} {\bibfnamefont {A.~I.}\ \bibnamefont {Fern{\'{a}}ndez-Dom{\'{i}}nguez}}, \bibinfo {author} {\bibfnamefont {P.~A.}\ \bibnamefont {Gon{\c{c}}alves}}, \bibinfo {author} {\bibfnamefont {F.}~\bibnamefont {Todisco}}, \bibinfo {author} {\bibfnamefont {J.~D.}\ \bibnamefont {Cox}}, \bibinfo {author} {\bibfnamefont {K.}~\bibnamefont {Busch}}, \bibinfo {author} {\bibfnamefont {N.}~\bibnamefont {Stenger}}, \bibinfo {author} {\bibfnamefont {S.~I.}\ \bibnamefont {Bozhevolnyi}}, \bibinfo {author} {\bibfnamefont {N.~A.}\ \bibnamefont {Mortensen}},\ and\ \bibinfo {author} {\bibfnamefont {C.}~\bibnamefont {Wolff}},\ }\bibfield  {title} {\bibinfo {title} {{On the applicability of quantum-optical concepts in strong-coupling nanophotonics}},\ }\href {https://doi.org/10.1088/1361-6633/aba348} {\bibfield  {journal} {\bibinfo  {journal} {Reports Prog. Phys.}\ }\textbf {\bibinfo {volume} {83}},\  (\bibinfo {year}
  {2020})}\BibitemShut {NoStop}%
\bibitem [{\citenamefont {Liu}\ \emph {et~al.}(2015)\citenamefont {Liu}, \citenamefont {Galfsky}, \citenamefont {Sun}, \citenamefont {Xia}, \citenamefont {Lin}, \citenamefont {Lee}, \citenamefont {K{\'{e}}na-Cohen},\ and\ \citenamefont {Menon}}]{Liu2015a}%
  \BibitemOpen
  \bibfield  {author} {\bibinfo {author} {\bibfnamefont {X.}~\bibnamefont {Liu}}, \bibinfo {author} {\bibfnamefont {T.}~\bibnamefont {Galfsky}}, \bibinfo {author} {\bibfnamefont {Z.}~\bibnamefont {Sun}}, \bibinfo {author} {\bibfnamefont {F.}~\bibnamefont {Xia}}, \bibinfo {author} {\bibfnamefont {E.-c.}\ \bibnamefont {Lin}}, \bibinfo {author} {\bibfnamefont {Y.-H.}\ \bibnamefont {Lee}}, \bibinfo {author} {\bibfnamefont {S.}~\bibnamefont {K{\'{e}}na-Cohen}},\ and\ \bibinfo {author} {\bibfnamefont {V.~M.}\ \bibnamefont {Menon}},\ }\bibfield  {title} {\bibinfo {title} {{Strong light–matter coupling in two-dimensional atomic crystals}},\ }\href {https://doi.org/10.1038/nphoton.2014.304} {\bibfield  {journal} {\bibinfo  {journal} {Nat. Photonics}\ }\textbf {\bibinfo {volume} {9}},\ \bibinfo {pages} {30} (\bibinfo {year} {2015})}\BibitemShut {NoStop}%
\bibitem [{\citenamefont {Dufferwiel}\ \emph {et~al.}(2015)\citenamefont {Dufferwiel}, \citenamefont {Schwarz}, \citenamefont {Withers}, \citenamefont {Trichet}, \citenamefont {Li}, \citenamefont {Sich}, \citenamefont {{Del Pozo-Zamudio}}, \citenamefont {Clark}, \citenamefont {Nalitov}, \citenamefont {Solnyshkov}, \citenamefont {Malpuech}, \citenamefont {Novoselov}, \citenamefont {Smith}, \citenamefont {Skolnick}, \citenamefont {Krizhanovskii},\ and\ \citenamefont {Tartakovskii}}]{Dufferwiel2015}%
  \BibitemOpen
  \bibfield  {author} {\bibinfo {author} {\bibfnamefont {S.}~\bibnamefont {Dufferwiel}}, \bibinfo {author} {\bibfnamefont {S.}~\bibnamefont {Schwarz}}, \bibinfo {author} {\bibfnamefont {F.}~\bibnamefont {Withers}}, \bibinfo {author} {\bibfnamefont {A.~A.~P.}\ \bibnamefont {Trichet}}, \bibinfo {author} {\bibfnamefont {F.}~\bibnamefont {Li}}, \bibinfo {author} {\bibfnamefont {M.}~\bibnamefont {Sich}}, \bibinfo {author} {\bibfnamefont {O.}~\bibnamefont {{Del Pozo-Zamudio}}}, \bibinfo {author} {\bibfnamefont {C.}~\bibnamefont {Clark}}, \bibinfo {author} {\bibfnamefont {A.}~\bibnamefont {Nalitov}}, \bibinfo {author} {\bibfnamefont {D.~D.}\ \bibnamefont {Solnyshkov}}, \bibinfo {author} {\bibfnamefont {G.}~\bibnamefont {Malpuech}}, \bibinfo {author} {\bibfnamefont {K.~S.}\ \bibnamefont {Novoselov}}, \bibinfo {author} {\bibfnamefont {J.~M.}\ \bibnamefont {Smith}}, \bibinfo {author} {\bibfnamefont {M.~S.}\ \bibnamefont {Skolnick}}, \bibinfo {author} {\bibfnamefont {D.~N.}\ \bibnamefont {Krizhanovskii}},\ and\ \bibinfo
  {author} {\bibfnamefont {A.~I.}\ \bibnamefont {Tartakovskii}},\ }\bibfield  {title} {\bibinfo {title} {{Exciton–polaritons in van der Waals heterostructures embedded in tunable microcavities}},\ }\href {https://doi.org/10.1038/ncomms9579} {\bibfield  {journal} {\bibinfo  {journal} {Nat. Commun.}\ }\textbf {\bibinfo {volume} {6}},\ \bibinfo {pages} {8579} (\bibinfo {year} {2015})}\BibitemShut {NoStop}%
\bibitem [{\citenamefont {Sidler}\ \emph {et~al.}(2017)\citenamefont {Sidler}, \citenamefont {Back}, \citenamefont {Cotlet}, \citenamefont {Srivastava}, \citenamefont {Fink}, \citenamefont {Kroner}, \citenamefont {Demler},\ and\ \citenamefont {Imamoglu}}]{Sidler2017}%
  \BibitemOpen
  \bibfield  {author} {\bibinfo {author} {\bibfnamefont {M.}~\bibnamefont {Sidler}}, \bibinfo {author} {\bibfnamefont {P.}~\bibnamefont {Back}}, \bibinfo {author} {\bibfnamefont {O.}~\bibnamefont {Cotlet}}, \bibinfo {author} {\bibfnamefont {A.}~\bibnamefont {Srivastava}}, \bibinfo {author} {\bibfnamefont {T.}~\bibnamefont {Fink}}, \bibinfo {author} {\bibfnamefont {M.}~\bibnamefont {Kroner}}, \bibinfo {author} {\bibfnamefont {E.}~\bibnamefont {Demler}},\ and\ \bibinfo {author} {\bibfnamefont {A.}~\bibnamefont {Imamoglu}},\ }\bibfield  {title} {\bibinfo {title} {{Fermi polaron-polaritons in charge-tunable atomically thin semiconductors}},\ }\href {https://doi.org/10.1038/nphys3949} {\bibfield  {journal} {\bibinfo  {journal} {Nat. Phys.}\ }\textbf {\bibinfo {volume} {13}},\ \bibinfo {pages} {255} (\bibinfo {year} {2017})}\BibitemShut {NoStop}%
\bibitem [{\citenamefont {Zhang}\ \emph {et~al.}(2018)\citenamefont {Zhang}, \citenamefont {Gogna}, \citenamefont {Burg}, \citenamefont {Tutuc},\ and\ \citenamefont {Deng}}]{Gogna}%
  \BibitemOpen
  \bibfield  {author} {\bibinfo {author} {\bibfnamefont {L.}~\bibnamefont {Zhang}}, \bibinfo {author} {\bibfnamefont {R.}~\bibnamefont {Gogna}}, \bibinfo {author} {\bibfnamefont {W.}~\bibnamefont {Burg}}, \bibinfo {author} {\bibfnamefont {E.}~\bibnamefont {Tutuc}},\ and\ \bibinfo {author} {\bibfnamefont {H.}~\bibnamefont {Deng}},\ }\bibfield  {title} {\bibinfo {title} {{Photonic-crystal exciton-polaritons in monolayer semiconductors}},\ }\href {https://doi.org/10.1038/s41467-018-03188-x} {\bibfield  {journal} {\bibinfo  {journal} {Nat. Commun.}\ }\textbf {\bibinfo {volume} {9}},\ \bibinfo {pages} {713} (\bibinfo {year} {2018})}\BibitemShut {NoStop}%
\bibitem [{\citenamefont {Lackner}\ \emph {et~al.}(2021)\citenamefont {Lackner}, \citenamefont {Dusel}, \citenamefont {Egorov}, \citenamefont {Han}, \citenamefont {Knopf}, \citenamefont {Eilenberger}, \citenamefont {Schr{\"{o}}der}, \citenamefont {Watanabe}, \citenamefont {Taniguchi}, \citenamefont {Tongay}, \citenamefont {Anton-Solanas}, \citenamefont {H{\"{o}}fling},\ and\ \citenamefont {Schneider}}]{Lackner2021}%
  \BibitemOpen
  \bibfield  {author} {\bibinfo {author} {\bibfnamefont {L.}~\bibnamefont {Lackner}}, \bibinfo {author} {\bibfnamefont {M.}~\bibnamefont {Dusel}}, \bibinfo {author} {\bibfnamefont {O.~A.}\ \bibnamefont {Egorov}}, \bibinfo {author} {\bibfnamefont {B.}~\bibnamefont {Han}}, \bibinfo {author} {\bibfnamefont {H.}~\bibnamefont {Knopf}}, \bibinfo {author} {\bibfnamefont {F.}~\bibnamefont {Eilenberger}}, \bibinfo {author} {\bibfnamefont {S.}~\bibnamefont {Schr{\"{o}}der}}, \bibinfo {author} {\bibfnamefont {K.}~\bibnamefont {Watanabe}}, \bibinfo {author} {\bibfnamefont {T.}~\bibnamefont {Taniguchi}}, \bibinfo {author} {\bibfnamefont {S.}~\bibnamefont {Tongay}}, \bibinfo {author} {\bibfnamefont {C.}~\bibnamefont {Anton-Solanas}}, \bibinfo {author} {\bibfnamefont {S.}~\bibnamefont {H{\"{o}}fling}},\ and\ \bibinfo {author} {\bibfnamefont {C.}~\bibnamefont {Schneider}},\ }\bibfield  {title} {\bibinfo {title} {{Tunable exciton-polaritons emerging from WS2 monolayer excitons in a photonic lattice at room temperature}},\
  }\href {https://doi.org/10.1038/s41467-021-24925-9} {\bibfield  {journal} {\bibinfo  {journal} {Nat. Commun.}\ }\textbf {\bibinfo {volume} {12}},\ \bibinfo {pages} {4933} (\bibinfo {year} {2021})}\BibitemShut {NoStop}%
\bibitem [{\citenamefont {Shan}\ \emph {et~al.}(2022)\citenamefont {Shan}, \citenamefont {Iorsh}, \citenamefont {Han}, \citenamefont {Rupprecht}, \citenamefont {Knopf}, \citenamefont {Eilenberger}, \citenamefont {Esmann}, \citenamefont {Yumigeta}, \citenamefont {Watanabe}, \citenamefont {Taniguchi}, \citenamefont {Klembt}, \citenamefont {H{\"{o}}fling}, \citenamefont {Tongay}, \citenamefont {Ant{\'{o}}n-Solanas}, \citenamefont {Shelykh},\ and\ \citenamefont {Schneider}}]{Shan2022}%
  \BibitemOpen
  \bibfield  {author} {\bibinfo {author} {\bibfnamefont {H.}~\bibnamefont {Shan}}, \bibinfo {author} {\bibfnamefont {I.}~\bibnamefont {Iorsh}}, \bibinfo {author} {\bibfnamefont {B.}~\bibnamefont {Han}}, \bibinfo {author} {\bibfnamefont {C.}~\bibnamefont {Rupprecht}}, \bibinfo {author} {\bibfnamefont {H.}~\bibnamefont {Knopf}}, \bibinfo {author} {\bibfnamefont {F.}~\bibnamefont {Eilenberger}}, \bibinfo {author} {\bibfnamefont {M.}~\bibnamefont {Esmann}}, \bibinfo {author} {\bibfnamefont {K.}~\bibnamefont {Yumigeta}}, \bibinfo {author} {\bibfnamefont {K.}~\bibnamefont {Watanabe}}, \bibinfo {author} {\bibfnamefont {T.}~\bibnamefont {Taniguchi}}, \bibinfo {author} {\bibfnamefont {S.}~\bibnamefont {Klembt}}, \bibinfo {author} {\bibfnamefont {S.}~\bibnamefont {H{\"{o}}fling}}, \bibinfo {author} {\bibfnamefont {S.}~\bibnamefont {Tongay}}, \bibinfo {author} {\bibfnamefont {C.}~\bibnamefont {Ant{\'{o}}n-Solanas}}, \bibinfo {author} {\bibfnamefont {I.~A.}\ \bibnamefont {Shelykh}},\ and\ \bibinfo {author} {\bibfnamefont
  {C.}~\bibnamefont {Schneider}},\ }\bibfield  {title} {\bibinfo {title} {{Brightening of a dark monolayer semiconductor via strong light-matter coupling in a cavity}},\ }\href {https://doi.org/10.1038/s41467-022-30645-5} {\bibfield  {journal} {\bibinfo  {journal} {Nat. Commun.}\ }\textbf {\bibinfo {volume} {13}},\ \bibinfo {pages} {3001} (\bibinfo {year} {2022})}\BibitemShut {NoStop}%
\bibitem [{\citenamefont {Qin}\ \emph {et~al.}(2022)\citenamefont {Qin}, \citenamefont {Duan}, \citenamefont {Xiao}, \citenamefont {Liu}, \citenamefont {Yu}, \citenamefont {Wang},\ and\ \citenamefont {Liao}}]{Qin2022}%
  \BibitemOpen
  \bibfield  {author} {\bibinfo {author} {\bibfnamefont {M.}~\bibnamefont {Qin}}, \bibinfo {author} {\bibfnamefont {J.}~\bibnamefont {Duan}}, \bibinfo {author} {\bibfnamefont {S.}~\bibnamefont {Xiao}}, \bibinfo {author} {\bibfnamefont {W.}~\bibnamefont {Liu}}, \bibinfo {author} {\bibfnamefont {T.}~\bibnamefont {Yu}}, \bibinfo {author} {\bibfnamefont {T.}~\bibnamefont {Wang}},\ and\ \bibinfo {author} {\bibfnamefont {Q.}~\bibnamefont {Liao}},\ }\bibfield  {title} {\bibinfo {title} {{Manipulating strong coupling between exciton and quasibound states in the continuum resonance}},\ }\href {https://doi.org/10.1103/PhysRevB.105.195425} {\bibfield  {journal} {\bibinfo  {journal} {Phys. Rev. B}\ }\textbf {\bibinfo {volume} {105}},\ \bibinfo {pages} {195425} (\bibinfo {year} {2022})}\BibitemShut {NoStop}%
\bibitem [{\citenamefont {Zheng}\ \emph {et~al.}(2023)\citenamefont {Zheng}, \citenamefont {Bai}, \citenamefont {Zhang},\ and\ \citenamefont {Liu}}]{Zheng2023}%
  \BibitemOpen
  \bibfield  {author} {\bibinfo {author} {\bibfnamefont {H.}~\bibnamefont {Zheng}}, \bibinfo {author} {\bibfnamefont {Y.}~\bibnamefont {Bai}}, \bibinfo {author} {\bibfnamefont {Q.}~\bibnamefont {Zhang}},\ and\ \bibinfo {author} {\bibfnamefont {S.}~\bibnamefont {Liu}},\ }\bibfield  {title} {\bibinfo {title} {{Multi-mode strong coupling in Fabry-Perot cavity-WS 2 photonic crystal hybrid structures}},\ }\href {https://doi.org/10.1364/OE.496305} {\bibfield  {journal} {\bibinfo  {journal} {Opt. Express}\ }\textbf {\bibinfo {volume} {31}},\ \bibinfo {pages} {24976} (\bibinfo {year} {2023})}\BibitemShut {NoStop}%
\bibitem [{\citenamefont {Maggiolini}\ \emph {et~al.}(2023)\citenamefont {Maggiolini}, \citenamefont {Polimeno}, \citenamefont {Todisco}, \citenamefont {{Di Renzo}}, \citenamefont {Han}, \citenamefont {{De Giorgi}}, \citenamefont {Ardizzone}, \citenamefont {Schneider}, \citenamefont {Mastria}, \citenamefont {Cannavale}, \citenamefont {Pugliese}, \citenamefont {{De Marco}}, \citenamefont {Rizzo}, \citenamefont {Maiorano}, \citenamefont {Gigli}, \citenamefont {Gerace}, \citenamefont {Sanvitto},\ and\ \citenamefont {Ballarini}}]{Maggiolini2023}%
  \BibitemOpen
  \bibfield  {author} {\bibinfo {author} {\bibfnamefont {E.}~\bibnamefont {Maggiolini}}, \bibinfo {author} {\bibfnamefont {L.}~\bibnamefont {Polimeno}}, \bibinfo {author} {\bibfnamefont {F.}~\bibnamefont {Todisco}}, \bibinfo {author} {\bibfnamefont {A.}~\bibnamefont {{Di Renzo}}}, \bibinfo {author} {\bibfnamefont {B.}~\bibnamefont {Han}}, \bibinfo {author} {\bibfnamefont {M.}~\bibnamefont {{De Giorgi}}}, \bibinfo {author} {\bibfnamefont {V.}~\bibnamefont {Ardizzone}}, \bibinfo {author} {\bibfnamefont {C.}~\bibnamefont {Schneider}}, \bibinfo {author} {\bibfnamefont {R.}~\bibnamefont {Mastria}}, \bibinfo {author} {\bibfnamefont {A.}~\bibnamefont {Cannavale}}, \bibinfo {author} {\bibfnamefont {M.}~\bibnamefont {Pugliese}}, \bibinfo {author} {\bibfnamefont {L.}~\bibnamefont {{De Marco}}}, \bibinfo {author} {\bibfnamefont {A.}~\bibnamefont {Rizzo}}, \bibinfo {author} {\bibfnamefont {V.}~\bibnamefont {Maiorano}}, \bibinfo {author} {\bibfnamefont {G.}~\bibnamefont {Gigli}}, \bibinfo {author} {\bibfnamefont
  {D.}~\bibnamefont {Gerace}}, \bibinfo {author} {\bibfnamefont {D.}~\bibnamefont {Sanvitto}},\ and\ \bibinfo {author} {\bibfnamefont {D.}~\bibnamefont {Ballarini}},\ }\bibfield  {title} {\bibinfo {title} {{Strongly enhanced light–matter coupling of monolayer WS2 from a bound state in the continuum}},\ }\href {https://doi.org/10.1038/s41563-023-01562-9} {\bibfield  {journal} {\bibinfo  {journal} {Nat. Mater.}\ }\textbf {\bibinfo {volume} {22}},\ \bibinfo {pages} {964} (\bibinfo {year} {2023})}\BibitemShut {NoStop}%
\bibitem [{\citenamefont {Danielsen}\ \emph {et~al.}(2025)\citenamefont {Danielsen}, \citenamefont {Lassaline}, \citenamefont {Linde}, \citenamefont {Nielsen}, \citenamefont {Zambrana-Puyalto}, \citenamefont {Sarbajna}, \citenamefont {Nguyen}, \citenamefont {Booth}, \citenamefont {Leitherer-Stenger},\ and\ \citenamefont {Raza}}]{Danielsen2025a}%
  \BibitemOpen
  \bibfield  {author} {\bibinfo {author} {\bibfnamefont {D.~R.}\ \bibnamefont {Danielsen}}, \bibinfo {author} {\bibfnamefont {N.}~\bibnamefont {Lassaline}}, \bibinfo {author} {\bibfnamefont {S.~J.}\ \bibnamefont {Linde}}, \bibinfo {author} {\bibfnamefont {M.~V.}\ \bibnamefont {Nielsen}}, \bibinfo {author} {\bibfnamefont {X.}~\bibnamefont {Zambrana-Puyalto}}, \bibinfo {author} {\bibfnamefont {A.}~\bibnamefont {Sarbajna}}, \bibinfo {author} {\bibfnamefont {D.~H.}\ \bibnamefont {Nguyen}}, \bibinfo {author} {\bibfnamefont {T.~J.}\ \bibnamefont {Booth}}, \bibinfo {author} {\bibfnamefont {N.}~\bibnamefont {Leitherer-Stenger}},\ and\ \bibinfo {author} {\bibfnamefont {S.}~\bibnamefont {Raza}},\ }\bibfield  {title} {\bibinfo {title} {{Fourier-Tailored Light–Matter Coupling in van der Waals Heterostructures}},\ }\bibfield  {journal} {\bibinfo  {journal} {ACS Nano}\ }\href {https://doi.org/10.1021/acsnano.5c02025} {10.1021/acsnano.5c02025} (\bibinfo {year} {2025})\BibitemShut {NoStop}%
\bibitem [{\citenamefont {Sortino}\ \emph {et~al.}()\citenamefont {Sortino}, \citenamefont {Biechteler}, \citenamefont {Lafeta}, \citenamefont {K{\"{u}}hner}, \citenamefont {Hartschuh}, \citenamefont {Menezes}, \citenamefont {Maier},\ and\ \citenamefont {Tittl}}]{Sortino}%
  \BibitemOpen
  \bibfield  {author} {\bibinfo {author} {\bibfnamefont {L.}~\bibnamefont {Sortino}}, \bibinfo {author} {\bibfnamefont {J.}~\bibnamefont {Biechteler}}, \bibinfo {author} {\bibfnamefont {L.}~\bibnamefont {Lafeta}}, \bibinfo {author} {\bibfnamefont {L.}~\bibnamefont {K{\"{u}}hner}}, \bibinfo {author} {\bibfnamefont {A.}~\bibnamefont {Hartschuh}}, \bibinfo {author} {\bibfnamefont {L.~d.~S.}\ \bibnamefont {Menezes}}, \bibinfo {author} {\bibfnamefont {S.~A.}\ \bibnamefont {Maier}},\ and\ \bibinfo {author} {\bibfnamefont {A.}~\bibnamefont {Tittl}},\ }\bibfield  {title} {\bibinfo {title} {{Van der Waals heterostructure metasurfaces: atomic-layer assembly of ultrathin optical cavities}},\ }\Eprint {https://arxiv.org/abs/2407.16480} {arXiv:2407.16480} \BibitemShut {NoStop}%
\bibitem [{\citenamefont {{B. Iyer}}\ \emph {et~al.}(2022)\citenamefont {{B. Iyer}}, \citenamefont {Luan}, \citenamefont {Shinar}, \citenamefont {Shinar},\ and\ \citenamefont {Fei}}]{B.Iyer2022}%
  \BibitemOpen
  \bibfield  {author} {\bibinfo {author} {\bibfnamefont {R.}~\bibnamefont {{B. Iyer}}}, \bibinfo {author} {\bibfnamefont {Y.}~\bibnamefont {Luan}}, \bibinfo {author} {\bibfnamefont {R.}~\bibnamefont {Shinar}}, \bibinfo {author} {\bibfnamefont {J.}~\bibnamefont {Shinar}},\ and\ \bibinfo {author} {\bibfnamefont {Z.}~\bibnamefont {Fei}},\ }\bibfield  {title} {\bibinfo {title} {{Nano-optical imaging of exciton–plasmon polaritons in WSe2/Au heterostructures}},\ }\href {https://doi.org/10.1039/D2NR04321A} {\bibfield  {journal} {\bibinfo  {journal} {Nanoscale}\ }\textbf {\bibinfo {volume} {14}},\ \bibinfo {pages} {15663} (\bibinfo {year} {2022})}\BibitemShut {NoStop}%
\bibitem [{\citenamefont {Casses}\ \emph {et~al.}(2024)\citenamefont {Casses}, \citenamefont {Zhou}, \citenamefont {Lin}, \citenamefont {Tan}, \citenamefont {{Bendixen-Fernex de Mongex}}, \citenamefont {Kaltenecker}, \citenamefont {Xiao}, \citenamefont {Wubs},\ and\ \citenamefont {Stenger}}]{Casses2024}%
  \BibitemOpen
  \bibfield  {author} {\bibinfo {author} {\bibfnamefont {L.~N.}\ \bibnamefont {Casses}}, \bibinfo {author} {\bibfnamefont {B.}~\bibnamefont {Zhou}}, \bibinfo {author} {\bibfnamefont {Q.}~\bibnamefont {Lin}}, \bibinfo {author} {\bibfnamefont {A.}~\bibnamefont {Tan}}, \bibinfo {author} {\bibfnamefont {D.-P.}\ \bibnamefont {{Bendixen-Fernex de Mongex}}}, \bibinfo {author} {\bibfnamefont {K.~J.}\ \bibnamefont {Kaltenecker}}, \bibinfo {author} {\bibfnamefont {S.}~\bibnamefont {Xiao}}, \bibinfo {author} {\bibfnamefont {M.}~\bibnamefont {Wubs}},\ and\ \bibinfo {author} {\bibfnamefont {N.}~\bibnamefont {Stenger}},\ }\bibfield  {title} {\bibinfo {title} {{Full Quantitative Near-Field Characterization of Strongly Coupled Exciton–Plasmon Polaritons in Thin-Layered WSe 2 on a Monocrystalline Gold Platelet}},\ }\href {https://doi.org/10.1021/acsphotonics.4c00580} {\bibfield  {journal} {\bibinfo  {journal} {ACS Photonics}\ }\textbf {\bibinfo {volume} {11}},\ \bibinfo {pages} {3593} (\bibinfo {year} {2024})}\BibitemShut
  {NoStop}%
\bibitem [{\citenamefont {Wen}\ \emph {et~al.}(2017)\citenamefont {Wen}, \citenamefont {Wang}, \citenamefont {Wang}, \citenamefont {Deng}, \citenamefont {Zhuang}, \citenamefont {Zhang}, \citenamefont {Liu}, \citenamefont {She}, \citenamefont {Chen}, \citenamefont {Chen}, \citenamefont {Deng},\ and\ \citenamefont {Xu}}]{Wen2017}%
  \BibitemOpen
  \bibfield  {author} {\bibinfo {author} {\bibfnamefont {J.}~\bibnamefont {Wen}}, \bibinfo {author} {\bibfnamefont {H.}~\bibnamefont {Wang}}, \bibinfo {author} {\bibfnamefont {W.}~\bibnamefont {Wang}}, \bibinfo {author} {\bibfnamefont {Z.}~\bibnamefont {Deng}}, \bibinfo {author} {\bibfnamefont {C.}~\bibnamefont {Zhuang}}, \bibinfo {author} {\bibfnamefont {Y.}~\bibnamefont {Zhang}}, \bibinfo {author} {\bibfnamefont {F.}~\bibnamefont {Liu}}, \bibinfo {author} {\bibfnamefont {J.}~\bibnamefont {She}}, \bibinfo {author} {\bibfnamefont {J.}~\bibnamefont {Chen}}, \bibinfo {author} {\bibfnamefont {H.}~\bibnamefont {Chen}}, \bibinfo {author} {\bibfnamefont {S.}~\bibnamefont {Deng}},\ and\ \bibinfo {author} {\bibfnamefont {N.}~\bibnamefont {Xu}},\ }\bibfield  {title} {\bibinfo {title} {{Room-Temperature Strong Light–Matter Interaction with Active Control in Single Plasmonic Nanorod Coupled with Two-Dimensional Atomic Crystals}},\ }\href {https://doi.org/10.1021/acs.nanolett.7b01344} {\bibfield  {journal} {\bibinfo
  {journal} {Nano Lett.}\ }\textbf {\bibinfo {volume} {17}},\ \bibinfo {pages} {4689} (\bibinfo {year} {2017})}\BibitemShut {NoStop}%
\bibitem [{\citenamefont {St{\"{u}}hrenberg}\ \emph {et~al.}(2018)\citenamefont {St{\"{u}}hrenberg}, \citenamefont {Munkhbat}, \citenamefont {Baranov}, \citenamefont {Cuadra}, \citenamefont {Yankovich}, \citenamefont {Antosiewicz}, \citenamefont {Olsson},\ and\ \citenamefont {Shegai}}]{Stuhrenberg2018}%
  \BibitemOpen
  \bibfield  {author} {\bibinfo {author} {\bibfnamefont {M.}~\bibnamefont {St{\"{u}}hrenberg}}, \bibinfo {author} {\bibfnamefont {B.}~\bibnamefont {Munkhbat}}, \bibinfo {author} {\bibfnamefont {D.~G.}\ \bibnamefont {Baranov}}, \bibinfo {author} {\bibfnamefont {J.}~\bibnamefont {Cuadra}}, \bibinfo {author} {\bibfnamefont {A.~B.}\ \bibnamefont {Yankovich}}, \bibinfo {author} {\bibfnamefont {T.~J.}\ \bibnamefont {Antosiewicz}}, \bibinfo {author} {\bibfnamefont {E.}~\bibnamefont {Olsson}},\ and\ \bibinfo {author} {\bibfnamefont {T.}~\bibnamefont {Shegai}},\ }\bibfield  {title} {\bibinfo {title} {{Strong Light-Matter Coupling between Plasmons in Individual Gold Bi-pyramids and Excitons in Mono- and Multilayer WSe2}},\ }\href {https://doi.org/10.1021/acs.nanolett.8b02652} {\bibfield  {journal} {\bibinfo  {journal} {Nano Lett.}\ }\textbf {\bibinfo {volume} {18}},\ \bibinfo {pages} {5938} (\bibinfo {year} {2018})}\BibitemShut {NoStop}%
\bibitem [{\citenamefont {Geisler}\ \emph {et~al.}(2019)\citenamefont {Geisler}, \citenamefont {Cui}, \citenamefont {Wang}, \citenamefont {Rindzevicius}, \citenamefont {Gammelgaard}, \citenamefont {Jessen}, \citenamefont {Gon{\c{c}}alves}, \citenamefont {Todisco}, \citenamefont {B{\o}ggild}, \citenamefont {Boisen}, \citenamefont {Wubs}, \citenamefont {Mortensen}, \citenamefont {Xiao},\ and\ \citenamefont {Stenger}}]{Geisler2019}%
  \BibitemOpen
  \bibfield  {author} {\bibinfo {author} {\bibfnamefont {M.}~\bibnamefont {Geisler}}, \bibinfo {author} {\bibfnamefont {X.}~\bibnamefont {Cui}}, \bibinfo {author} {\bibfnamefont {J.}~\bibnamefont {Wang}}, \bibinfo {author} {\bibfnamefont {T.}~\bibnamefont {Rindzevicius}}, \bibinfo {author} {\bibfnamefont {L.}~\bibnamefont {Gammelgaard}}, \bibinfo {author} {\bibfnamefont {B.~S.}\ \bibnamefont {Jessen}}, \bibinfo {author} {\bibfnamefont {P.~A.~D.}\ \bibnamefont {Gon{\c{c}}alves}}, \bibinfo {author} {\bibfnamefont {F.}~\bibnamefont {Todisco}}, \bibinfo {author} {\bibfnamefont {P.}~\bibnamefont {B{\o}ggild}}, \bibinfo {author} {\bibfnamefont {A.}~\bibnamefont {Boisen}}, \bibinfo {author} {\bibfnamefont {M.}~\bibnamefont {Wubs}}, \bibinfo {author} {\bibfnamefont {N.~A.}\ \bibnamefont {Mortensen}}, \bibinfo {author} {\bibfnamefont {S.}~\bibnamefont {Xiao}},\ and\ \bibinfo {author} {\bibfnamefont {N.}~\bibnamefont {Stenger}},\ }\bibfield  {title} {\bibinfo {title} {{Single-Crystalline Gold Nanodisks on WS 2 Mono-
  and Multilayers for Strong Coupling at Room Temperature}},\ }\href {https://doi.org/10.1021/acsphotonics.8b01766} {\bibfield  {journal} {\bibinfo  {journal} {ACS Photonics}\ }\textbf {\bibinfo {volume} {6}},\ \bibinfo {pages} {994} (\bibinfo {year} {2019})}\BibitemShut {NoStop}%
\bibitem [{\citenamefont {Munkhbat}\ \emph {et~al.}(2020)\citenamefont {Munkhbat}, \citenamefont {Baranov}, \citenamefont {Bisht}, \citenamefont {Hoque}, \citenamefont {Karpiak}, \citenamefont {Dash},\ and\ \citenamefont {Shegai}}]{Munkhbat2020}%
  \BibitemOpen
  \bibfield  {author} {\bibinfo {author} {\bibfnamefont {B.}~\bibnamefont {Munkhbat}}, \bibinfo {author} {\bibfnamefont {D.~G.}\ \bibnamefont {Baranov}}, \bibinfo {author} {\bibfnamefont {A.}~\bibnamefont {Bisht}}, \bibinfo {author} {\bibfnamefont {M.~A.}\ \bibnamefont {Hoque}}, \bibinfo {author} {\bibfnamefont {B.}~\bibnamefont {Karpiak}}, \bibinfo {author} {\bibfnamefont {S.~P.}\ \bibnamefont {Dash}},\ and\ \bibinfo {author} {\bibfnamefont {T.}~\bibnamefont {Shegai}},\ }\bibfield  {title} {\bibinfo {title} {{Electrical Control of Hybrid Monolayer Tungsten Disulfide–Plasmonic Nanoantenna Light–Matter States at Cryogenic and Room Temperatures}},\ }\href {https://doi.org/10.1021/acsnano.9b09684} {\bibfield  {journal} {\bibinfo  {journal} {ACS Nano}\ }\textbf {\bibinfo {volume} {14}},\ \bibinfo {pages} {1196} (\bibinfo {year} {2020})}\BibitemShut {NoStop}%
\bibitem [{\citenamefont {Painter}\ \emph {et~al.}(1999)\citenamefont {Painter}, \citenamefont {Lee}, \citenamefont {Scherer}, \citenamefont {Yariv}, \citenamefont {O'Brien}, \citenamefont {Dapkus},\ and\ \citenamefont {Kim}}]{Painter1999}%
  \BibitemOpen
  \bibfield  {author} {\bibinfo {author} {\bibfnamefont {O.}~\bibnamefont {Painter}}, \bibinfo {author} {\bibfnamefont {R.~K.}\ \bibnamefont {Lee}}, \bibinfo {author} {\bibfnamefont {A.}~\bibnamefont {Scherer}}, \bibinfo {author} {\bibfnamefont {A.}~\bibnamefont {Yariv}}, \bibinfo {author} {\bibfnamefont {J.~D.}\ \bibnamefont {O'Brien}}, \bibinfo {author} {\bibfnamefont {P.~D.}\ \bibnamefont {Dapkus}},\ and\ \bibinfo {author} {\bibfnamefont {I.}~\bibnamefont {Kim}},\ }\bibfield  {title} {\bibinfo {title} {{Two-dimensional photonic band-gap defect mode laser}},\ }\href {https://doi.org/10.1126/science.284.5421.1819} {\bibfield  {journal} {\bibinfo  {journal} {Science}\ }\textbf {\bibinfo {volume} {284}},\ \bibinfo {pages} {1819} (\bibinfo {year} {1999})}\BibitemShut {NoStop}%
\bibitem [{\citenamefont {Akahane}\ \emph {et~al.}(2003)\citenamefont {Akahane}, \citenamefont {Asano}, \citenamefont {Song},\ and\ \citenamefont {Noda}}]{Akahane2003}%
  \BibitemOpen
  \bibfield  {author} {\bibinfo {author} {\bibfnamefont {Y.}~\bibnamefont {Akahane}}, \bibinfo {author} {\bibfnamefont {T.}~\bibnamefont {Asano}}, \bibinfo {author} {\bibfnamefont {B.~S.}\ \bibnamefont {Song}},\ and\ \bibinfo {author} {\bibfnamefont {S.}~\bibnamefont {Noda}},\ }\bibfield  {title} {\bibinfo {title} {{High-Q photonic nanocavity in a two-dimensional photonic crystal}},\ }\href {https://doi.org/10.1038/nature02063} {\bibfield  {journal} {\bibinfo  {journal} {Nature}\ }\textbf {\bibinfo {volume} {425}},\ \bibinfo {pages} {944} (\bibinfo {year} {2003})}\BibitemShut {NoStop}%
\bibitem [{\citenamefont {Ota}\ \emph {et~al.}(2018)\citenamefont {Ota}, \citenamefont {Katsumi}, \citenamefont {Watanabe}, \citenamefont {Iwamoto},\ and\ \citenamefont {Arakawa}}]{Ota2018}%
  \BibitemOpen
  \bibfield  {author} {\bibinfo {author} {\bibfnamefont {Y.}~\bibnamefont {Ota}}, \bibinfo {author} {\bibfnamefont {R.}~\bibnamefont {Katsumi}}, \bibinfo {author} {\bibfnamefont {K.}~\bibnamefont {Watanabe}}, \bibinfo {author} {\bibfnamefont {S.}~\bibnamefont {Iwamoto}},\ and\ \bibinfo {author} {\bibfnamefont {Y.}~\bibnamefont {Arakawa}},\ }\bibfield  {title} {\bibinfo {title} {{Topological photonic crystal nanocavity laser}},\ }\href {https://doi.org/10.1038/s42005-018-0083-7} {\bibfield  {journal} {\bibinfo  {journal} {Commun. Phys.}\ }\textbf {\bibinfo {volume} {1}},\ \bibinfo {pages} {86} (\bibinfo {year} {2018})}\BibitemShut {NoStop}%
\bibitem [{\citenamefont {Ruppert}\ \emph {et~al.}(2014)\citenamefont {Ruppert}, \citenamefont {Aslan},\ and\ \citenamefont {Heinz}}]{Ruppert2014}%
  \BibitemOpen
  \bibfield  {author} {\bibinfo {author} {\bibfnamefont {C.}~\bibnamefont {Ruppert}}, \bibinfo {author} {\bibfnamefont {B.}~\bibnamefont {Aslan}},\ and\ \bibinfo {author} {\bibfnamefont {T.~F.}\ \bibnamefont {Heinz}},\ }\bibfield  {title} {\bibinfo {title} {{Optical Properties and Band Gap of Single- and Few-Layer MoTe 2 Crystals}},\ }\href {https://doi.org/10.1021/nl502557g} {\bibfield  {journal} {\bibinfo  {journal} {Nano Lett.}\ }\textbf {\bibinfo {volume} {14}},\ \bibinfo {pages} {6231} (\bibinfo {year} {2014})}\BibitemShut {NoStop}%
\bibitem [{\citenamefont {Fang}\ \emph {et~al.}(2019)\citenamefont {Fang}, \citenamefont {Liu}, \citenamefont {Lin}, \citenamefont {Su}, \citenamefont {Wei}, \citenamefont {Krauss}, \citenamefont {Li}, \citenamefont {Wang},\ and\ \citenamefont {Wang}}]{Fang2019}%
  \BibitemOpen
  \bibfield  {author} {\bibinfo {author} {\bibfnamefont {H.}~\bibnamefont {Fang}}, \bibinfo {author} {\bibfnamefont {J.}~\bibnamefont {Liu}}, \bibinfo {author} {\bibfnamefont {Q.}~\bibnamefont {Lin}}, \bibinfo {author} {\bibfnamefont {R.}~\bibnamefont {Su}}, \bibinfo {author} {\bibfnamefont {Y.}~\bibnamefont {Wei}}, \bibinfo {author} {\bibfnamefont {T.~F.}\ \bibnamefont {Krauss}}, \bibinfo {author} {\bibfnamefont {J.}~\bibnamefont {Li}}, \bibinfo {author} {\bibfnamefont {Y.}~\bibnamefont {Wang}},\ and\ \bibinfo {author} {\bibfnamefont {X.}~\bibnamefont {Wang}},\ }\bibfield  {title} {\bibinfo {title} {{Laser-Like Emission from a Sandwiched MoTe2 Heterostructure on a Silicon Single-Mode Resonator}},\ }\href {https://doi.org/10.1002/adom.201900538} {\bibfield  {journal} {\bibinfo  {journal} {Adv. Opt. Mater.}\ }\textbf {\bibinfo {volume} {7}},\ \bibinfo {pages} {5} (\bibinfo {year} {2019})}\BibitemShut {NoStop}%
\bibitem [{\citenamefont {Rosser}\ \emph {et~al.}(2022)\citenamefont {Rosser}, \citenamefont {Gerace}, \citenamefont {Chen}, \citenamefont {Liu}, \citenamefont {Whitehead}, \citenamefont {Ryou}, \citenamefont {Andreani},\ and\ \citenamefont {Majumdar}}]{Rosser2022}%
  \BibitemOpen
  \bibfield  {author} {\bibinfo {author} {\bibfnamefont {D.}~\bibnamefont {Rosser}}, \bibinfo {author} {\bibfnamefont {D.}~\bibnamefont {Gerace}}, \bibinfo {author} {\bibfnamefont {Y.}~\bibnamefont {Chen}}, \bibinfo {author} {\bibfnamefont {Y.}~\bibnamefont {Liu}}, \bibinfo {author} {\bibfnamefont {J.}~\bibnamefont {Whitehead}}, \bibinfo {author} {\bibfnamefont {A.}~\bibnamefont {Ryou}}, \bibinfo {author} {\bibfnamefont {L.~C.}\ \bibnamefont {Andreani}},\ and\ \bibinfo {author} {\bibfnamefont {A.}~\bibnamefont {Majumdar}},\ }\bibfield  {title} {\bibinfo {title} {{Dispersive coupling between MoSe2 and an integrated zero-dimensional nanocavity}},\ }\href {https://doi.org/10.1364/OME.443536} {\bibfield  {journal} {\bibinfo  {journal} {Opt. Mater. Express}\ }\textbf {\bibinfo {volume} {12}},\ \bibinfo {pages} {59} (\bibinfo {year} {2022})}\BibitemShut {NoStop}%
\bibitem [{\citenamefont {Qian}\ \emph {et~al.}(2022)\citenamefont {Qian}, \citenamefont {Villafa{\~{n}}e}, \citenamefont {Soubelet}, \citenamefont {H{\"{o}}tger}, \citenamefont {Taniguchi}, \citenamefont {Watanabe}, \citenamefont {Wilson}, \citenamefont {Stier}, \citenamefont {Holleitner},\ and\ \citenamefont {Finley}}]{Qian2022}%
  \BibitemOpen
  \bibfield  {author} {\bibinfo {author} {\bibfnamefont {C.}~\bibnamefont {Qian}}, \bibinfo {author} {\bibfnamefont {V.}~\bibnamefont {Villafa{\~{n}}e}}, \bibinfo {author} {\bibfnamefont {P.}~\bibnamefont {Soubelet}}, \bibinfo {author} {\bibfnamefont {A.}~\bibnamefont {H{\"{o}}tger}}, \bibinfo {author} {\bibfnamefont {T.}~\bibnamefont {Taniguchi}}, \bibinfo {author} {\bibfnamefont {K.}~\bibnamefont {Watanabe}}, \bibinfo {author} {\bibfnamefont {N.~P.}\ \bibnamefont {Wilson}}, \bibinfo {author} {\bibfnamefont {A.~V.}\ \bibnamefont {Stier}}, \bibinfo {author} {\bibfnamefont {A.~W.}\ \bibnamefont {Holleitner}},\ and\ \bibinfo {author} {\bibfnamefont {J.~J.}\ \bibnamefont {Finley}},\ }\bibfield  {title} {\bibinfo {title} {{Nonlocal Exciton-Photon Interactions in Hybrid High-Q Beam Nanocavities with Encapsulated MoS2 Monolayers}},\ }\href {https://doi.org/10.1103/PhysRevLett.128.237403} {\bibfield  {journal} {\bibinfo  {journal} {Phys. Rev. Lett.}\ }\textbf {\bibinfo {volume} {128}},\ \bibinfo {pages} {237403}
  (\bibinfo {year} {2022})}\BibitemShut {NoStop}%
\bibitem [{\citenamefont {Robinson}\ \emph {et~al.}(2005)\citenamefont {Robinson}, \citenamefont {Manolatou}, \citenamefont {Chen},\ and\ \citenamefont {Lipson}}]{Robinson2005}%
  \BibitemOpen
  \bibfield  {author} {\bibinfo {author} {\bibfnamefont {J.~T.}\ \bibnamefont {Robinson}}, \bibinfo {author} {\bibfnamefont {C.}~\bibnamefont {Manolatou}}, \bibinfo {author} {\bibfnamefont {L.}~\bibnamefont {Chen}},\ and\ \bibinfo {author} {\bibfnamefont {M.}~\bibnamefont {Lipson}},\ }\bibfield  {title} {\bibinfo {title} {{Ultrasmall Mode Volumes in Dielectric Optical Microcavities}},\ }\href {https://doi.org/10.1103/PhysRevLett.95.143901} {\bibfield  {journal} {\bibinfo  {journal} {Phys. Rev. Lett.}\ }\textbf {\bibinfo {volume} {95}},\ \bibinfo {pages} {143901} (\bibinfo {year} {2005})}\BibitemShut {NoStop}%
\bibitem [{\citenamefont {Hu}\ and\ \citenamefont {Weiss}(2016)}]{Hu2016}%
  \BibitemOpen
  \bibfield  {author} {\bibinfo {author} {\bibfnamefont {S.}~\bibnamefont {Hu}}\ and\ \bibinfo {author} {\bibfnamefont {S.~M.}\ \bibnamefont {Weiss}},\ }\bibfield  {title} {\bibinfo {title} {{Design of Photonic Crystal Cavities for Extreme Light Concentration}},\ }\href {https://doi.org/10.1021/acsphotonics.6b00219} {\bibfield  {journal} {\bibinfo  {journal} {ACS Photonics}\ }\textbf {\bibinfo {volume} {3}},\ \bibinfo {pages} {1647} (\bibinfo {year} {2016})}\BibitemShut {NoStop}%
\bibitem [{\citenamefont {Choi}\ \emph {et~al.}(2017)\citenamefont {Choi}, \citenamefont {Heuck},\ and\ \citenamefont {Englund}}]{Choi2017}%
  \BibitemOpen
  \bibfield  {author} {\bibinfo {author} {\bibfnamefont {H.}~\bibnamefont {Choi}}, \bibinfo {author} {\bibfnamefont {M.}~\bibnamefont {Heuck}},\ and\ \bibinfo {author} {\bibfnamefont {D.}~\bibnamefont {Englund}},\ }\bibfield  {title} {\bibinfo {title} {{Self-Similar Nanocavity Design with Ultrasmall Mode Volume for Single-Photon Nonlinearities}},\ }\href {https://doi.org/10.1103/PhysRevLett.118.223605} {\bibfield  {journal} {\bibinfo  {journal} {Phys. Rev. Lett.}\ }\textbf {\bibinfo {volume} {118}},\ \bibinfo {pages} {223605} (\bibinfo {year} {2017})}\BibitemShut {NoStop}%
\bibitem [{\citenamefont {Hu}\ \emph {et~al.}(2018)\citenamefont {Hu}, \citenamefont {Khater}, \citenamefont {Salas-Montiel}, \citenamefont {Kratschmer}, \citenamefont {Engelmann}, \citenamefont {Green},\ and\ \citenamefont {Weiss}}]{Hu2018}%
  \BibitemOpen
  \bibfield  {author} {\bibinfo {author} {\bibfnamefont {S.}~\bibnamefont {Hu}}, \bibinfo {author} {\bibfnamefont {M.}~\bibnamefont {Khater}}, \bibinfo {author} {\bibfnamefont {R.}~\bibnamefont {Salas-Montiel}}, \bibinfo {author} {\bibfnamefont {E.}~\bibnamefont {Kratschmer}}, \bibinfo {author} {\bibfnamefont {S.}~\bibnamefont {Engelmann}}, \bibinfo {author} {\bibfnamefont {W.~M.}\ \bibnamefont {Green}},\ and\ \bibinfo {author} {\bibfnamefont {S.~M.}\ \bibnamefont {Weiss}},\ }\bibfield  {title} {\bibinfo {title} {{Experimental realization of deep-subwavelength confinement in dielectric optical resonators}},\ }\href {https://doi.org/10.1126/sciadv.aat2355} {\bibfield  {journal} {\bibinfo  {journal} {Sci. Adv.}\ }\textbf {\bibinfo {volume} {4}},\  (\bibinfo {year} {2018})}\BibitemShut {NoStop}%
\bibitem [{\citenamefont {Babar}\ \emph {et~al.}(2023)\citenamefont {Babar}, \citenamefont {Weis}, \citenamefont {Tsoukalas}, \citenamefont {Kadkhodazadeh}, \citenamefont {Arregui}, \citenamefont {{Vosoughi Lahijani}},\ and\ \citenamefont {Stobbe}}]{Babar2023}%
  \BibitemOpen
  \bibfield  {author} {\bibinfo {author} {\bibfnamefont {A.~N.}\ \bibnamefont {Babar}}, \bibinfo {author} {\bibfnamefont {T.~A.~S.}\ \bibnamefont {Weis}}, \bibinfo {author} {\bibfnamefont {K.}~\bibnamefont {Tsoukalas}}, \bibinfo {author} {\bibfnamefont {S.}~\bibnamefont {Kadkhodazadeh}}, \bibinfo {author} {\bibfnamefont {G.}~\bibnamefont {Arregui}}, \bibinfo {author} {\bibfnamefont {B.}~\bibnamefont {{Vosoughi Lahijani}}},\ and\ \bibinfo {author} {\bibfnamefont {S.}~\bibnamefont {Stobbe}},\ }\bibfield  {title} {\bibinfo {title} {{Self-assembled photonic cavities with atomic-scale confinement}},\ }\href {https://doi.org/10.1038/s41586-023-06736-8} {\bibfield  {journal} {\bibinfo  {journal} {Nature}\ }\textbf {\bibinfo {volume} {624}},\ \bibinfo {pages} {57} (\bibinfo {year} {2023})}\BibitemShut {NoStop}%
\bibitem [{\citenamefont {Albrechtsen}\ \emph {et~al.}(2022)\citenamefont {Albrechtsen}, \citenamefont {{Vosoughi Lahijani}}, \citenamefont {Christiansen}, \citenamefont {Nguyen}, \citenamefont {Casses}, \citenamefont {Hansen}, \citenamefont {Stenger}, \citenamefont {Sigmund}, \citenamefont {Jansen}, \citenamefont {M{\o}rk},\ and\ \citenamefont {Stobbe}}]{Albrechtsen2022}%
  \BibitemOpen
  \bibfield  {author} {\bibinfo {author} {\bibfnamefont {M.}~\bibnamefont {Albrechtsen}}, \bibinfo {author} {\bibfnamefont {B.}~\bibnamefont {{Vosoughi Lahijani}}}, \bibinfo {author} {\bibfnamefont {R.~E.}\ \bibnamefont {Christiansen}}, \bibinfo {author} {\bibfnamefont {V.~T.~H.}\ \bibnamefont {Nguyen}}, \bibinfo {author} {\bibfnamefont {L.~N.}\ \bibnamefont {Casses}}, \bibinfo {author} {\bibfnamefont {S.~E.}\ \bibnamefont {Hansen}}, \bibinfo {author} {\bibfnamefont {N.}~\bibnamefont {Stenger}}, \bibinfo {author} {\bibfnamefont {O.}~\bibnamefont {Sigmund}}, \bibinfo {author} {\bibfnamefont {H.}~\bibnamefont {Jansen}}, \bibinfo {author} {\bibfnamefont {J.}~\bibnamefont {M{\o}rk}},\ and\ \bibinfo {author} {\bibfnamefont {S.}~\bibnamefont {Stobbe}},\ }\bibfield  {title} {\bibinfo {title} {{Nanometer-scale photon confinement in topology-optimized dielectric cavities}},\ }\href {https://doi.org/10.1038/s41467-022-33874-w} {\bibfield  {journal} {\bibinfo  {journal} {Nat. Commun.}\ }\textbf {\bibinfo {volume}
  {13}},\ \bibinfo {pages} {6281} (\bibinfo {year} {2022})}\BibitemShut {NoStop}%
\bibitem [{\citenamefont {Xiong}\ \emph {et~al.}(2024)\citenamefont {Xiong}, \citenamefont {Christiansen}, \citenamefont {Schr{\"{o}}der}, \citenamefont {Yu}, \citenamefont {Casses}, \citenamefont {Semenova}, \citenamefont {Yvind}, \citenamefont {Stenger}, \citenamefont {Sigmund},\ and\ \citenamefont {M{\o}rk}}]{Xiong2024}%
  \BibitemOpen
  \bibfield  {author} {\bibinfo {author} {\bibfnamefont {M.}~\bibnamefont {Xiong}}, \bibinfo {author} {\bibfnamefont {R.~E.}\ \bibnamefont {Christiansen}}, \bibinfo {author} {\bibfnamefont {F.}~\bibnamefont {Schr{\"{o}}der}}, \bibinfo {author} {\bibfnamefont {Y.}~\bibnamefont {Yu}}, \bibinfo {author} {\bibfnamefont {L.~N.}\ \bibnamefont {Casses}}, \bibinfo {author} {\bibfnamefont {E.}~\bibnamefont {Semenova}}, \bibinfo {author} {\bibfnamefont {K.}~\bibnamefont {Yvind}}, \bibinfo {author} {\bibfnamefont {N.}~\bibnamefont {Stenger}}, \bibinfo {author} {\bibfnamefont {O.}~\bibnamefont {Sigmund}},\ and\ \bibinfo {author} {\bibfnamefont {J.}~\bibnamefont {M{\o}rk}},\ }\bibfield  {title} {\bibinfo {title} {{Experimental realization of deep sub-wavelength confinement of light in a topology-optimized InP nanocavity}},\ }\href {https://doi.org/10.1364/OME.513625} {\bibfield  {journal} {\bibinfo  {journal} {Opt. Mater. Express}\ }\textbf {\bibinfo {volume} {14}},\ \bibinfo {pages} {397} (\bibinfo {year}
  {2024})}\BibitemShut {NoStop}%
\bibitem [{\citenamefont {Jensen}\ and\ \citenamefont {Sigmund}(2011)}]{Jensen2011}%
  \BibitemOpen
  \bibfield  {author} {\bibinfo {author} {\bibfnamefont {J.~S.}\ \bibnamefont {Jensen}}\ and\ \bibinfo {author} {\bibfnamefont {O.}~\bibnamefont {Sigmund}},\ }\bibfield  {title} {\bibinfo {title} {{Topology optimization for nano-photonics}},\ }\href {https://doi.org/10.1002/lpor.201000014} {\bibfield  {journal} {\bibinfo  {journal} {Laser Photonics Rev.}\ }\textbf {\bibinfo {volume} {5}},\ \bibinfo {pages} {308} (\bibinfo {year} {2011})}\BibitemShut {NoStop}%
\bibitem [{\citenamefont {Molesky}\ \emph {et~al.}(2018)\citenamefont {Molesky}, \citenamefont {Lin}, \citenamefont {Piggott}, \citenamefont {Jin}, \citenamefont {Vuckovi{\'{c}}},\ and\ \citenamefont {Rodriguez}}]{Molesky2018}%
  \BibitemOpen
  \bibfield  {author} {\bibinfo {author} {\bibfnamefont {S.}~\bibnamefont {Molesky}}, \bibinfo {author} {\bibfnamefont {Z.}~\bibnamefont {Lin}}, \bibinfo {author} {\bibfnamefont {A.~Y.}\ \bibnamefont {Piggott}}, \bibinfo {author} {\bibfnamefont {W.}~\bibnamefont {Jin}}, \bibinfo {author} {\bibfnamefont {J.}~\bibnamefont {Vuckovi{\'{c}}}},\ and\ \bibinfo {author} {\bibfnamefont {A.~W.}\ \bibnamefont {Rodriguez}},\ }\bibfield  {title} {\bibinfo {title} {{Inverse design in nanophotonics}},\ }\href {https://doi.org/10.1038/s41566-018-0246-9} {\bibfield  {journal} {\bibinfo  {journal} {Nat. Photonics}\ }\textbf {\bibinfo {volume} {12}},\ \bibinfo {pages} {659} (\bibinfo {year} {2018})}\BibitemShut {NoStop}%
\bibitem [{\citenamefont {Wang}\ \emph {et~al.}(2018)\citenamefont {Wang}, \citenamefont {Christiansen}, \citenamefont {Yu}, \citenamefont {M{\o}rk},\ and\ \citenamefont {Sigmund}}]{Wang2018b}%
  \BibitemOpen
  \bibfield  {author} {\bibinfo {author} {\bibfnamefont {F.}~\bibnamefont {Wang}}, \bibinfo {author} {\bibfnamefont {R.~E.}\ \bibnamefont {Christiansen}}, \bibinfo {author} {\bibfnamefont {Y.}~\bibnamefont {Yu}}, \bibinfo {author} {\bibfnamefont {J.}~\bibnamefont {M{\o}rk}},\ and\ \bibinfo {author} {\bibfnamefont {O.}~\bibnamefont {Sigmund}},\ }\bibfield  {title} {\bibinfo {title} {{Maximizing the quality factor to mode volume ratio for ultra-small photonic crystal cavities}},\ }\href {https://doi.org/10.1063/1.5064468} {\bibfield  {journal} {\bibinfo  {journal} {Appl. Phys. Lett.}\ }\textbf {\bibinfo {volume} {113}},\  (\bibinfo {year} {2018})}\BibitemShut {NoStop}%
\bibitem [{\citenamefont {Wang}\ and\ \citenamefont {Shen}(2006)}]{Wang2006}%
  \BibitemOpen
  \bibfield  {author} {\bibinfo {author} {\bibfnamefont {F.}~\bibnamefont {Wang}}\ and\ \bibinfo {author} {\bibfnamefont {Y.~R.}\ \bibnamefont {Shen}},\ }\bibfield  {title} {\bibinfo {title} {{General properties of local plasmons in metal nanostructures}},\ }\href {https://doi.org/10.1103/PhysRevLett.97.206806} {\bibfield  {journal} {\bibinfo  {journal} {Phys. Rev. Lett.}\ }\textbf {\bibinfo {volume} {97}},\ \bibinfo {pages} {1} (\bibinfo {year} {2006})}\BibitemShut {NoStop}%
\bibitem [{\citenamefont {Naik}\ \emph {et~al.}(2013)\citenamefont {Naik}, \citenamefont {Shalaev},\ and\ \citenamefont {Boltasseva}}]{Naik2013}%
  \BibitemOpen
  \bibfield  {author} {\bibinfo {author} {\bibfnamefont {G.~V.}\ \bibnamefont {Naik}}, \bibinfo {author} {\bibfnamefont {V.~M.}\ \bibnamefont {Shalaev}},\ and\ \bibinfo {author} {\bibfnamefont {A.}~\bibnamefont {Boltasseva}},\ }\bibfield  {title} {\bibinfo {title} {{Alternative plasmonic materials: Beyond gold and silver}},\ }\href {https://doi.org/10.1002/adma.201205076} {\bibfield  {journal} {\bibinfo  {journal} {Adv. Mater.}\ }\textbf {\bibinfo {volume} {25}},\ \bibinfo {pages} {3264} (\bibinfo {year} {2013})}\BibitemShut {NoStop}%
\bibitem [{\citenamefont {Khurgin}(2015)}]{Khurgin2015}%
  \BibitemOpen
  \bibfield  {author} {\bibinfo {author} {\bibfnamefont {J.~B.}\ \bibnamefont {Khurgin}},\ }\bibfield  {title} {\bibinfo {title} {{How to deal with the loss in plasmonics and metamaterials}},\ }\href {https://doi.org/10.1038/nnano.2014.310} {\bibfield  {journal} {\bibinfo  {journal} {Nat. Nanotechnol.}\ }\textbf {\bibinfo {volume} {10}},\ \bibinfo {pages} {2} (\bibinfo {year} {2015})}\BibitemShut {NoStop}%
\bibitem [{\citenamefont {Gurrieri}\ \emph {et~al.}(2024)\citenamefont {Gurrieri}, \citenamefont {Denning}, \citenamefont {Seegert}, \citenamefont {Kristensen},\ and\ \citenamefont {M{\o}rk}}]{Gurrieri2024}%
  \BibitemOpen
  \bibfield  {author} {\bibinfo {author} {\bibfnamefont {M.~V.}\ \bibnamefont {Gurrieri}}, \bibinfo {author} {\bibfnamefont {E.~V.}\ \bibnamefont {Denning}}, \bibinfo {author} {\bibfnamefont {K.}~\bibnamefont {Seegert}}, \bibinfo {author} {\bibfnamefont {P.~T.}\ \bibnamefont {Kristensen}},\ and\ \bibinfo {author} {\bibfnamefont {J.}~\bibnamefont {M{\o}rk}},\ }\bibfield  {title} {\bibinfo {title} {{Dynamics and condensation of polaritons in an optical nanocavity coupled to two-dimensional materials}},\ }\href {https://doi.org/10.1103/PhysRevB.109.155432} {\bibfield  {journal} {\bibinfo  {journal} {Phys. Rev. B}\ }\textbf {\bibinfo {volume} {109}},\ \bibinfo {pages} {1} (\bibinfo {year} {2024})}\BibitemShut {NoStop}%
\bibitem [{\citenamefont {Abutoama}\ \emph {et~al.}(2024)\citenamefont {Abutoama}, \citenamefont {Kountouris}, \citenamefont {M{\o}rk},\ and\ \citenamefont {Kristensen}}]{Abutoama2024}%
  \BibitemOpen
  \bibfield  {author} {\bibinfo {author} {\bibfnamefont {M.}~\bibnamefont {Abutoama}}, \bibinfo {author} {\bibfnamefont {G.}~\bibnamefont {Kountouris}}, \bibinfo {author} {\bibfnamefont {J.}~\bibnamefont {M{\o}rk}},\ and\ \bibinfo {author} {\bibfnamefont {P.~T.}\ \bibnamefont {Kristensen}},\ }\bibfield  {title} {\bibinfo {title} {{Modal approach to the coupling strength of quantum emitters in electromagnetic resonators}},\ }\href {https://doi.org/10.1103/PhysRevB.110.195434} {\bibfield  {journal} {\bibinfo  {journal} {Phys. Rev. B}\ }\textbf {\bibinfo {volume} {110}},\ \bibinfo {pages} {195434} (\bibinfo {year} {2024})}\BibitemShut {NoStop}%
\bibitem [{\citenamefont {Dong}\ \emph {et~al.}()\citenamefont {Dong}, \citenamefont {Babar}, \citenamefont {Christiansen}, \citenamefont {Hansen}, \citenamefont {Stobbe}, \citenamefont {Yu},\ and\ \citenamefont {M{\o}rk}}]{Dong2024}%
  \BibitemOpen
  \bibfield  {author} {\bibinfo {author} {\bibfnamefont {G.}~\bibnamefont {Dong}}, \bibinfo {author} {\bibfnamefont {A.~N.}\ \bibnamefont {Babar}}, \bibinfo {author} {\bibfnamefont {R.~E.}\ \bibnamefont {Christiansen}}, \bibinfo {author} {\bibfnamefont {S.~E.}\ \bibnamefont {Hansen}}, \bibinfo {author} {\bibfnamefont {S.}~\bibnamefont {Stobbe}}, \bibinfo {author} {\bibfnamefont {Y.}~\bibnamefont {Yu}},\ and\ \bibinfo {author} {\bibfnamefont {J.}~\bibnamefont {M{\o}rk}},\ }\bibfield  {title} {\bibinfo {title} {{Enhancement and speed-up of carrier dynamics in a dielectric nanocavity with deep sub-wavelength confinement}},\ }\Eprint {https://arxiv.org/abs/2412.08471} {arXiv:2412.08471} \BibitemShut {NoStop}%
\bibitem [{\citenamefont {Kountouris}\ \emph {et~al.}(2024)\citenamefont {Kountouris}, \citenamefont {Darket}, \citenamefont {Vestergaard}, \citenamefont {Denning}, \citenamefont {M{\o}rk},\ and\ \citenamefont {Kristensen}}]{Kountouris2024}%
  \BibitemOpen
  \bibfield  {author} {\bibinfo {author} {\bibfnamefont {G.}~\bibnamefont {Kountouris}}, \bibinfo {author} {\bibfnamefont {A.~S.}\ \bibnamefont {Darket}}, \bibinfo {author} {\bibfnamefont {L.}~\bibnamefont {Vestergaard}}, \bibinfo {author} {\bibfnamefont {E.~V.}\ \bibnamefont {Denning}}, \bibinfo {author} {\bibfnamefont {J.}~\bibnamefont {M{\o}rk}},\ and\ \bibinfo {author} {\bibfnamefont {P.~T.}\ \bibnamefont {Kristensen}},\ }\bibfield  {title} {\bibinfo {title} {{Lithographically defined quantum dot with subwavelength confinement of light}},\ }\href {https://doi.org/10.1103/PhysRevB.110.L241301} {\bibfield  {journal} {\bibinfo  {journal} {Phys. Rev. B}\ }\textbf {\bibinfo {volume} {110}},\ \bibinfo {pages} {L241301} (\bibinfo {year} {2024})}\BibitemShut {NoStop}%
\bibitem [{\citenamefont {Xiong}\ \emph {et~al.}()\citenamefont {Xiong}, \citenamefont {Yu}, \citenamefont {Berdnikov}, \citenamefont {Borregaard}, \citenamefont {Dubr{\'{e}}}, \citenamefont {Semenova}, \citenamefont {Yvind},\ and\ \citenamefont {M{\o}rk}}]{Xiong2024a}%
  \BibitemOpen
  \bibfield  {author} {\bibinfo {author} {\bibfnamefont {M.}~\bibnamefont {Xiong}}, \bibinfo {author} {\bibfnamefont {Y.}~\bibnamefont {Yu}}, \bibinfo {author} {\bibfnamefont {Y.}~\bibnamefont {Berdnikov}}, \bibinfo {author} {\bibfnamefont {S.~K.}\ \bibnamefont {Borregaard}}, \bibinfo {author} {\bibfnamefont {A.~H.}\ \bibnamefont {Dubr{\'{e}}}}, \bibinfo {author} {\bibfnamefont {E.}~\bibnamefont {Semenova}}, \bibinfo {author} {\bibfnamefont {K.}~\bibnamefont {Yvind}},\ and\ \bibinfo {author} {\bibfnamefont {J.}~\bibnamefont {M{\o}rk}},\ }\bibfield  {title} {\bibinfo {title} {{A nanolaser with extreme dielectric confinement}},\ }\Eprint {https://arxiv.org/abs/2412.02844} {arXiv:2412.02844} \BibitemShut {NoStop}%
\bibitem [{\citenamefont {Momma}\ and\ \citenamefont {Izumi}(2011)}]{Momma2011}%
  \BibitemOpen
  \bibfield  {author} {\bibinfo {author} {\bibfnamefont {K.}~\bibnamefont {Momma}}\ and\ \bibinfo {author} {\bibfnamefont {F.}~\bibnamefont {Izumi}},\ }\bibfield  {title} {\bibinfo {title} {{VESTA 3 for three-dimensional visualization of crystal, volumetric and morphology data}},\ }\href {https://doi.org/10.1107/S0021889811038970} {\bibfield  {journal} {\bibinfo  {journal} {J. Appl. Crystallogr.}\ }\textbf {\bibinfo {volume} {44}},\ \bibinfo {pages} {1272} (\bibinfo {year} {2011})}\BibitemShut {NoStop}%
\bibitem [{\citenamefont {Jain}\ \emph {et~al.}(2013)\citenamefont {Jain}, \citenamefont {Ong}, \citenamefont {Hautier}, \citenamefont {Chen}, \citenamefont {Richards}, \citenamefont {Dacek}, \citenamefont {Cholia}, \citenamefont {Gunter}, \citenamefont {Skinner}, \citenamefont {Ceder},\ and\ \citenamefont {Persson}}]{Jain2013}%
  \BibitemOpen
  \bibfield  {author} {\bibinfo {author} {\bibfnamefont {A.}~\bibnamefont {Jain}}, \bibinfo {author} {\bibfnamefont {S.~P.}\ \bibnamefont {Ong}}, \bibinfo {author} {\bibfnamefont {G.}~\bibnamefont {Hautier}}, \bibinfo {author} {\bibfnamefont {W.}~\bibnamefont {Chen}}, \bibinfo {author} {\bibfnamefont {W.~D.}\ \bibnamefont {Richards}}, \bibinfo {author} {\bibfnamefont {S.}~\bibnamefont {Dacek}}, \bibinfo {author} {\bibfnamefont {S.}~\bibnamefont {Cholia}}, \bibinfo {author} {\bibfnamefont {D.}~\bibnamefont {Gunter}}, \bibinfo {author} {\bibfnamefont {D.}~\bibnamefont {Skinner}}, \bibinfo {author} {\bibfnamefont {G.}~\bibnamefont {Ceder}},\ and\ \bibinfo {author} {\bibfnamefont {K.~A.}\ \bibnamefont {Persson}},\ }\bibfield  {title} {\bibinfo {title} {{Commentary: The Materials Project: A materials genome approach to accelerating materials innovation}},\ }\href {https://doi.org/10.1063/1.4812323} {\bibfield  {journal} {\bibinfo  {journal} {APL Mater.}\ }\textbf {\bibinfo {volume} {1}},\  (\bibinfo {year}
  {2013})}\BibitemShut {NoStop}%
\bibitem [{\citenamefont {Besga}\ \emph {et~al.}(2015)\citenamefont {Besga}, \citenamefont {Vaneph}, \citenamefont {Reichel}, \citenamefont {Est{\`{e}}ve}, \citenamefont {Reinhard}, \citenamefont {Miguel-S{\'{a}}nchez}, \citenamefont {Imamoğlu},\ and\ \citenamefont {Volz}}]{Besga2015}%
  \BibitemOpen
  \bibfield  {author} {\bibinfo {author} {\bibfnamefont {B.}~\bibnamefont {Besga}}, \bibinfo {author} {\bibfnamefont {C.}~\bibnamefont {Vaneph}}, \bibinfo {author} {\bibfnamefont {J.}~\bibnamefont {Reichel}}, \bibinfo {author} {\bibfnamefont {J.}~\bibnamefont {Est{\`{e}}ve}}, \bibinfo {author} {\bibfnamefont {A.}~\bibnamefont {Reinhard}}, \bibinfo {author} {\bibfnamefont {J.}~\bibnamefont {Miguel-S{\'{a}}nchez}}, \bibinfo {author} {\bibfnamefont {A.}~\bibnamefont {Imamoğlu}},\ and\ \bibinfo {author} {\bibfnamefont {T.}~\bibnamefont {Volz}},\ }\bibfield  {title} {\bibinfo {title} {{Polariton Boxes in a Tunable Fiber Cavity}},\ }\href {https://doi.org/10.1103/PhysRevApplied.3.014008} {\bibfield  {journal} {\bibinfo  {journal} {Phys. Rev. Appl.}\ }\textbf {\bibinfo {volume} {3}},\ \bibinfo {pages} {014008} (\bibinfo {year} {2015})}\BibitemShut {NoStop}%
\bibitem [{\citenamefont {Mu{\~{n}}oz-Matutano}\ \emph {et~al.}(2019)\citenamefont {Mu{\~{n}}oz-Matutano}, \citenamefont {Wood}, \citenamefont {Johnsson}, \citenamefont {Vidal}, \citenamefont {Baragiola}, \citenamefont {Reinhard}, \citenamefont {Lema{\^{i}}tre}, \citenamefont {Bloch}, \citenamefont {Amo}, \citenamefont {Nogues}, \citenamefont {Besga}, \citenamefont {Richard},\ and\ \citenamefont {Volz}}]{Munoz-Matutano2019}%
  \BibitemOpen
  \bibfield  {author} {\bibinfo {author} {\bibfnamefont {G.}~\bibnamefont {Mu{\~{n}}oz-Matutano}}, \bibinfo {author} {\bibfnamefont {A.}~\bibnamefont {Wood}}, \bibinfo {author} {\bibfnamefont {M.}~\bibnamefont {Johnsson}}, \bibinfo {author} {\bibfnamefont {X.}~\bibnamefont {Vidal}}, \bibinfo {author} {\bibfnamefont {B.~Q.}\ \bibnamefont {Baragiola}}, \bibinfo {author} {\bibfnamefont {A.}~\bibnamefont {Reinhard}}, \bibinfo {author} {\bibfnamefont {A.}~\bibnamefont {Lema{\^{i}}tre}}, \bibinfo {author} {\bibfnamefont {J.}~\bibnamefont {Bloch}}, \bibinfo {author} {\bibfnamefont {A.}~\bibnamefont {Amo}}, \bibinfo {author} {\bibfnamefont {G.}~\bibnamefont {Nogues}}, \bibinfo {author} {\bibfnamefont {B.}~\bibnamefont {Besga}}, \bibinfo {author} {\bibfnamefont {M.}~\bibnamefont {Richard}},\ and\ \bibinfo {author} {\bibfnamefont {T.}~\bibnamefont {Volz}},\ }\bibfield  {title} {\bibinfo {title} {{Emergence of quantum correlations from interacting fibre-cavity polaritons}},\ }\href
  {https://doi.org/10.1038/s41563-019-0281-z} {\bibfield  {journal} {\bibinfo  {journal} {Nat. Mater.}\ }\textbf {\bibinfo {volume} {18}},\ \bibinfo {pages} {213} (\bibinfo {year} {2019})}\BibitemShut {NoStop}%
\bibitem [{\citenamefont {Denning}\ \emph {et~al.}(2022{\natexlab{a}})\citenamefont {Denning}, \citenamefont {Wubs}, \citenamefont {Stenger}, \citenamefont {M{\o}rk},\ and\ \citenamefont {Kristensen}}]{Denning2022a}%
  \BibitemOpen
  \bibfield  {author} {\bibinfo {author} {\bibfnamefont {E.~V.}\ \bibnamefont {Denning}}, \bibinfo {author} {\bibfnamefont {M.}~\bibnamefont {Wubs}}, \bibinfo {author} {\bibfnamefont {N.}~\bibnamefont {Stenger}}, \bibinfo {author} {\bibfnamefont {J.}~\bibnamefont {M{\o}rk}},\ and\ \bibinfo {author} {\bibfnamefont {P.~T.}\ \bibnamefont {Kristensen}},\ }\bibfield  {title} {\bibinfo {title} {{Quantum theory of two-dimensional materials coupled to electromagnetic resonators}},\ }\href {https://doi.org/10.1103/PhysRevB.105.085306} {\bibfield  {journal} {\bibinfo  {journal} {Phys. Rev. B}\ }\textbf {\bibinfo {volume} {105}},\ \bibinfo {pages} {85306} (\bibinfo {year} {2022}{\natexlab{a}})}\BibitemShut {NoStop}%
\bibitem [{\citenamefont {Bamba}\ \emph {et~al.}(2011)\citenamefont {Bamba}, \citenamefont {Imamoğlu}, \citenamefont {Carusotto},\ and\ \citenamefont {Ciuti}}]{Bamba2011}%
  \BibitemOpen
  \bibfield  {author} {\bibinfo {author} {\bibfnamefont {M.}~\bibnamefont {Bamba}}, \bibinfo {author} {\bibfnamefont {A.}~\bibnamefont {Imamoğlu}}, \bibinfo {author} {\bibfnamefont {I.}~\bibnamefont {Carusotto}},\ and\ \bibinfo {author} {\bibfnamefont {C.}~\bibnamefont {Ciuti}},\ }\bibfield  {title} {\bibinfo {title} {{Origin of strong photon antibunching in weakly nonlinear photonic molecules}},\ }\href {https://doi.org/10.1103/PhysRevA.83.021802} {\bibfield  {journal} {\bibinfo  {journal} {Phys. Rev. A}\ }\textbf {\bibinfo {volume} {83}},\ \bibinfo {pages} {21802} (\bibinfo {year} {2011})}\BibitemShut {NoStop}%
\bibitem [{\citenamefont {Ryou}\ \emph {et~al.}(2018)\citenamefont {Ryou}, \citenamefont {Rosser}, \citenamefont {Saxena}, \citenamefont {Fryett},\ and\ \citenamefont {Majumdar}}]{Ryou2018}%
  \BibitemOpen
  \bibfield  {author} {\bibinfo {author} {\bibfnamefont {A.}~\bibnamefont {Ryou}}, \bibinfo {author} {\bibfnamefont {D.}~\bibnamefont {Rosser}}, \bibinfo {author} {\bibfnamefont {A.}~\bibnamefont {Saxena}}, \bibinfo {author} {\bibfnamefont {T.}~\bibnamefont {Fryett}},\ and\ \bibinfo {author} {\bibfnamefont {A.}~\bibnamefont {Majumdar}},\ }\bibfield  {title} {\bibinfo {title} {{Strong photon antibunching in weakly nonlinear two-dimensional exciton-polaritons}},\ }\href {https://doi.org/10.1103/PhysRevB.97.235307} {\bibfield  {journal} {\bibinfo  {journal} {Phys. Rev. B}\ }\textbf {\bibinfo {volume} {97}},\ \bibinfo {pages} {235307} (\bibinfo {year} {2018})}\BibitemShut {NoStop}%
\bibitem [{\citenamefont {Denning}\ \emph {et~al.}(2022{\natexlab{b}})\citenamefont {Denning}, \citenamefont {Wubs}, \citenamefont {Stenger}, \citenamefont {M{\o}rk},\ and\ \citenamefont {Kristensen}}]{Denning2022}%
  \BibitemOpen
  \bibfield  {author} {\bibinfo {author} {\bibfnamefont {E.~V.}\ \bibnamefont {Denning}}, \bibinfo {author} {\bibfnamefont {M.}~\bibnamefont {Wubs}}, \bibinfo {author} {\bibfnamefont {N.}~\bibnamefont {Stenger}}, \bibinfo {author} {\bibfnamefont {J.}~\bibnamefont {M{\o}rk}},\ and\ \bibinfo {author} {\bibfnamefont {P.~T.}\ \bibnamefont {Kristensen}},\ }\bibfield  {title} {\bibinfo {title} {{Cavity-induced exciton localization and polariton blockade in two-dimensional semiconductors coupled to an electromagnetic resonator}},\ }\href {https://doi.org/10.1103/PhysRevResearch.4.L012020} {\bibfield  {journal} {\bibinfo  {journal} {Phys. Rev. Res.}\ }\textbf {\bibinfo {volume} {4}},\ \bibinfo {pages} {L012020} (\bibinfo {year} {2022}{\natexlab{b}})}\BibitemShut {NoStop}%
\bibitem [{\citenamefont {Kyriienko}\ \emph {et~al.}(2020)\citenamefont {Kyriienko}, \citenamefont {Krizhanovskii},\ and\ \citenamefont {Shelykh}}]{Kyriienko2020}%
  \BibitemOpen
  \bibfield  {author} {\bibinfo {author} {\bibfnamefont {O.}~\bibnamefont {Kyriienko}}, \bibinfo {author} {\bibfnamefont {D.~N.}\ \bibnamefont {Krizhanovskii}},\ and\ \bibinfo {author} {\bibfnamefont {I.~A.}\ \bibnamefont {Shelykh}},\ }\bibfield  {title} {\bibinfo {title} {{Nonlinear Quantum Optics with Trion Polaritons in 2D Monolayers: Conventional and Unconventional Photon Blockade}},\ }\href {https://doi.org/10.1103/PhysRevLett.125.197402} {\bibfield  {journal} {\bibinfo  {journal} {Phys. Rev. Lett.}\ }\textbf {\bibinfo {volume} {125}},\ \bibinfo {pages} {197402} (\bibinfo {year} {2020})}\BibitemShut {NoStop}%
\bibitem [{\citenamefont {Delteil}\ \emph {et~al.}(2019)\citenamefont {Delteil}, \citenamefont {Fink}, \citenamefont {Schade}, \citenamefont {H{\"{o}}fling}, \citenamefont {Schneider},\ and\ \citenamefont {İmamoğlu}}]{Delteil2019a}%
  \BibitemOpen
  \bibfield  {author} {\bibinfo {author} {\bibfnamefont {A.}~\bibnamefont {Delteil}}, \bibinfo {author} {\bibfnamefont {T.}~\bibnamefont {Fink}}, \bibinfo {author} {\bibfnamefont {A.}~\bibnamefont {Schade}}, \bibinfo {author} {\bibfnamefont {S.}~\bibnamefont {H{\"{o}}fling}}, \bibinfo {author} {\bibfnamefont {C.}~\bibnamefont {Schneider}},\ and\ \bibinfo {author} {\bibfnamefont {A.}~\bibnamefont {İmamoğlu}},\ }\bibfield  {title} {\bibinfo {title} {{Towards polariton blockade of confined exciton–polaritons}},\ }\href {https://doi.org/10.1038/s41563-019-0282-y https://www.nature.com/articles/s41563-019-0282-y} {\bibfield  {journal} {\bibinfo  {journal} {Nat. Mater.}\ }\textbf {\bibinfo {volume} {18}},\ \bibinfo {pages} {219} (\bibinfo {year} {2019})}\BibitemShut {NoStop}%
\bibitem [{\citenamefont {Xiao}\ \emph {et~al.}(2012)\citenamefont {Xiao}, \citenamefont {Liu}, \citenamefont {Feng}, \citenamefont {Xu},\ and\ \citenamefont {Yao}}]{Xiao2012}%
  \BibitemOpen
  \bibfield  {author} {\bibinfo {author} {\bibfnamefont {D.}~\bibnamefont {Xiao}}, \bibinfo {author} {\bibfnamefont {G.~B.}\ \bibnamefont {Liu}}, \bibinfo {author} {\bibfnamefont {W.}~\bibnamefont {Feng}}, \bibinfo {author} {\bibfnamefont {X.}~\bibnamefont {Xu}},\ and\ \bibinfo {author} {\bibfnamefont {W.}~\bibnamefont {Yao}},\ }\bibfield  {title} {\bibinfo {title} {{Coupled spin and valley physics in monolayers of MoS 2 and other group-VI dichalcogenides}},\ }\href {https://doi.org/10.1103/PhysRevLett.108.196802} {\bibfield  {journal} {\bibinfo  {journal} {Phys. Rev. Lett.}\ }\textbf {\bibinfo {volume} {108}},\ \bibinfo {pages} {1} (\bibinfo {year} {2012})}\BibitemShut {NoStop}%
\bibitem [{\citenamefont {Mak}\ \emph {et~al.}(2012)\citenamefont {Mak}, \citenamefont {He}, \citenamefont {Shan},\ and\ \citenamefont {Heinz}}]{Mak2012a}%
  \BibitemOpen
  \bibfield  {author} {\bibinfo {author} {\bibfnamefont {K.~F.}\ \bibnamefont {Mak}}, \bibinfo {author} {\bibfnamefont {K.}~\bibnamefont {He}}, \bibinfo {author} {\bibfnamefont {J.}~\bibnamefont {Shan}},\ and\ \bibinfo {author} {\bibfnamefont {T.~F.}\ \bibnamefont {Heinz}},\ }\bibfield  {title} {\bibinfo {title} {{Control of valley polarization in monolayer MoS2 by optical helicity}},\ }\href {https://doi.org/10.1038/nnano.2012.96} {\bibfield  {journal} {\bibinfo  {journal} {Nat. Nanotechnol.}\ }\textbf {\bibinfo {volume} {7}},\ \bibinfo {pages} {494} (\bibinfo {year} {2012})}\BibitemShut {NoStop}%
\bibitem [{\citenamefont {Yang}\ \emph {et~al.}(2015)\citenamefont {Yang}, \citenamefont {Sinitsyn}, \citenamefont {Chen}, \citenamefont {Yuan}, \citenamefont {Zhang}, \citenamefont {Lou},\ and\ \citenamefont {Crooker}}]{Yang2015a}%
  \BibitemOpen
  \bibfield  {author} {\bibinfo {author} {\bibfnamefont {L.}~\bibnamefont {Yang}}, \bibinfo {author} {\bibfnamefont {N.~A.}\ \bibnamefont {Sinitsyn}}, \bibinfo {author} {\bibfnamefont {W.}~\bibnamefont {Chen}}, \bibinfo {author} {\bibfnamefont {J.}~\bibnamefont {Yuan}}, \bibinfo {author} {\bibfnamefont {J.}~\bibnamefont {Zhang}}, \bibinfo {author} {\bibfnamefont {J.}~\bibnamefont {Lou}},\ and\ \bibinfo {author} {\bibfnamefont {S.~A.}\ \bibnamefont {Crooker}},\ }\bibfield  {title} {\bibinfo {title} {{Long-lived nanosecond spin relaxation and spin coherence of electrons in monolayer MoS2 and WS2}},\ }\href {https://doi.org/10.1038/nphys3419} {\bibfield  {journal} {\bibinfo  {journal} {Nat. Phys.}\ }\textbf {\bibinfo {volume} {11}},\ \bibinfo {pages} {830} (\bibinfo {year} {2015})}\BibitemShut {NoStop}%
\bibitem [{\citenamefont {Fong}\ \emph {et~al.}(2021)\citenamefont {Fong}, \citenamefont {Ota}, \citenamefont {Arakawa}, \citenamefont {Iwamoto},\ and\ \citenamefont {Kato}}]{Fong2021}%
  \BibitemOpen
  \bibfield  {author} {\bibinfo {author} {\bibfnamefont {C.~F.}\ \bibnamefont {Fong}}, \bibinfo {author} {\bibfnamefont {Y.}~\bibnamefont {Ota}}, \bibinfo {author} {\bibfnamefont {Y.}~\bibnamefont {Arakawa}}, \bibinfo {author} {\bibfnamefont {S.}~\bibnamefont {Iwamoto}},\ and\ \bibinfo {author} {\bibfnamefont {Y.~K.}\ \bibnamefont {Kato}},\ }\bibfield  {title} {\bibinfo {title} {{Chiral modes near exceptional points in symmetry broken H1 photonic crystal cavities}},\ }\href {https://doi.org/10.1103/PhysRevResearch.3.043096} {\bibfield  {journal} {\bibinfo  {journal} {Phys. Rev. Res.}\ }\textbf {\bibinfo {volume} {3}},\ \bibinfo {pages} {043096} (\bibinfo {year} {2021})}\BibitemShut {NoStop}%
\bibitem [{\citenamefont {Cheben}\ \emph {et~al.}(2018)\citenamefont {Cheben}, \citenamefont {Halir}, \citenamefont {Schmid}, \citenamefont {Atwater},\ and\ \citenamefont {Smith}}]{Cheben2018}%
  \BibitemOpen
  \bibfield  {author} {\bibinfo {author} {\bibfnamefont {P.}~\bibnamefont {Cheben}}, \bibinfo {author} {\bibfnamefont {R.}~\bibnamefont {Halir}}, \bibinfo {author} {\bibfnamefont {J.~H.}\ \bibnamefont {Schmid}}, \bibinfo {author} {\bibfnamefont {H.~A.}\ \bibnamefont {Atwater}},\ and\ \bibinfo {author} {\bibfnamefont {D.~R.}\ \bibnamefont {Smith}},\ }\bibfield  {title} {\bibinfo {title} {{Subwavelength integrated photonics}},\ }\href {https://doi.org/10.1038/s41586-018-0421-7} {\bibfield  {journal} {\bibinfo  {journal} {Nature}\ }\textbf {\bibinfo {volume} {560}},\ \bibinfo {pages} {565} (\bibinfo {year} {2018})}\BibitemShut {NoStop}%
\bibitem [{\citenamefont {Wang}\ \emph {et~al.}(2020)\citenamefont {Wang}, \citenamefont {Sciarrino}, \citenamefont {Laing},\ and\ \citenamefont {Thompson}}]{Wang2020}%
  \BibitemOpen
  \bibfield  {author} {\bibinfo {author} {\bibfnamefont {J.}~\bibnamefont {Wang}}, \bibinfo {author} {\bibfnamefont {F.}~\bibnamefont {Sciarrino}}, \bibinfo {author} {\bibfnamefont {A.}~\bibnamefont {Laing}},\ and\ \bibinfo {author} {\bibfnamefont {M.~G.}\ \bibnamefont {Thompson}},\ }\bibfield  {title} {\bibinfo {title} {{Integrated photonic quantum technologies}},\ }\href {https://doi.org/10.1038/s41566-019-0532-1} {\bibfield  {journal} {\bibinfo  {journal} {Nat. Photonics}\ }\textbf {\bibinfo {volume} {14}},\ \bibinfo {pages} {273} (\bibinfo {year} {2020})}\BibitemShut {NoStop}%
\bibitem [{\citenamefont {Chen}\ \emph {et~al.}(2015)\citenamefont {Chen}, \citenamefont {Sahin}, \citenamefont {Suslu}, \citenamefont {Ding}, \citenamefont {Bertoni}, \citenamefont {Peeters},\ and\ \citenamefont {Tongay}}]{Chen2015}%
  \BibitemOpen
  \bibfield  {author} {\bibinfo {author} {\bibfnamefont {B.}~\bibnamefont {Chen}}, \bibinfo {author} {\bibfnamefont {H.}~\bibnamefont {Sahin}}, \bibinfo {author} {\bibfnamefont {A.}~\bibnamefont {Suslu}}, \bibinfo {author} {\bibfnamefont {L.}~\bibnamefont {Ding}}, \bibinfo {author} {\bibfnamefont {M.~I.}\ \bibnamefont {Bertoni}}, \bibinfo {author} {\bibfnamefont {F.~M.}\ \bibnamefont {Peeters}},\ and\ \bibinfo {author} {\bibfnamefont {S.}~\bibnamefont {Tongay}},\ }\bibfield  {title} {\bibinfo {title} {{Environmental Changes in MoTe 2 Excitonic Dynamics by Defects-Activated Molecular Interaction}},\ }\href {https://doi.org/10.1021/acsnano.5b00985} {\bibfield  {journal} {\bibinfo  {journal} {ACS Nano}\ }\textbf {\bibinfo {volume} {9}},\ \bibinfo {pages} {5326} (\bibinfo {year} {2015})}\BibitemShut {NoStop}%
\bibitem [{\citenamefont {Ajayi}\ \emph {et~al.}(2017)\citenamefont {Ajayi}, \citenamefont {Ardelean}, \citenamefont {Shepard}, \citenamefont {Wang}, \citenamefont {Antony}, \citenamefont {Taniguchi}, \citenamefont {Watanabe}, \citenamefont {Heinz}, \citenamefont {Strauf}, \citenamefont {Zhu},\ and\ \citenamefont {Hone}}]{Ajayi2017}%
  \BibitemOpen
  \bibfield  {author} {\bibinfo {author} {\bibfnamefont {O.~A.}\ \bibnamefont {Ajayi}}, \bibinfo {author} {\bibfnamefont {J.~V.}\ \bibnamefont {Ardelean}}, \bibinfo {author} {\bibfnamefont {G.~D.}\ \bibnamefont {Shepard}}, \bibinfo {author} {\bibfnamefont {J.}~\bibnamefont {Wang}}, \bibinfo {author} {\bibfnamefont {A.}~\bibnamefont {Antony}}, \bibinfo {author} {\bibfnamefont {T.}~\bibnamefont {Taniguchi}}, \bibinfo {author} {\bibfnamefont {K.}~\bibnamefont {Watanabe}}, \bibinfo {author} {\bibfnamefont {T.~F.}\ \bibnamefont {Heinz}}, \bibinfo {author} {\bibfnamefont {S.}~\bibnamefont {Strauf}}, \bibinfo {author} {\bibfnamefont {X.~Y.}\ \bibnamefont {Zhu}},\ and\ \bibinfo {author} {\bibfnamefont {J.~C.}\ \bibnamefont {Hone}},\ }\bibfield  {title} {\bibinfo {title} {{Approaching the intrinsic photoluminescence linewidth in transition metal dichalcogenide monolayers}},\ }\href {https://doi.org/10.1088/2053-1583/aa6aa1} {\bibfield  {journal} {\bibinfo  {journal} {2D Mater.}\ }\textbf {\bibinfo {volume} {4}},\
  (\bibinfo {year} {2017})}\BibitemShut {NoStop}%
\bibitem [{\citenamefont {Cadiz}\ \emph {et~al.}(2017)\citenamefont {Cadiz}, \citenamefont {Courtade}, \citenamefont {Robert}, \citenamefont {Wang}, \citenamefont {Shen}, \citenamefont {Cai}, \citenamefont {Taniguchi}, \citenamefont {Watanabe}, \citenamefont {Carrere}, \citenamefont {Lagarde}, \citenamefont {Manca}, \citenamefont {Amand}, \citenamefont {Renucci}, \citenamefont {Tongay}, \citenamefont {Marie},\ and\ \citenamefont {Urbaszek}}]{Cadiz2017}%
  \BibitemOpen
  \bibfield  {author} {\bibinfo {author} {\bibfnamefont {F.}~\bibnamefont {Cadiz}}, \bibinfo {author} {\bibfnamefont {E.}~\bibnamefont {Courtade}}, \bibinfo {author} {\bibfnamefont {C.}~\bibnamefont {Robert}}, \bibinfo {author} {\bibfnamefont {G.}~\bibnamefont {Wang}}, \bibinfo {author} {\bibfnamefont {Y.}~\bibnamefont {Shen}}, \bibinfo {author} {\bibfnamefont {H.}~\bibnamefont {Cai}}, \bibinfo {author} {\bibfnamefont {T.}~\bibnamefont {Taniguchi}}, \bibinfo {author} {\bibfnamefont {K.}~\bibnamefont {Watanabe}}, \bibinfo {author} {\bibfnamefont {H.}~\bibnamefont {Carrere}}, \bibinfo {author} {\bibfnamefont {D.}~\bibnamefont {Lagarde}}, \bibinfo {author} {\bibfnamefont {M.}~\bibnamefont {Manca}}, \bibinfo {author} {\bibfnamefont {T.}~\bibnamefont {Amand}}, \bibinfo {author} {\bibfnamefont {P.}~\bibnamefont {Renucci}}, \bibinfo {author} {\bibfnamefont {S.}~\bibnamefont {Tongay}}, \bibinfo {author} {\bibfnamefont {X.}~\bibnamefont {Marie}},\ and\ \bibinfo {author} {\bibfnamefont {B.}~\bibnamefont {Urbaszek}},\
  }\bibfield  {title} {\bibinfo {title} {{Excitonic Linewidth Approaching the Homogeneous Limit in MoS2-Based van der Waals Heterostructures}},\ }\href {https://doi.org/10.1103/PhysRevX.7.021026} {\bibfield  {journal} {\bibinfo  {journal} {Phys. Rev. X}\ }\textbf {\bibinfo {volume} {7}},\ \bibinfo {pages} {021026} (\bibinfo {year} {2017})}\BibitemShut {NoStop}%
\bibitem [{\citenamefont {Han}\ \emph {et~al.}(2018)\citenamefont {Han}, \citenamefont {Robert}, \citenamefont {Courtade}, \citenamefont {Manca}, \citenamefont {Shree}, \citenamefont {Amand}, \citenamefont {Renucci}, \citenamefont {Taniguchi}, \citenamefont {Watanabe}, \citenamefont {Marie}, \citenamefont {Golub}, \citenamefont {Glazov},\ and\ \citenamefont {Urbaszek}}]{Han2018}%
  \BibitemOpen
  \bibfield  {author} {\bibinfo {author} {\bibfnamefont {B.}~\bibnamefont {Han}}, \bibinfo {author} {\bibfnamefont {C.}~\bibnamefont {Robert}}, \bibinfo {author} {\bibfnamefont {E.}~\bibnamefont {Courtade}}, \bibinfo {author} {\bibfnamefont {M.}~\bibnamefont {Manca}}, \bibinfo {author} {\bibfnamefont {S.}~\bibnamefont {Shree}}, \bibinfo {author} {\bibfnamefont {T.}~\bibnamefont {Amand}}, \bibinfo {author} {\bibfnamefont {P.}~\bibnamefont {Renucci}}, \bibinfo {author} {\bibfnamefont {T.}~\bibnamefont {Taniguchi}}, \bibinfo {author} {\bibfnamefont {K.}~\bibnamefont {Watanabe}}, \bibinfo {author} {\bibfnamefont {X.}~\bibnamefont {Marie}}, \bibinfo {author} {\bibfnamefont {L.~E.}\ \bibnamefont {Golub}}, \bibinfo {author} {\bibfnamefont {M.~M.}\ \bibnamefont {Glazov}},\ and\ \bibinfo {author} {\bibfnamefont {B.}~\bibnamefont {Urbaszek}},\ }\bibfield  {title} {\bibinfo {title} {{Exciton States in Monolayer MoSe2 and MoTe2 Probed by Upconversion Spectroscopy}},\ }\href {https://doi.org/10.1103/PhysRevX.8.031073}
  {\bibfield  {journal} {\bibinfo  {journal} {Phys. Rev. X}\ }\textbf {\bibinfo {volume} {8}},\ \bibinfo {pages} {1} (\bibinfo {year} {2018})}\BibitemShut {NoStop}%
\bibitem [{\citenamefont {Kountouris}\ \emph {et~al.}(2022)\citenamefont {Kountouris}, \citenamefont {M{\o}rk}, \citenamefont {Denning},\ and\ \citenamefont {Kristensen}}]{Kountouris2022}%
  \BibitemOpen
  \bibfield  {author} {\bibinfo {author} {\bibfnamefont {G.}~\bibnamefont {Kountouris}}, \bibinfo {author} {\bibfnamefont {J.}~\bibnamefont {M{\o}rk}}, \bibinfo {author} {\bibfnamefont {E.~V.}\ \bibnamefont {Denning}},\ and\ \bibinfo {author} {\bibfnamefont {P.~T.}\ \bibnamefont {Kristensen}},\ }\bibfield  {title} {\bibinfo {title} {{Modal properties of dielectric bowtie cavities with deep sub-wavelength confinement}},\ }\href {https://doi.org/10.1364/OE.472793} {\bibfield  {journal} {\bibinfo  {journal} {Opt. Express}\ }\textbf {\bibinfo {volume} {30}},\ \bibinfo {pages} {40367} (\bibinfo {year} {2022})}\BibitemShut {NoStop}%
\bibitem [{\citenamefont {Ching}\ \emph {et~al.}(1998)\citenamefont {Ching}, \citenamefont {Leung}, \citenamefont {{Maassen van den Brink}}, \citenamefont {Suen}, \citenamefont {Tong},\ and\ \citenamefont {Young}}]{Ching1998}%
  \BibitemOpen
  \bibfield  {author} {\bibinfo {author} {\bibfnamefont {E.~S.~C.}\ \bibnamefont {Ching}}, \bibinfo {author} {\bibfnamefont {P.~T.}\ \bibnamefont {Leung}}, \bibinfo {author} {\bibfnamefont {A.}~\bibnamefont {{Maassen van den Brink}}}, \bibinfo {author} {\bibfnamefont {W.~M.}\ \bibnamefont {Suen}}, \bibinfo {author} {\bibfnamefont {S.~S.}\ \bibnamefont {Tong}},\ and\ \bibinfo {author} {\bibfnamefont {K.}~\bibnamefont {Young}},\ }\bibfield  {title} {\bibinfo {title} {{Quasinormal-mode expansion for waves in open systems}},\ }\href {https://doi.org/10.1103/RevModPhys.70.1545} {\bibfield  {journal} {\bibinfo  {journal} {Rev. Mod. Phys.}\ }\textbf {\bibinfo {volume} {70}},\ \bibinfo {pages} {1545} (\bibinfo {year} {1998})}\BibitemShut {NoStop}%
\bibitem [{\citenamefont {Muljarov}\ \emph {et~al.}(2010)\citenamefont {Muljarov}, \citenamefont {Langbein},\ and\ \citenamefont {Zimmermann}}]{Muljarov2010}%
  \BibitemOpen
  \bibfield  {author} {\bibinfo {author} {\bibfnamefont {E.~A.}\ \bibnamefont {Muljarov}}, \bibinfo {author} {\bibfnamefont {W.}~\bibnamefont {Langbein}},\ and\ \bibinfo {author} {\bibfnamefont {R.}~\bibnamefont {Zimmermann}},\ }\bibfield  {title} {\bibinfo {title} {{Brillouin-Wigner perturbation theory in open electromagnetic systems}},\ }\href {https://doi.org/10.1209/0295-5075/92/50010} {\bibfield  {journal} {\bibinfo  {journal} {EPL (Europhysics Lett.}\ }\textbf {\bibinfo {volume} {92}},\ \bibinfo {pages} {50010} (\bibinfo {year} {2010})}\BibitemShut {NoStop}%
\bibitem [{\citenamefont {Kristensen}\ and\ \citenamefont {Hughes}(2013)}]{Kristensen2013}%
  \BibitemOpen
  \bibfield  {author} {\bibinfo {author} {\bibfnamefont {P.~T.}\ \bibnamefont {Kristensen}}\ and\ \bibinfo {author} {\bibfnamefont {S.}~\bibnamefont {Hughes}},\ }\bibfield  {title} {\bibinfo {title} {{Modes and Mode Volumes of Leaky Optical Cavities and Plasmonic Nanoresonators}},\ }\href {https://doi.org/10.1021/ph400114e} {\bibfield  {journal} {\bibinfo  {journal} {ACS Photonics}\ }\textbf {\bibinfo {volume} {1}},\ \bibinfo {pages} {2} (\bibinfo {year} {2013})}\BibitemShut {NoStop}%
\bibitem [{\citenamefont {Lalanne}\ \emph {et~al.}(2018)\citenamefont {Lalanne}, \citenamefont {Yan}, \citenamefont {Vynck}, \citenamefont {Sauvan},\ and\ \citenamefont {Hugonin}}]{Lalanne2018}%
  \BibitemOpen
  \bibfield  {author} {\bibinfo {author} {\bibfnamefont {P.}~\bibnamefont {Lalanne}}, \bibinfo {author} {\bibfnamefont {W.}~\bibnamefont {Yan}}, \bibinfo {author} {\bibfnamefont {K.}~\bibnamefont {Vynck}}, \bibinfo {author} {\bibfnamefont {C.}~\bibnamefont {Sauvan}},\ and\ \bibinfo {author} {\bibfnamefont {J.}~\bibnamefont {Hugonin}},\ }\bibfield  {title} {\bibinfo {title} {{Light Interaction with Photonic and Plasmonic Resonances}},\ }\href {https://doi.org/10.1002/lpor.201700113} {\bibfield  {journal} {\bibinfo  {journal} {Laser Photon. Rev.}\ }\textbf {\bibinfo {volume} {12}},\ \bibinfo {pages} {1} (\bibinfo {year} {2018})}\BibitemShut {NoStop}%
\bibitem [{\citenamefont {Kristensen}\ \emph {et~al.}(2020)\citenamefont {Kristensen}, \citenamefont {Herrmann}, \citenamefont {Intravaia},\ and\ \citenamefont {Busch}}]{Kristensen2020}%
  \BibitemOpen
  \bibfield  {author} {\bibinfo {author} {\bibfnamefont {P.~T.}\ \bibnamefont {Kristensen}}, \bibinfo {author} {\bibfnamefont {K.}~\bibnamefont {Herrmann}}, \bibinfo {author} {\bibfnamefont {F.}~\bibnamefont {Intravaia}},\ and\ \bibinfo {author} {\bibfnamefont {K.}~\bibnamefont {Busch}},\ }\bibfield  {title} {\bibinfo {title} {{Modeling electromagnetic resonators using quasinormal modes}},\ }\href {https://doi.org/10.1364/AOP.377940} {\bibfield  {journal} {\bibinfo  {journal} {Adv. Opt. Photonics}\ }\textbf {\bibinfo {volume} {12}},\ \bibinfo {pages} {612} (\bibinfo {year} {2020})}\BibitemShut {NoStop}%
\bibitem [{\citenamefont {Both}\ and\ \citenamefont {Weiss}(2021)}]{Both2022}%
  \BibitemOpen
  \bibfield  {author} {\bibinfo {author} {\bibfnamefont {S.}~\bibnamefont {Both}}\ and\ \bibinfo {author} {\bibfnamefont {T.}~\bibnamefont {Weiss}},\ }\bibfield  {title} {\bibinfo {title} {{Resonant states and their role in nanophotonics}},\ }\href {https://doi.org/10.1088/1361-6641/ac3290} {\bibfield  {journal} {\bibinfo  {journal} {Semicond. Sci. Technol.}\ }\textbf {\bibinfo {volume} {37}},\ \bibinfo {pages} {013002} (\bibinfo {year} {2021})}\BibitemShut {NoStop}%
\bibitem [{\citenamefont {Dimopoulos}\ \emph {et~al.}(2022)\citenamefont {Dimopoulos}, \citenamefont {Sakanas}, \citenamefont {Marchevsky}, \citenamefont {Xiong}, \citenamefont {Yu}, \citenamefont {Semenova}, \citenamefont {M{\o}rk},\ and\ \citenamefont {Yvind}}]{Dimopoulos2022}%
  \BibitemOpen
  \bibfield  {author} {\bibinfo {author} {\bibfnamefont {E.}~\bibnamefont {Dimopoulos}}, \bibinfo {author} {\bibfnamefont {A.}~\bibnamefont {Sakanas}}, \bibinfo {author} {\bibfnamefont {A.}~\bibnamefont {Marchevsky}}, \bibinfo {author} {\bibfnamefont {M.}~\bibnamefont {Xiong}}, \bibinfo {author} {\bibfnamefont {Y.}~\bibnamefont {Yu}}, \bibinfo {author} {\bibfnamefont {E.}~\bibnamefont {Semenova}}, \bibinfo {author} {\bibfnamefont {J.}~\bibnamefont {M{\o}rk}},\ and\ \bibinfo {author} {\bibfnamefont {K.}~\bibnamefont {Yvind}},\ }\bibfield  {title} {\bibinfo {title} {{Electrically‐Driven Photonic Crystal Lasers with Ultra‐low Threshold}},\ }\href {https://doi.org/10.1002/lpor.202200109} {\bibfield  {journal} {\bibinfo  {journal} {Laser Photon. Rev.}\ }\textbf {\bibinfo {volume} {16}},\ \bibinfo {pages} {1} (\bibinfo {year} {2022})}\BibitemShut {NoStop}%
\bibitem [{\citenamefont {Schr{\"{o}}der}\ \emph {et~al.}(2025)\citenamefont {Schr{\"{o}}der}, \citenamefont {van Exter}, \citenamefont {Xiong}, \citenamefont {Kountouris}, \citenamefont {Wubs}, \citenamefont {Kristensen},\ and\ \citenamefont {Stenger}}]{Schroder2025a}%
  \BibitemOpen
  \bibfield  {author} {\bibinfo {author} {\bibfnamefont {F.}~\bibnamefont {Schr{\"{o}}der}}, \bibinfo {author} {\bibfnamefont {M.~P.}\ \bibnamefont {van Exter}}, \bibinfo {author} {\bibfnamefont {M.}~\bibnamefont {Xiong}}, \bibinfo {author} {\bibfnamefont {G.}~\bibnamefont {Kountouris}}, \bibinfo {author} {\bibfnamefont {M.}~\bibnamefont {Wubs}}, \bibinfo {author} {\bibfnamefont {P.~T.}\ \bibnamefont {Kristensen}},\ and\ \bibinfo {author} {\bibfnamefont {N.}~\bibnamefont {Stenger}},\ }\bibfield  {title} {\bibinfo {title} {{Confocal polarization tomography of dielectric nanocavities}},\ }\bibfield  {journal} {\bibinfo  {journal} {Nanophotonics}\ }\href {https://doi.org/10.1515/nanoph-2024-0744} {10.1515/nanoph-2024-0744} (\bibinfo {year} {2025})\BibitemShut {NoStop}%
\bibitem [{\citenamefont {Carlson}\ \emph {et~al.}(2021)\citenamefont {Carlson}, \citenamefont {Salzwedel}, \citenamefont {Selig}, \citenamefont {Knorr},\ and\ \citenamefont {Hughes}}]{Carlson2021}%
  \BibitemOpen
  \bibfield  {author} {\bibinfo {author} {\bibfnamefont {C.}~\bibnamefont {Carlson}}, \bibinfo {author} {\bibfnamefont {R.}~\bibnamefont {Salzwedel}}, \bibinfo {author} {\bibfnamefont {M.}~\bibnamefont {Selig}}, \bibinfo {author} {\bibfnamefont {A.}~\bibnamefont {Knorr}},\ and\ \bibinfo {author} {\bibfnamefont {S.}~\bibnamefont {Hughes}},\ }\bibfield  {title} {\bibinfo {title} {{Strong coupling regime and hybrid quasinormal modes from a single plasmonic resonator coupled to a transition metal dichalcogenide monolayer}},\ }\href {https://doi.org/10.1103/PhysRevB.104.125424} {\bibfield  {journal} {\bibinfo  {journal} {Phys. Rev. B}\ }\textbf {\bibinfo {volume} {104}},\ \bibinfo {pages} {125424} (\bibinfo {year} {2021})}\BibitemShut {NoStop}%
\bibitem [{\citenamefont {Pettit}\ and\ \citenamefont {Turner}(1965)}]{Pettit1965}%
  \BibitemOpen
  \bibfield  {author} {\bibinfo {author} {\bibfnamefont {G.~D.}\ \bibnamefont {Pettit}}\ and\ \bibinfo {author} {\bibfnamefont {W.~J.}\ \bibnamefont {Turner}},\ }\bibfield  {title} {\bibinfo {title} {{Refractive Index of InP}},\ }\href {https://doi.org/10.1063/1.1714410} {\bibfield  {journal} {\bibinfo  {journal} {J. Appl. Phys.}\ }\textbf {\bibinfo {volume} {36}},\ \bibinfo {pages} {2081} (\bibinfo {year} {1965})}\BibitemShut {NoStop}%
\bibitem [{\citenamefont {McCaulley}\ \emph {et~al.}(1994)\citenamefont {McCaulley}, \citenamefont {Donnelly}, \citenamefont {Vernon},\ and\ \citenamefont {Taha}}]{McCaulley1994}%
  \BibitemOpen
  \bibfield  {author} {\bibinfo {author} {\bibfnamefont {J.~A.}\ \bibnamefont {McCaulley}}, \bibinfo {author} {\bibfnamefont {V.~M.}\ \bibnamefont {Donnelly}}, \bibinfo {author} {\bibfnamefont {M.}~\bibnamefont {Vernon}},\ and\ \bibinfo {author} {\bibfnamefont {I.}~\bibnamefont {Taha}},\ }\bibfield  {title} {\bibinfo {title} {{Temperature dependence ofthe near-infrared refractive index of silicon, gallium arsenide, and indium phosphide}},\ }\href@noop {} {\bibfield  {journal} {\bibinfo  {journal} {Phys. Rev. B}\ }\textbf {\bibinfo {volume} {49}},\ \bibinfo {pages} {7408} (\bibinfo {year} {1994})}\BibitemShut {NoStop}%
\bibitem [{\citenamefont {Gerber}\ and\ \citenamefont {Marie}(2018)}]{Gerber2018}%
  \BibitemOpen
  \bibfield  {author} {\bibinfo {author} {\bibfnamefont {I.~C.}\ \bibnamefont {Gerber}}\ and\ \bibinfo {author} {\bibfnamefont {X.}~\bibnamefont {Marie}},\ }\bibfield  {title} {\bibinfo {title} {{Dependence of band structure and exciton properties of encapsulated WSe2 monolayers on the hBN-layer thickness}},\ }\href {https://doi.org/10.1103/PhysRevB.98.245126} {\bibfield  {journal} {\bibinfo  {journal} {Phys. Rev. B}\ }\textbf {\bibinfo {volume} {98}},\ \bibinfo {pages} {245126} (\bibinfo {year} {2018})}\BibitemShut {NoStop}%
\bibitem [{\citenamefont {Meckbach}\ \emph {et~al.}(2020)\citenamefont {Meckbach}, \citenamefont {Hader}, \citenamefont {Huttner}, \citenamefont {Neuhaus}, \citenamefont {Steiner}, \citenamefont {Stroucken}, \citenamefont {Moloney},\ and\ \citenamefont {Koch}}]{Meckbach2020}%
  \BibitemOpen
  \bibfield  {author} {\bibinfo {author} {\bibfnamefont {L.}~\bibnamefont {Meckbach}}, \bibinfo {author} {\bibfnamefont {J.}~\bibnamefont {Hader}}, \bibinfo {author} {\bibfnamefont {U.}~\bibnamefont {Huttner}}, \bibinfo {author} {\bibfnamefont {J.}~\bibnamefont {Neuhaus}}, \bibinfo {author} {\bibfnamefont {J.~T.}\ \bibnamefont {Steiner}}, \bibinfo {author} {\bibfnamefont {T.}~\bibnamefont {Stroucken}}, \bibinfo {author} {\bibfnamefont {J.~V.}\ \bibnamefont {Moloney}},\ and\ \bibinfo {author} {\bibfnamefont {S.~W.}\ \bibnamefont {Koch}},\ }\bibfield  {title} {\bibinfo {title} {{Ultrafast band-gap renormalization and build-up of optical gain in monolayer MoTe2}},\ }\href {https://doi.org/10.1103/PhysRevB.101.075401} {\bibfield  {journal} {\bibinfo  {journal} {Phys. Rev. B}\ }\textbf {\bibinfo {volume} {101}},\ \bibinfo {pages} {1} (\bibinfo {year} {2020})}\BibitemShut {NoStop}%
\bibitem [{\citenamefont {Edalati-Boostan}\ \emph {et~al.}(2020)\citenamefont {Edalati-Boostan}, \citenamefont {Cocchi},\ and\ \citenamefont {Draxl}}]{Edalati-Boostan2020}%
  \BibitemOpen
  \bibfield  {author} {\bibinfo {author} {\bibfnamefont {S.}~\bibnamefont {Edalati-Boostan}}, \bibinfo {author} {\bibfnamefont {C.}~\bibnamefont {Cocchi}},\ and\ \bibinfo {author} {\bibfnamefont {C.}~\bibnamefont {Draxl}},\ }\bibfield  {title} {\bibinfo {title} {{MoTe2 as a natural hyperbolic material across the visible and the ultraviolet region}},\ }\href {https://doi.org/10.1103/PhysRevMaterials.4.085202} {\bibfield  {journal} {\bibinfo  {journal} {Phys. Rev. Mater.}\ }\textbf {\bibinfo {volume} {4}},\ \bibinfo {pages} {1} (\bibinfo {year} {2020})}\BibitemShut {NoStop}%
\bibitem [{\citenamefont {Grudinin}\ \emph {et~al.}(2023)\citenamefont {Grudinin}, \citenamefont {Ermolaev}, \citenamefont {Baranov}, \citenamefont {Toksumakov}, \citenamefont {Voronin}, \citenamefont {Slavich}, \citenamefont {Vyshnevyy}, \citenamefont {Mazitov}, \citenamefont {Kruglov}, \citenamefont {Ghazaryan}, \citenamefont {Arsenin}, \citenamefont {Novoselov},\ and\ \citenamefont {Volkov}}]{Grudinin2023}%
  \BibitemOpen
  \bibfield  {author} {\bibinfo {author} {\bibfnamefont {D.~V.}\ \bibnamefont {Grudinin}}, \bibinfo {author} {\bibfnamefont {G.~A.}\ \bibnamefont {Ermolaev}}, \bibinfo {author} {\bibfnamefont {D.~G.}\ \bibnamefont {Baranov}}, \bibinfo {author} {\bibfnamefont {A.~N.}\ \bibnamefont {Toksumakov}}, \bibinfo {author} {\bibfnamefont {K.~V.}\ \bibnamefont {Voronin}}, \bibinfo {author} {\bibfnamefont {A.~S.}\ \bibnamefont {Slavich}}, \bibinfo {author} {\bibfnamefont {A.~A.}\ \bibnamefont {Vyshnevyy}}, \bibinfo {author} {\bibfnamefont {A.~B.}\ \bibnamefont {Mazitov}}, \bibinfo {author} {\bibfnamefont {I.~A.}\ \bibnamefont {Kruglov}}, \bibinfo {author} {\bibfnamefont {D.~A.}\ \bibnamefont {Ghazaryan}}, \bibinfo {author} {\bibfnamefont {A.~V.}\ \bibnamefont {Arsenin}}, \bibinfo {author} {\bibfnamefont {K.~S.}\ \bibnamefont {Novoselov}},\ and\ \bibinfo {author} {\bibfnamefont {V.~S.}\ \bibnamefont {Volkov}},\ }\bibfield  {title} {\bibinfo {title} {{Hexagonal boron nitride nanophotonics: a record-breaking material for the
  ultraviolet and visible spectral ranges}},\ }\href {https://doi.org/10.1039/D3MH00215B} {\bibfield  {journal} {\bibinfo  {journal} {Mater. Horizons}\ }\textbf {\bibinfo {volume} {10}},\ \bibinfo {pages} {2427} (\bibinfo {year} {2023})}\BibitemShut {NoStop}%
\bibitem [{\citenamefont {Kutrowska-Girzycka}\ \emph {et~al.}(2022)\citenamefont {Kutrowska-Girzycka}, \citenamefont {Zieba-Ost{\'{o}}j}, \citenamefont {Biega{\'{n}}ska}, \citenamefont {Florian}, \citenamefont {Steinhoff}, \citenamefont {Rogowicz}, \citenamefont {Mrowi{\'{n}}ski}, \citenamefont {Watanabe}, \citenamefont {Taniguchi}, \citenamefont {Gies}, \citenamefont {Tongay}, \citenamefont {Schneider},\ and\ \citenamefont {Syperek}}]{Kutrowska-Girzycka2022}%
  \BibitemOpen
  \bibfield  {author} {\bibinfo {author} {\bibfnamefont {J.}~\bibnamefont {Kutrowska-Girzycka}}, \bibinfo {author} {\bibfnamefont {E.}~\bibnamefont {Zieba-Ost{\'{o}}j}}, \bibinfo {author} {\bibfnamefont {D.}~\bibnamefont {Biega{\'{n}}ska}}, \bibinfo {author} {\bibfnamefont {M.}~\bibnamefont {Florian}}, \bibinfo {author} {\bibfnamefont {A.}~\bibnamefont {Steinhoff}}, \bibinfo {author} {\bibfnamefont {E.}~\bibnamefont {Rogowicz}}, \bibinfo {author} {\bibfnamefont {P.}~\bibnamefont {Mrowi{\'{n}}ski}}, \bibinfo {author} {\bibfnamefont {K.}~\bibnamefont {Watanabe}}, \bibinfo {author} {\bibfnamefont {T.}~\bibnamefont {Taniguchi}}, \bibinfo {author} {\bibfnamefont {C.}~\bibnamefont {Gies}}, \bibinfo {author} {\bibfnamefont {S.}~\bibnamefont {Tongay}}, \bibinfo {author} {\bibfnamefont {C.}~\bibnamefont {Schneider}},\ and\ \bibinfo {author} {\bibfnamefont {M.}~\bibnamefont {Syperek}},\ }\bibfield  {title} {\bibinfo {title} {{Exploring the effect of dielectric screening on neutral and charged-exciton properties in
  monolayer and bilayer MoTe2}},\ }\href {https://doi.org/10.1063/5.0089192} {\bibfield  {journal} {\bibinfo  {journal} {Appl. Phys. Rev.}\ }\textbf {\bibinfo {volume} {9}},\ \bibinfo {pages} {041410} (\bibinfo {year} {2022})}\BibitemShut {NoStop}%
\bibitem [{\citenamefont {Andreani}\ \emph {et~al.}(1999)\citenamefont {Andreani}, \citenamefont {Panzarini},\ and\ \citenamefont {G{\'{e}}rard}}]{Andreani1999}%
  \BibitemOpen
  \bibfield  {author} {\bibinfo {author} {\bibfnamefont {L.~C.}\ \bibnamefont {Andreani}}, \bibinfo {author} {\bibfnamefont {G.}~\bibnamefont {Panzarini}},\ and\ \bibinfo {author} {\bibfnamefont {J.-M.}\ \bibnamefont {G{\'{e}}rard}},\ }\bibfield  {title} {\bibinfo {title} {{Strong-coupling regime for quantum boxes in pillar microcavities: Theory}},\ }\href {https://doi.org/10.1103/PhysRevB.60.13276 M4 - Citavi} {\bibfield  {journal} {\bibinfo  {journal} {Phys. Rev. B}\ }\textbf {\bibinfo {volume} {60}},\ \bibinfo {pages} {13276} (\bibinfo {year} {1999})}\BibitemShut {NoStop}%
\bibitem [{\citenamefont {Todisco}\ \emph {et~al.}(2020)\citenamefont {Todisco}, \citenamefont {Malureanu}, \citenamefont {Wolff}, \citenamefont {Gon{\c{c}}alves}, \citenamefont {Roberts}, \citenamefont {Mortensen},\ and\ \citenamefont {Tserkezis}}]{Todisco2020}%
  \BibitemOpen
  \bibfield  {author} {\bibinfo {author} {\bibfnamefont {F.}~\bibnamefont {Todisco}}, \bibinfo {author} {\bibfnamefont {R.}~\bibnamefont {Malureanu}}, \bibinfo {author} {\bibfnamefont {C.}~\bibnamefont {Wolff}}, \bibinfo {author} {\bibfnamefont {P.~A.~D.}\ \bibnamefont {Gon{\c{c}}alves}}, \bibinfo {author} {\bibfnamefont {A.~S.}\ \bibnamefont {Roberts}}, \bibinfo {author} {\bibfnamefont {N.~A.}\ \bibnamefont {Mortensen}},\ and\ \bibinfo {author} {\bibfnamefont {C.}~\bibnamefont {Tserkezis}},\ }\bibfield  {title} {\bibinfo {title} {{Magnetic and electric Mie-exciton polaritons in silicon nanodisks}},\ }\href {https://doi.org/10.1515/nanoph-2019-0444} {\bibfield  {journal} {\bibinfo  {journal} {Nanophotonics}\ }\textbf {\bibinfo {volume} {9}},\ \bibinfo {pages} {803} (\bibinfo {year} {2020})}\BibitemShut {NoStop}%
\bibitem [{\citenamefont {Bleu}\ \emph {et~al.}(2024)\citenamefont {Bleu}, \citenamefont {Choo}, \citenamefont {Levinsen},\ and\ \citenamefont {Parish}}]{Bleu2024}%
  \BibitemOpen
  \bibfield  {author} {\bibinfo {author} {\bibfnamefont {O.}~\bibnamefont {Bleu}}, \bibinfo {author} {\bibfnamefont {K.}~\bibnamefont {Choo}}, \bibinfo {author} {\bibfnamefont {J.}~\bibnamefont {Levinsen}},\ and\ \bibinfo {author} {\bibfnamefont {M.~M.}\ \bibnamefont {Parish}},\ }\bibfield  {title} {\bibinfo {title} {{Dissipative light-matter coupling and anomalous dispersion in nonideal cavities}},\ }\href {https://doi.org/10.1103/PhysRevA.109.023707} {\bibfield  {journal} {\bibinfo  {journal} {Phys. Rev. A}\ }\textbf {\bibinfo {volume} {109}},\ \bibinfo {pages} {1} (\bibinfo {year} {2024})}\BibitemShut {NoStop}%
\bibitem [{\citenamefont {Helmrich}\ \emph {et~al.}(2018)\citenamefont {Helmrich}, \citenamefont {Schneider}, \citenamefont {Achtstein}, \citenamefont {Arora}, \citenamefont {Herzog}, \citenamefont {de~Vasconcellos}, \citenamefont {Kolarczik}, \citenamefont {Sch{\"{o}}ps}, \citenamefont {Bratschitsch}, \citenamefont {Woggon},\ and\ \citenamefont {Owschimikow}}]{Helmrich2018}%
  \BibitemOpen
  \bibfield  {author} {\bibinfo {author} {\bibfnamefont {S.}~\bibnamefont {Helmrich}}, \bibinfo {author} {\bibfnamefont {R.}~\bibnamefont {Schneider}}, \bibinfo {author} {\bibfnamefont {A.~W.}\ \bibnamefont {Achtstein}}, \bibinfo {author} {\bibfnamefont {A.}~\bibnamefont {Arora}}, \bibinfo {author} {\bibfnamefont {B.}~\bibnamefont {Herzog}}, \bibinfo {author} {\bibfnamefont {S.~M.}\ \bibnamefont {de~Vasconcellos}}, \bibinfo {author} {\bibfnamefont {M.}~\bibnamefont {Kolarczik}}, \bibinfo {author} {\bibfnamefont {O.}~\bibnamefont {Sch{\"{o}}ps}}, \bibinfo {author} {\bibfnamefont {R.}~\bibnamefont {Bratschitsch}}, \bibinfo {author} {\bibfnamefont {U.}~\bibnamefont {Woggon}},\ and\ \bibinfo {author} {\bibfnamefont {N.}~\bibnamefont {Owschimikow}},\ }\bibfield  {title} {\bibinfo {title} {{Exciton–phonon coupling in mono- and bilayer MoTe2}},\ }\href {https://doi.org/10.1088/2053-1583/aacfb7} {\bibfield  {journal} {\bibinfo  {journal} {2D Mater.}\ }\textbf {\bibinfo {volume} {5}},\ \bibinfo {pages} {45007}
  (\bibinfo {year} {2018})}\BibitemShut {NoStop}%
\bibitem [{\citenamefont {Gorbachev}\ \emph {et~al.}(2011)\citenamefont {Gorbachev}, \citenamefont {Riaz}, \citenamefont {Nair}, \citenamefont {Jalil}, \citenamefont {Britnell}, \citenamefont {Belle}, \citenamefont {Hill}, \citenamefont {Novoselov}, \citenamefont {Watanabe}, \citenamefont {Taniguchi}, \citenamefont {Geim},\ and\ \citenamefont {Blake}}]{Gorbachev2011a}%
  \BibitemOpen
  \bibfield  {author} {\bibinfo {author} {\bibfnamefont {R.~V.}\ \bibnamefont {Gorbachev}}, \bibinfo {author} {\bibfnamefont {I.}~\bibnamefont {Riaz}}, \bibinfo {author} {\bibfnamefont {R.~R.}\ \bibnamefont {Nair}}, \bibinfo {author} {\bibfnamefont {R.}~\bibnamefont {Jalil}}, \bibinfo {author} {\bibfnamefont {L.}~\bibnamefont {Britnell}}, \bibinfo {author} {\bibfnamefont {B.~D.}\ \bibnamefont {Belle}}, \bibinfo {author} {\bibfnamefont {E.~W.}\ \bibnamefont {Hill}}, \bibinfo {author} {\bibfnamefont {K.~S.}\ \bibnamefont {Novoselov}}, \bibinfo {author} {\bibfnamefont {K.}~\bibnamefont {Watanabe}}, \bibinfo {author} {\bibfnamefont {T.}~\bibnamefont {Taniguchi}}, \bibinfo {author} {\bibfnamefont {A.~K.}\ \bibnamefont {Geim}},\ and\ \bibinfo {author} {\bibfnamefont {P.}~\bibnamefont {Blake}},\ }\bibfield  {title} {\bibinfo {title} {{Hunting for Monolayer Boron Nitride: Optical and Raman Signatures}},\ }\href {https://doi.org/10.1002/smll.201001628} {\bibfield  {journal} {\bibinfo  {journal} {Small}\ }\textbf
  {\bibinfo {volume} {7}},\ \bibinfo {pages} {465} (\bibinfo {year} {2011})}\BibitemShut {NoStop}%
\bibitem [{\citenamefont {Holm}(2024)}]{Holm2024}%
  \BibitemOpen
  \bibfield  {author} {\bibinfo {author} {\bibfnamefont {P.}~\bibnamefont {Holm}},\ }\emph {\bibinfo {title} {{Atomically precise patterning by thermal scanning probe lithography}}},\ \href@noop {} {\bibinfo {type} {Master's thesis}},\ \bibinfo  {school} {Technical University of Denmark} (\bibinfo {year} {2024})\BibitemShut {NoStop}%
\bibitem [{\citenamefont {Castellanos-Gomez}\ \emph {et~al.}(2014)\citenamefont {Castellanos-Gomez}, \citenamefont {Buscema}, \citenamefont {Molenaar}, \citenamefont {Singh}, \citenamefont {Janssen}, \citenamefont {{Van Der Zant}},\ and\ \citenamefont {Steele}}]{Castellanos-Gomez2014}%
  \BibitemOpen
  \bibfield  {author} {\bibinfo {author} {\bibfnamefont {A.}~\bibnamefont {Castellanos-Gomez}}, \bibinfo {author} {\bibfnamefont {M.}~\bibnamefont {Buscema}}, \bibinfo {author} {\bibfnamefont {R.}~\bibnamefont {Molenaar}}, \bibinfo {author} {\bibfnamefont {V.}~\bibnamefont {Singh}}, \bibinfo {author} {\bibfnamefont {L.}~\bibnamefont {Janssen}}, \bibinfo {author} {\bibfnamefont {H.~S.}\ \bibnamefont {{Van Der Zant}}},\ and\ \bibinfo {author} {\bibfnamefont {G.~A.}\ \bibnamefont {Steele}},\ }\bibfield  {title} {\bibinfo {title} {{Deterministic transfer of two-dimensional materials by all-dry viscoelastic stamping}},\ }\bibfield  {journal} {\bibinfo  {journal} {2D Mater.}\ }\textbf {\bibinfo {volume} {1}},\ \href {https://doi.org/10.1088/2053-1583/1/1/011002} {10.1088/2053-1583/1/1/011002} (\bibinfo {year} {2014})\BibitemShut {NoStop}%
\bibitem [{\citenamefont {Zomer}\ \emph {et~al.}(2014)\citenamefont {Zomer}, \citenamefont {Guimar{\~{a}}es}, \citenamefont {Brant}, \citenamefont {Tombros},\ and\ \citenamefont {{Van Wees}}}]{Zomer2014}%
  \BibitemOpen
  \bibfield  {author} {\bibinfo {author} {\bibfnamefont {P.~J.}\ \bibnamefont {Zomer}}, \bibinfo {author} {\bibfnamefont {M.~H.}\ \bibnamefont {Guimar{\~{a}}es}}, \bibinfo {author} {\bibfnamefont {J.~C.}\ \bibnamefont {Brant}}, \bibinfo {author} {\bibfnamefont {N.}~\bibnamefont {Tombros}},\ and\ \bibinfo {author} {\bibfnamefont {B.~J.}\ \bibnamefont {{Van Wees}}},\ }\bibfield  {title} {\bibinfo {title} {{Fast pick up technique for high quality heterostructures of bilayer graphene and hexagonal boron nitride}},\ }\bibfield  {journal} {\bibinfo  {journal} {Appl. Phys. Lett.}\ }\textbf {\bibinfo {volume} {105}},\ \href {https://doi.org/10.1063/1.4886096} {10.1063/1.4886096} (\bibinfo {year} {2014})\BibitemShut {NoStop}%
\bibitem [{\citenamefont {Pizzocchero}\ \emph {et~al.}(2016)\citenamefont {Pizzocchero}, \citenamefont {Gammelgaard}, \citenamefont {Jessen}, \citenamefont {Caridad}, \citenamefont {Wang}, \citenamefont {Hone}, \citenamefont {B{\o}ggild},\ and\ \citenamefont {Booth}}]{Pizzocchero2016}%
  \BibitemOpen
  \bibfield  {author} {\bibinfo {author} {\bibfnamefont {F.}~\bibnamefont {Pizzocchero}}, \bibinfo {author} {\bibfnamefont {L.}~\bibnamefont {Gammelgaard}}, \bibinfo {author} {\bibfnamefont {B.~S.}\ \bibnamefont {Jessen}}, \bibinfo {author} {\bibfnamefont {J.~M.}\ \bibnamefont {Caridad}}, \bibinfo {author} {\bibfnamefont {L.}~\bibnamefont {Wang}}, \bibinfo {author} {\bibfnamefont {J.}~\bibnamefont {Hone}}, \bibinfo {author} {\bibfnamefont {P.}~\bibnamefont {B{\o}ggild}},\ and\ \bibinfo {author} {\bibfnamefont {T.~J.}\ \bibnamefont {Booth}},\ }\bibfield  {title} {\bibinfo {title} {{The hot pick-up technique for batch assembly of van der Waals heterostructures}},\ }\bibfield  {journal} {\bibinfo  {journal} {Nat. Commun.}\ }\textbf {\bibinfo {volume} {7}},\ \href {https://doi.org/10.1038/ncomms11894} {10.1038/ncomms11894} (\bibinfo {year} {2016})\BibitemShut {NoStop}%
\bibitem [{\citenamefont {Andrews}(1997)}]{Andrews1997}%
  \BibitemOpen
  \bibfield  {author} {\bibinfo {author} {\bibfnamefont {L.~C.}\ \bibnamefont {Andrews}},\ }\href {https://doi.org/10.1117/3.270709} {\emph {\bibinfo {title} {{Special Functions of Mathematics for Engineers}}}},\ \bibinfo {edition} {2nd}\ ed.\ (\bibinfo  {publisher} {SPIE—The International Society for Optical Engineering and Oxford University Press},\ \bibinfo {year} {1997})\BibitemShut {NoStop}%
\bibitem [{\citenamefont {Kristensen}\ \emph {et~al.}(2015)\citenamefont {Kristensen}, \citenamefont {Ge},\ and\ \citenamefont {Hughes}}]{Kristensen2015}%
  \BibitemOpen
  \bibfield  {author} {\bibinfo {author} {\bibfnamefont {P.~T.}\ \bibnamefont {Kristensen}}, \bibinfo {author} {\bibfnamefont {R.-C.}\ \bibnamefont {Ge}},\ and\ \bibinfo {author} {\bibfnamefont {S.}~\bibnamefont {Hughes}},\ }\bibfield  {title} {\bibinfo {title} {{Normalization of quasinormal modes in leaky optical cavities and plasmonic resonators}},\ }\href {https://doi.org/10.1103/PhysRevA.92.053810} {\bibfield  {journal} {\bibinfo  {journal} {Phys. Rev. A}\ }\textbf {\bibinfo {volume} {92}},\ \bibinfo {pages} {053810} (\bibinfo {year} {2015})}\BibitemShut {NoStop}%
\bibitem [{\citenamefont {Haastrup}\ \emph {et~al.}(2018)\citenamefont {Haastrup}, \citenamefont {Strange}, \citenamefont {Pandey}, \citenamefont {Deilmann}, \citenamefont {Schmidt}, \citenamefont {Hinsche}, \citenamefont {Gjerding}, \citenamefont {Torelli}, \citenamefont {Larsen}, \citenamefont {Riis-Jensen}, \citenamefont {Gath}, \citenamefont {Jacobsen}, \citenamefont {{J{\o}rgen Mortensen}}, \citenamefont {Olsen},\ and\ \citenamefont {Thygesen}}]{Haastrup2018}%
  \BibitemOpen
  \bibfield  {author} {\bibinfo {author} {\bibfnamefont {S.}~\bibnamefont {Haastrup}}, \bibinfo {author} {\bibfnamefont {M.}~\bibnamefont {Strange}}, \bibinfo {author} {\bibfnamefont {M.}~\bibnamefont {Pandey}}, \bibinfo {author} {\bibfnamefont {T.}~\bibnamefont {Deilmann}}, \bibinfo {author} {\bibfnamefont {P.~S.}\ \bibnamefont {Schmidt}}, \bibinfo {author} {\bibfnamefont {N.~F.}\ \bibnamefont {Hinsche}}, \bibinfo {author} {\bibfnamefont {M.~N.}\ \bibnamefont {Gjerding}}, \bibinfo {author} {\bibfnamefont {D.}~\bibnamefont {Torelli}}, \bibinfo {author} {\bibfnamefont {P.~M.}\ \bibnamefont {Larsen}}, \bibinfo {author} {\bibfnamefont {A.~C.}\ \bibnamefont {Riis-Jensen}}, \bibinfo {author} {\bibfnamefont {J.}~\bibnamefont {Gath}}, \bibinfo {author} {\bibfnamefont {K.~W.}\ \bibnamefont {Jacobsen}}, \bibinfo {author} {\bibfnamefont {J.}~\bibnamefont {{J{\o}rgen Mortensen}}}, \bibinfo {author} {\bibfnamefont {T.}~\bibnamefont {Olsen}},\ and\ \bibinfo {author} {\bibfnamefont {K.~S.}\ \bibnamefont {Thygesen}},\
  }\bibfield  {title} {\bibinfo {title} {{The Computational 2D Materials Database: high-throughput modeling and discovery of atomically thin crystals}},\ }\href {https://doi.org/10.1088/2053-1583/aacfc1} {\bibfield  {journal} {\bibinfo  {journal} {2D Mater.}\ }\textbf {\bibinfo {volume} {5}},\ \bibinfo {pages} {042002} (\bibinfo {year} {2018})}\BibitemShut {NoStop}%
\bibitem [{\citenamefont {Hu}\ and\ \citenamefont {Fei}(2020)}]{Hu2020}%
  \BibitemOpen
  \bibfield  {author} {\bibinfo {author} {\bibfnamefont {F.}~\bibnamefont {Hu}}\ and\ \bibinfo {author} {\bibfnamefont {Z.}~\bibnamefont {Fei}},\ }\bibfield  {title} {\bibinfo {title} {{Recent Progress on Exciton Polaritons in Layered Transition‐Metal Dichalcogenides}},\ }\href {https://doi.org/10.1002/adom.201901003} {\bibfield  {journal} {\bibinfo  {journal} {Adv. Opt. Mater.}\ }\textbf {\bibinfo {volume} {8}},\ \bibinfo {pages} {1} (\bibinfo {year} {2020})}\BibitemShut {NoStop}%
\end{thebibliography}%
\end{document}